\documentclass[]{gji}
\usepackage{graphicx}
\usepackage{timet}
\usepackage{amsmath}
\usepackage{amssymb}
\usepackage{multicol}
\usepackage{sistyle}
\usepackage{subcaption}
\usepackage{color}

\let\bibhang\relax
\usepackage{natbib}
\title[Regimes of rotating convection in an experimental model of the Earth's Tangent Cylinder]
  {Regimes of rotating convection in an experimental model\\of the Earth's tangent cylinder}
\author[Rishav Agrawal, Martin Holdsworth, Alban Poth\'{e}rat]
  {Rishav Agrawal\textsuperscript{1}, Martin Holdsworth\textsuperscript{2} and Alban Poth\'{e}rat\textsuperscript{2*}\\
  \textsuperscript{1} School of Engineering, University of Liverpool, UK \\
	\textsuperscript{2} Centre for Fluid and Complex systems, Coventry University, UK }

\date{Received XXX; in original form XXX}
\pagerange{\pageref{firstpage}--\pageref{lastpage}}
\volume{xx}
\pubyear{xx}

\usepackage[percent]{overpic}
\usepackage{sistyle}
\usepackage{xcolor}

\usepackage{rotating}
\usepackage{tikz}
\usetikzlibrary{arrows.meta}

\newcommand{\unitz}{\boldsymbol{\hat e}_z}
\newcommand{\Ek}{E\!k}
\newcommand{\Ra}{R\!a}
\newcommand{\Ro}{R\!o}

\newcommand{\Nu}{N\!u}
\newcommand{\Prt}{P\!r}
\newcommand{\Fr}{F\!r}
\newcommand{\Bi}{B\!i}
\newcommand{\Rt}{\widetilde{R\!a}}
\newcommand{\Rtwns}{\widetilde{R\!a}_{\rm ws}}

\newcommand{\vel}{\boldsymbol u}
\newcommand{\up}{\boldsymbol u_\perp}

\newcommand{\Fp}{\boldsymbol F_{\rm P}}
\newcommand{\Fc}{\boldsymbol F_{\rm C}}
\newcommand{\Fi}{\boldsymbol F_{\rm I}}
\newcommand{\Fv}{\boldsymbol F_{\rm V}}

\newcommand{\mIn}{\boldsymbol m_{\rm In}}
\newcommand{\mOut}{\boldsymbol m_{\rm Out}}
\newcommand{\Pconv}{\boldsymbol P_{\rm conv}}
\begin{document}

\label{firstpage}

\maketitle

\begin{summary}

The flow in the liquid core of the Earth is controlled by the interplay between buoyancy and the Coriolis force due to planetary rotation. Fast rotation imposes the Taylor-Proudman Constraint (TPC) that opposes fluid motion across an imaginary cylindrical surface called the Tangent Cylinder (TC) obtained by extruding the equatorial perimeter of the solid inner core along the rotation direction, and up to the core-mantle boundary (CMB). Because this boundary is imaginary, however, mass and heat transfer through the TC, as well as the regimes of convection within the TC may differ from those inside a solid cylinder with well-defined boundary conditions. To date however, the influence of this peculiar boundary is unknown and this impedes our understanding of the flow in the polar regions of the core. To clarify this question, we reproduce the TC geometry experimentally in the Little Earth Experiment 2 (LEE2), where the CMB is modelled as a cold, cylindrical vessel filled with water, with a hot cylinder inside it acting as the inner solid core. The vessel is filled with water so as to optically map the velocity field in regimes of criticality and rotational constraint consistent with those of the Earth (Rayleigh number up to $\Rt=191$ times critical and ratio of inertial to Coriolis forces in the range $10^{-3}$ to $10^{-1}$). We find that although the regimes of convection within the TC broadly resemble those within a cylinder with solid impermeable insulating walls (CSIIW), the TC boundary introduces differences of particular significance in the context of the Earth's core. 

	The main new mechanism arises out of the {inertia} near the cold lateral boundary of the vessel, which drives inertia at the outer boundary of the TC, as convection in the equatorial regions of the Earth's core does. The baroclinicity just outside the TC suppresses the classical wall modes found in CSIIW and the inertia there causes an early breakup of the TPC at the TC boundary. The breakup appears locally at low criticality $\Rt\simeq3$ and then over the entire TC surface at $\Rt\simeq36$. The flow remains dominated by the Coriolis force even up to $\Rt=191$, but because of inertia near the TC boundary, geostrophic turbulence appears at much lower criticality than in other settings ($\Rt\simeq36$ instead of typically $\Rt\simeq80$). It also promotes the emergence of a few large structures, of similar topology to the flux patches observed in geomagnetic data. The heat flux escapes increasingly through the TC boundary as the TPC there becomes weaker. It eventually bypasses the axial heat flux and homogenises temperature laterally. This translates into a diffusivity-free scaling $\Nu^*\sim (\Ra_q^*)^{1/3}$ independent of rotation. Hence inertia driven by baroclinicity outside the TC provides a convenient shortcut to geostrophic turbulence, which is otherwise difficult to reach in experiments. These results also highlight a process whereby the convection outside the TC may control turbulence inside it and bypass the axial heat transfer. We finally discuss how Earth's conditions, especially its magnetic field may change how this process acts within the Earth's core.

\end{summary}


\section{Introduction}
\label{sec:introduction}

The flow in the outer core of the Earth is often modelled as being driven by convection due to a radial temperature gradient, representing either the super-adiabatic temperature or compositional gradient there \citep{cardin1992grl,cardin2015tg}. The dominating force is, however, the Coriolis force due to planetary rotation, rather than buoyancy.
The main effect of rotation is to suppress both variations of physical quantities along the rotation axis, and fluid motion that are not solenoidal in planes normal to it. The Taylor-Proudman Constraint (TPC) expresses that to remain 2D and 2D solenoidal in the limit of fast rotation, the flow must follow geostrophic contours aligned with surfaces intercepting the core that are of constant depth along the rotation axis \citep{proudman1916prsa,taylor1917prsa,greenspan1969}.
Since planetary interiors are practically axisymmetric, these surfaces are cylinders aligned with the rotation axis. In the case of the Earth and other planets with a solid core, the \emph{Tangent Cylinder} (TC) extruded from the solid core along the rotation axis plays a prominent role: because the depth of fluid just inside the TC is double of that just outside it, the TPC imposes that in the regimes of extremely fast rotation of the Earth, the TC should act as an impermeable mechanical barrier preventing any exchange of mass between these two regions. Indeed the polar regions inside the TC and the equatorial regions outside it feature very different convective flows: in the equatorial region, the nonlinear dynamics of geostrophic ``Busse" columns \citep{busse1970jfm} is driven by radial buoyancy in the equatorial plane and these columns are stretched to the Core-Mantle Boundary (CMB) as a result of the TPC, where their evolution is constrained by the curvature of the CMB (see reviews by \cite{olson2011pepi,olson2013areps,cardin2015tg,potherat2024crphys}). In the polar regions within the TC, by contrast, gravity and rotation are mostly aligned and the topographic effects are less pronounced. \textcolor{black}{It follows that the convection there may resemble Rotating Rayleigh-B\'enard Convection (RRBC) between two parallel boundaries normal to both temperature gradient and gravity and is expected to set in at higher values of the Rayleigh numbers $Ra=\alpha g (T_{\rm ICB}-T_{\rm CMB})H^3/(\nu\kappa)$, where $g$ is a typical gravity in the outer core, $\alpha$, $\nu$ and $\kappa$ are the thermal expansion coefficient, viscosity and thermal diffusivity of the liquid metal there, $T_{\rm ICB}-T_{\rm CMB}$ is the super-adiabatic temperature difference between the Inner Core Boundary (ICB) and the CMB, and $H$ is the outer core thickness (see the recent reviews by \cite{kunnen2021} and \cite{eckeshishkina2023}).}
{Numerical simulations in spherical shells showed that this difference in the nature of convection inside and outside the TC persists in turbulent regimes far from onset 
\citep{konoroberts2002,gubbinsemilio2007,schaefferetal2017}}. 
{Evidence for the TC in the Earth's core was found in geomagnetic data from the 1980 Magsat and 2000 Oersed missions, showing a strong westward swirl peaking at $\simeq\SI{1.5}{deg/yr}$ at the high latitudes that coincide with the TC \citep{hulot2002_nat} and confirmed by the flow inferred from the 1999-2021 Swarm satellite data \citep{gillet2022pnas}.}
Yet, there is mounting evidence from geomagnetic data that the TPC is violated at the Earth's TC boundary and that the flow there may locally exhibit a significant radial component \citep{pais2008gji,pais2014gji,gillet2015jgr,gillet2019gji,finlay2023nat}: A strong planetary gyre connects the upper region of the TC to the equatorial region near the CMB. This gyre is currently accelerating as part of what is believed to be an oscillating process \citep{livermore2017natgeo}.

Since the TPC relies mainly on the assumption that only pressure forces can balance the Coriolis forces, another force must be able to compete to induce
these flows to invalidate it. 
The Ekman number, based on the Earth rotation $\Omega$, the outer core thickness $H$ and the viscosity of the liquid metal therein $\nu$, that measures their ratio to the Coriolis force is of the order of $\Ek=\nu/(2\Omega H) \simeq10^{-15}$, where $\Omega$ is the background rotation \citep{schubert2011pepi}, so viscous forces are too insignificant to play a role in breaking the TPC. The two main possibilities are the Lorentz forces due to the electric currents within the core interacting with the Earth's magnetic field, and inertia, driven by buoyancy.
\cite{livermore2017natgeo} argue that the oscillatory nature of the gyre and the short timescale of these oscillations points to magnetohydrodynamic waves or temporal variations of the magnetic field. 
Indeed magnetohydrodynamic waves provide a mechanism for fast oscillations in the core \citep{gillet2010nat,gillet2022pnas} and the Lorentz force due to the Earth magnetic field can induce velocities across the TC that break the TPC \citep{hollerbach1994_pf,dormy1998_epsl, sakuraba2002gafd,cao2018pnas,hotta2018apjl,potherat2024prl}.

Whether or not any of these ideas explains the flow across the Earth's TC, the question of whether buoyancy drives sufficient inertia to overcome the TPC is yet to be clarified. First, the Convective Rossby number
$\Ro=U_{\rm ff}/(2\Omega H)$
based on the free-fall velocity
$U_{\rm ff}=(\alpha g (T_{\rm ICB}-T_{\rm CMB})H)^{1/2}$ provides a measure of buoyancy-driven inertia, and is thought to be rather small at the scale of the core, between $10^{-6}$ and $10^{-3}$ \citep{aurnou2020prf,schubert2011pepi}. Its value
strongly depends on how far beyond the onset convection operates: 
Inside the TC, criticalities are expected in the range 10 to $10^3$. Based on RRBC studies, geostrophic turbulence may exist at criticalities greater than around $10^2$ and produce intense structures such as the Large Scale Vortices (LSV) that appear in numerical simulations of RRBC \citep{aguirre-guzman2021jfm,maffei_et_al_2021}. \textcolor{black}{These could locally increase the global Rossby number }and would make it possible for buoyancy-driven inertia to locally compete with the Coriolis force and break the TPC. Unfortunately,  \textcolor{black}{LSV were} never observed in experiments.
{Outside the Tangent Cylinder, numerical simulations show that the nonlinear evolution of the Busse columns leads to intense columnar vortices in the close vicinity of the TC \citep{gillet2006_jfm,schaefferetal2017,lin2021_jfm}, consistent with geomagnetic data \citep{gillet2022pnas,finlay2023nat}. These provides a direct source of inertia near the TC boundary susceptible of breaking the TPC, at least locally.
}
\begin{figure}
	\centering
	\begin{overpic}[width=0.45\textwidth]{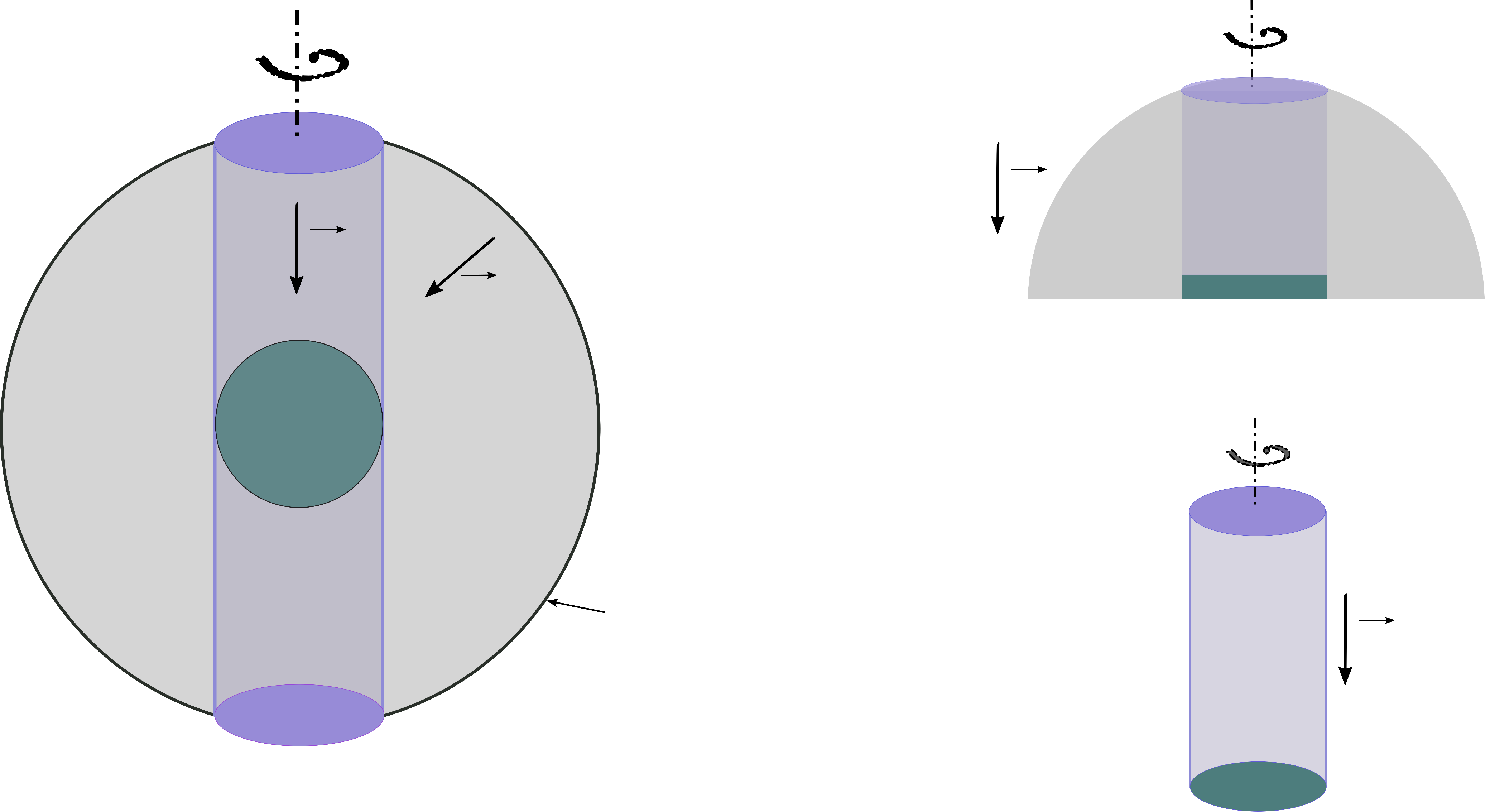}
		\put(0,52){\normalsize (a)}
		\put(66,52){\normalsize (b)}
		\put(66,24){\normalsize (c)}
		
		\put(31,33.5){\normalsize $g$}
		\put(21,36.5){\normalsize $g$}
		\put(68,40.8){\normalsize $g$}
		\put(91.1,10.5){\normalsize $g$}
		
		\put(81,25){\normalsize $\Omega$}
		\put(80,53){\normalsize $\Omega$}		
		
		\put(82,36.5){\normalsize $T_H$}
		\put(82,44.5){\normalsize $T_C$}		
		\put(82,14.9){\normalsize $T_H$}
		\put(82,3.9){\normalsize $T_C$}

		\put(21,52){\small Earth's rotation}			
		\put(41,11){\small Core-Mantle}
		\put(43,7.5){\small Boundary}	
		\put(64,11){\small Solid wall}	
		\put(16,26){\small Inner}
		\put(16.2,22){\small Core}
		\put(18,12){\small TC}
		\put(3,26){\small Outer}
		\put(3.2,22){\small Core}
		
	\end{overpic}	
	\caption{(a) Schematic of the Earth’s core with tangent cylinder (blue shaded region). (b) Schematic of the experimental apparatus used previously dedicated to study convection in TC \citep{aurnouetal2003,aujogueetal2018}. This set up consists of a hemispherical dome with a cylindrical heater and a horizontal heating surface protruded at the centre. The blue shaded region highlight the TC. (c) Schematic of the convection cell consisting of cylinder with solid impermeable insulating walls (CSIIW) commonly used to study rotating convection. This cylindrical cell is surrounded by solid walls, heated from below and rotated about the vertical axis. For more information on rotating convection in CSIIW, see recent reviews by \cite{kunnen2021,eckeshishkina2023}. }
	\label{fig:TC_Schematic}
	
\end{figure}

Considering the TC boundary is unlikely to be purely adiabatic and may not remain impermeable, the question arises of how fluxes of heat and mass through it may affect the convection inside the TC. Indeed most of the current knowledge on RRBC relies on cylindrical domains with solid impermeable and insulating walls (CSIIW). 
This question motivated 
two experimental studies where convection in TCs was characterized using custom-built experimental setups, 
whose principle is shown in figure \ref{fig:TC_Schematic}(b) \citep{aurnouetal2003,aujogue2016rsi,aujogueetal2018}. These experiments utilised a cylindrical heater placed at the centre of a hemispherical dome (shown in figure \ref{fig:TC_Schematic}b). 
Using dye visualisation, \cite{aurnouetal2003} observed four transient convection states with increasing Rayleigh number, (a) subcritical regime with no convection, (b) rim instability leading  to helical plumes around the edge of the tangent cylinder, (c) quasi-geostrophic convection with helical plumes throughout the tangent cylinder, and finally  (d) 3D turbulent convection. \cite{aujogueetal2018} studied convection in a TC using PIV measurements, calling their experiment `Little Earth Experiment (LEE)', a precursor of the device we use in this paper. The critical scalings for the onset Rayleigh numbers and wavelength of convection in a TC were similar to those known for plane convection, but with a greater prefactor for the critical wavelength. They observed similarities in the flow structures of supercritical plumes akin to those found in rotating convection in CSIIW, suggesting that the confinement within the TC affects convection in a similar way as solid lateral walls. Interestingly, they found that wall-modes, which appear below the onset of bulk convection in CSIIW \citep{Ning1993,Zhong1991,Zhong1993,eckeshishkina2023}, only appeared \emph{after} the onset of bulk convection, suggesting that the TC boundary does not behave as a solid wall in the subcritical regime.

Much in the spirit of the early annulus and spherical shell experiments on convection in equatorial region \citep{cardin2015tg,potherat2024crphys}, these two experiments produced crucial insight on the onset of convection in the TC and moderately supercritical regimes (below 20 times the critical Rayleigh number). In light of the discussion above, the questions raised by the study of the Earth's core now concern the breakup of the TPC and the nature of the flow at  higher criticality. While the successive regimes of RRBC in plane layers and CSIIW have been precisely mapped in the past three decades, a corresponding level of insight is still lacking for the quite unique configuration of planetary TCs. Here we set out to answer three TC-specific questions in view of understanding how these processes may express in planetary cores:
\begin{enumerate}
	\item  What are the regimes of convection inside a TC where {inertia} may exist along the TC boundary?
	\item  What level of criticality is required to break the TPC at the TC and what are the consequences for the convection inside the TC when this happens?
	\item  What part of the heat flux is diverted to the lateral TC boundary, as opposed to axially?
\end{enumerate}

\textcolor{black}{Note that instances where the TPC was found broken at the TC boundary suggest that the TC boundary does not collapse at once, as a result of a single event or an instability. Strictly speaking, the TPC is broken for arbitrary small amount of inertial, viscous or other forces since it is only valid in the limit $\Ek\rightarrow0$, $\Ro\rightarrow0$ etc. Instead, the TPC may be locally broken by isolated events such as waves or intense vortices, while the TC boundary remains impermeable elsewhere. For example, as inertia increases with $\Ro$, an ever larger of such events may occur that compromise the TC boundary's impermeability on an ever greater proportion of its surface. For this reason, the radial velocity at the TC boundary offers a good measure of how much the TC boundary has broken down.}

We pursue an experimental approach, building on the ideas and technical advances of \cite{aurnouetal2003} and \cite{aujogueetal2018}, to undertake a more systematic analysis of the regimes inside the TC, using a significantly upgraded version of LEE: LEE2. LEE2 is built to operate both in non-magnetic and magnetic regimes but we first address these questions in the absence of magnetic field, leaving aside the influence of the Lorentz force. We target a range of criticalities spanning the subcritical regime to criticalities over $10^2$, closer to the condition expected in the Earth core. 
We first provide the detail of the experimental setup and the experimental techniques deployed to achieve a step change in precision, control and parameter range compared to LEE (section \ref{sec:experiment}). We then proceed to identify the different regimes using PIV measurements informed by an estimate of the global force balance in five different slices of our experimental TC (section \ref{sec:Regimes of convection}). From these, we identify the impact of the different regimes on the axial and lateral heat transfer (section \ref{sec:Heat Transfer Scalings}). We then assess how these are related to the breakdown of the TPC (section \ref{sec:tpc}) and discuss possible consequences for the Earth Core (section \ref{sec:discussion}).

\section{Experimental setup and system parameters}
\label{sec:experiment}
\subsection{Experimental setup}
We investigate rotating convection in a TC using temperature and instantaneous velocity measurements with water as the working fluid. Experiments are based on the Little Earth Experiment-2 (LEE2), represented on figures \ref{fig:Schematic_combined}a and \ref{fig:Schematic_combined}b. LEE2 is built on the same principle as its predecessor LEE1 \citep{aujogue2016rsi,aujogueetal2018}, but to much higher accuracy in terms of measurements and control of the experimental conditions. 
Although we study rotating-convection in water in the present paper, one of the main purposes of this device is to study magneto-rotating-convection: in magnetohydrodynamic (MHD) experiments, with  30$\%$ sulphuric as a working fluid, chosen for its transparency and electric conductivity. For that purpose,  the setup must be able to be placed inside the 376 mm diameter bore of a large solenoidal magnet of up to 12 T at the High Magnetic Field Laboratory in CNRS, Grenoble, which severely constrained our design choices.
We use a transparent cylindrical vessel of inner diameter ($2R_D$) of 220 mm made of 5 mm thick transparent acrylic as the convection cell. At the centre of its bottom wall, a cylindrical heater (of diameter $2R_{T\!C}$ = 150 mm) protrudes into the vessel (shown in figure \ref{fig:Schematic_heater}). The heater fulfils two functions: First, its upper surface is kept at fixed hot temperature $T_H$, while the top and side outer walls of the cylinder are kept at a cold temperature $T_C$ to create an unstable temperature gradient prone to drive convection within the vessel. Second, its edge incurs a radial jump in domain height akin to the equatorial edge of the Earth's inner core, where a TC develops under fast rotation. The height above the heater is $H$ = 143 mm, providing an aspect ratio ($\Gamma = 2R_{\rm T \! C}/H$) of 1.05 for the Tangent Cylinder above it. {\color{black} The radius and the volumetric ratios between the TC and the outer cylinder remain around $\chi = R_{TC}/R_D = 0.68$ and $\Gamma_{\rm Vol} = Vol_{\rm InTC}/Vol_{\rm OutTC} = 0.78$, which are approximately twice and five times the values for the Earth, respectively. This makes the cold vertical walls in LEE2 quite close to the TC which incurs significant baroclinicity near the TC boundary. {Although driven through a different mechanism than the convection outside the Earth's TC, the baroclinic flow is topologically similar to it and acts as a similar source of inertia near the TC boundary.}

\begin{figure}
	\centering
	\begin{subfigure}[t]{0.46\linewidth}
		\centering
	\begin{overpic}[width=\linewidth]{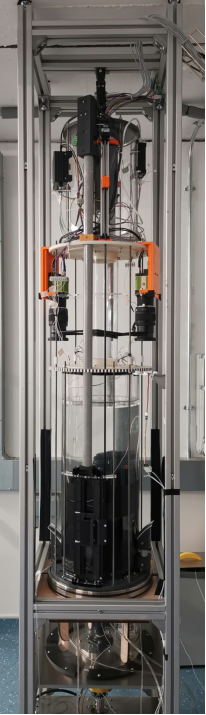}
		\end{overpic}
	\end{subfigure}
	\hfill
	\begin{subfigure}[t]{0.5\linewidth}
		\centering
		\raisebox{0.5cm}{
		\includegraphics[width=\linewidth]{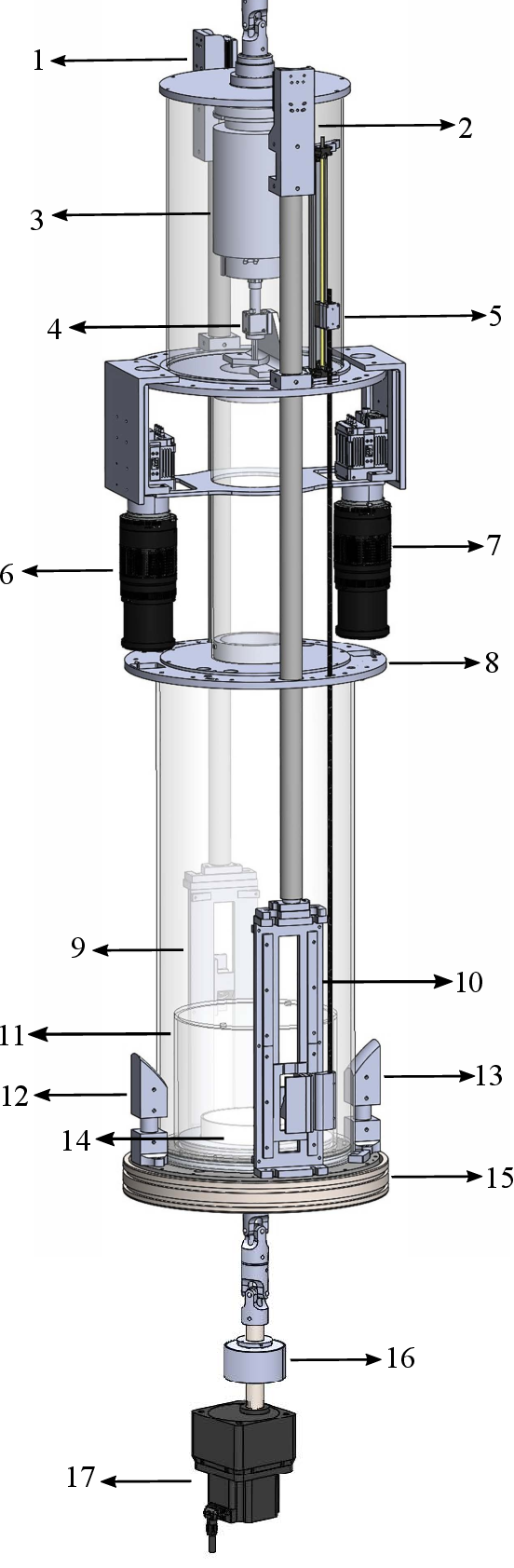}
	}
	\end{subfigure}
	\caption{(a) Photograph and (b) schematic of the LEE2 experimental set-up for the (non-magnetic) rotating-convection experiments with water. 1, 2: lasers for horizontal and vertical PIV, respectively, 3: upper slip ring, 4: camera for horizontal PIV, 5: stepper motor, 6, 7: cameras for vertical PIV, 8: location for optical speed sensor, 9, 10: optical carriages for vertical and horizontal PIV, respectively, 11: Main vessel, 12, 13: mirrors for vertical PIV, 14: heater, 15: static plate with bearing on top to aid smooth rotation, 16: lower slip ring, 17: motor to drive rotation.}
	\label{fig:Schematic_combined}
\end{figure}

\begin{figure}
	\centering
	\begin{minipage}[t]{0.6\linewidth}
		\centering
		\begin{overpic}[width=\linewidth]{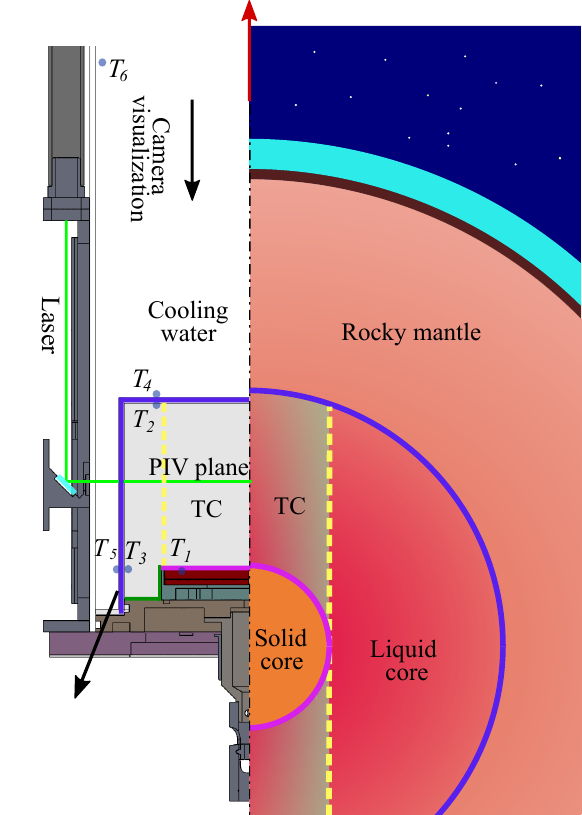}
			\put(3,98){\small (a)}
			\put(21,98){\small $\Omega \hat{e}_z$}
		\end{overpic}
	\end{minipage}
	\hfill
\raisebox{10.6\baselineskip}{
		\begin{minipage}[t]{0.38\linewidth}
		\centering
		\begin{overpic}[width=\linewidth]{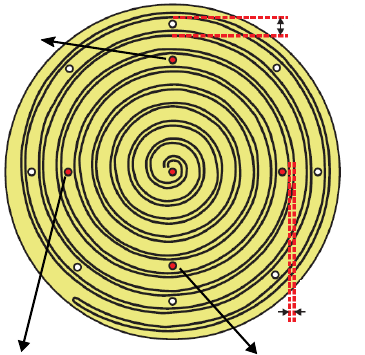}
			\put(3,99){\small (b)}
			\put(-5,85){\small $T_{1,1}$}
			\put(-5,-5){\small $T_{1,2}$}
			\put(60,-5){\small $T_{1,3}$}
			
		\end{overpic}		
  \par\vspace{2.2em}
		\begin{overpic}[width=\linewidth]{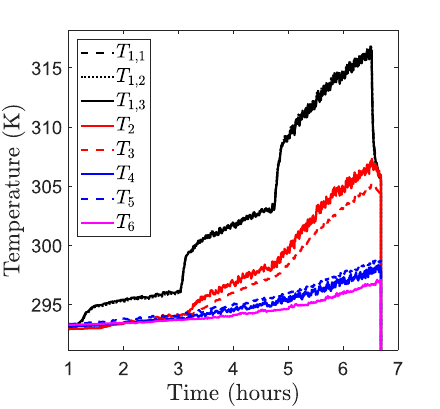}
			\put(3,85){\small (c)}
			\put(20,28){\footnotesize $2K$}
			\put(44,50){\footnotesize $5K$}
			\put(57,76){\footnotesize $10K$}
		\end{overpic}
	\end{minipage}}
	
	\caption{(a) Sketch of the Earth’s structure with liquid outer core and solid inner core (right) and schematic of the LEE2 setup, highlighting the main vessel, protruding heater and the PIV system (left). The thermal boundary conditions for the main vessel are as follows: imposed cold temperature on the top and side walls (shown in blue), imposed hot temperature at the top of the heater (shown in purple), and thermally insulating BC at the side of the heater and the bottom wall around the base of heater (shown in green). The temperature probes used to estimate the axial and radial heat fluxes are indicated as $T_1$, $T_2$ and $T_3$. $\Delta \! T_{\rm z} = T_1 - T_2$ and $\Omega \hat{e}_z$ are dimensional control parameters in this study. (b) Heater, consisting of a PEEK plate supporting two adjoining spirals of the metal wire. The temperature measurements on the heater are carried out using three thermal probes $T_{1,1}$, $T_{1,2}$, and $T_{1,3}$ located at around two-third of the radius away from the centre. (c) Raw temperature data collected for one of the runs at $\Ek = 1.4 \times 10^{-5}$. Here, we can see that the three thermal probes at the heater provide very similar results, confirming the temperature homogeneity in the azimuthal direction.}
	\label{fig:Schematic_heater}
\end{figure}

The electrical heater generates heat using current running through a metal wire. 
It consists of a PEEK plate supporting two adjoining spirals of resistive wire with respective electric currents moving radially inward and outward. 
	\textcolor{black}{The thermal boundary conditions are imposed cold temperature on the top and side walls of the main vessel, imposed hot temperature at the top of the heater, and thermally insulating at the side of the heater and the bottom wall around the base of heater (figure \ref{fig:Schematic_heater}a). The temperature inhomogeneity over the heater surface is measured through simultaneous temperature measurements probes at three different azimuthal locations at the top of the heater (shown as $T_{1,1}$, $T_{1,2}$ and $T_{1,3}$ in figure \ref{fig:Schematic_heater}b) of the convection cell. The maximum temperature inhomogeneity over the heater surface was found within \SI{0.2}{K}. The time-averaged temperature of these three probes $T_1 = \sum_{i = 1}^{i = 3} T_{1,i}/3$ is used to calculate temperature differences between the boundaries of the vessel.} These probes are negative temperature coefficient (NTC) thermistors (TE Connectivity, Product no.: GA10K3MBD1) with \SI{10}{k\Omega} resistance, probe diameter of 1.1 mm and tolerance 
$\pm$0.2K. The temperature probes on the cylinder walls (shown as $T_2$ and $T_3$ in figure \ref{fig:Schematic_heater}a) are  NTC thermistors (Amphenol Advanced Sensors 
MM100GG103A) with 10 k$\Omega$ resistance, a diameter of 2.03 mm and 
tolerance $\pm$\SI{0.05}{K}. 
From these, we monitor two temperature differences to analyse axial and radial heat transfer, respectively defined as:
\begin{eqnarray}
	\Delta \! T_{\rm z} &=& T_1 - T_2,\\
	\Delta \! T_{\rm r} &=& T_1 - T_3.	
\end{eqnarray}
$\Delta \! T_{\rm z}$ is controlled to a precision of around \SI{\pm0.05}{K} using a PID controller. 
For all results presented in this paper, the working fluid is water, 
of density $\rho=\SI{998}{kg/m^3}$, viscosity $\nu=\SI{0.9\times 10^{-6}}{m^2/s}$, thermal diffusivity $\kappa =\SI{1.4\times 10^{-7}}{m^2/s}$ and thermal expansion coefficient $\alpha =\SI{2.1 \times 10^{-4}}{/K}$.
The 
controlled temperature difference $\Delta T_{\rm z}$ was in the range [\SI{0.5}{K}, \SI{23}{K}], for which the radial temperature difference $\Delta \! T_{\rm r}$ laid in the range [\SI{0.5}{K}, \SI{35}{K}]. {\color{black} The Biot number $\Bi$ (measure of the ratio of the thermal resistance of the solid body to the liquid above it) is calculated to provide us with an indication of the relative temperature non-uniformity across the heater and at the top lid of the cylinder. For a 10mm thick ceramic plate and 5mm thick acrylic,
\begin{eqnarray}
	\Bi_{\rm heater} &=& \frac{L_{ceramic}/k_{ceramic}}{H/k_{H_2O}} 
	=   4.5 \times 10^{-4};\\
	\Bi_{\rm top \space lid} &=& \frac{L_{acrylic}/k_{acrylic}}{L_{H_2O}/k_{H_2O}} 
	= 0.43.
\end{eqnarray}
Near the heater, it is sufficiently small to guarantee good temperature homogeneity over the heater surface. However, the lower conductivity of the acrylic lid means that the Biot number is larger at the top (0.43) and so its value does not give a guarantee that the temperature is reasonably homogeneous.
To assess inhomogeneity, we therefore monitor the temperature difference between two locations at the inner wall of the tank, which are expected to differ most: inside top and inside side boundary near the heater. At high supercriticalities, the heat flux escapes through both boundaries, the bottom side wall can be at a similar temperature to the inside wall outside the TC.  Indeed, at high enough criticality, the inhomogeneity remains below 10\%.

Additionally, the acrylic used for the top and side walls is not a very good conductor, which leads the averaged temperature of the fluid to drift up during the measurements. However the temperature regulation prevents heat from building up at the top 
by correspondingly increasing the heating at the bottom to keep the temperature difference constant.
As a consequence, the temperature of the whole fluid inside the vessel drifts up slightly. This can ben seen from the evolution of temperature in time on figure \ref{fig:Schematic_heater}c illustrating this effect. Using thermistors placed inside the TC and outside the TC, we evaluated the temperature drift in both these regions. This enabled us to assess the part of the heat flux extracted from the heater that is lost in heating the fluid. 
respectively as $P_{\rm in}=\mIn C_p \partial T_2/\partial t$ and $P_{\rm out}=\mOut C_p \partial T_3/\partial t$, where $\mIn$ and $\mOut$ are the masses of water inside and outside the TC. Therefore, the ``effective" heater power used to drive convection is given by $\Pconv = P - P_{\rm in}-P_{\rm out}$.
For most cases, $\Pconv$ remains over 60\% of the injected power, $P$. 
{Figures showing heat flux data with both the power with and without a correction removing $P_{\rm in}$ and $P_{\rm out}$ are given in appendix \ref{appendix:Nu_Correction}}: they show that the correction does not affect the scaling behaviours discussed in this paper. Given that the raw data is more precise, we chose not to apply the correction for the reminder of the paper.} 

The vessel is rotated using a DC motor placed at the bottom of the set-up (figure \ref{fig:Schematic_combined}). The torque is transmitted from the motor to the experiment via a drive shaft. The weight of the experiment is supported by metal ball bearing. 
The angular velocity of the experiment $\Omega$ 
is continuously monitored with a built-in optical system, made of an alternating black and white (B/W) teeth wheel placed on the edge plate (shown in figure \ref{fig:Schematic_combined}a) and an optical detector fitted on one of the vertical supporting profiles. $\Omega$ was varied in the range 0 rpm to 60 $\pm$0.2 rpm.

Instantaneous velocity fields are recorded using a bespoke Particle Image Velocimetry (PIV) system. 
Cameras and lasers are kept far away from the region of high magnetic field where the convection cell sits in MHD runs (see figure \ref{fig:Schematic_combined}a,b). The laser and optics generate a single horizontal laser sheet intercepting the vessel at adjustable height, controlled by a stepper motor operated through slip ring while rotation is in progress. We visualize 5 successive horizontal planes inside the convection cell at heights $z=H/6$, $z=H/3$, $z=1/2$, $z=2H/3$ and $z=5H/6$ from the top of the heater (at $z=0$). Thanks to this system, once the flow has reached equilibrium for set rotation speed and temperature difference, the velocity fields in all 5 planes is recorded in sequence. This saves start up time and eliminates uncertainties incurred by restarting the experiment from a state of rest between measurement in different planes. This is a significant improvement over LEE1 \citep{aujogueetal2018} where the rotation had to be stopped to be able to change the height of the light sheet. To generate the light sheets, a continuous laser (Direct Diode Green Laser Module, Product no.: OFL420-G1000-TTL-ANA) emits green light of \SI{520}{nm} wavelength with a maximum power of \SI{1}{W}. The laser sheet of thickness around 1 mm is generated using a line generator disc placed outside but close to the convection cell. The fluid is seeded with highly reflective silver-coated hollow glass spheres of mean size \SI{10}{\mu m} (from Dantec Dynamics). The images are acquired at 20 frames per second (fps) with a USB3-operated Flea3 camera of resolution 2080$\times$1552 pixels. The spatial resolution is about 6.7 pixels for 1 mm of the visualised area. Velocity fields are computed using the LaVision software, where the adaptive cross-correlation method is employed for the velocity calculations.

For the PIV as well as the temperature measurements, all the signals are transferred from the rotating frame to the static frame using MOFLON slip rings (one located near the top and the other near the bottom of the set-up, shown in figure \ref{fig:Schematic_combined}b). The experimental protocol consists of first checking that all the temperature probes read the same value with the fluid at rest and the heater switched off (with uncertainty of around \SI{0.1}{K}). Then, the rotation is started with the desired $\Omega$ value and we wait until the flow reaches a solid body rotation (typically around 30 min, confirmed with PIV measurements). Then the heater is turned on with the prescribed $\Delta \! T_{\rm z}$ value. 
The maximum power and temperature of the heater are around \SI{350}{W} and \SI{343}{K}, respectively. Only when the system has reached a statistically steady mechanical and thermal state ($\approx$ 60 min in total) are the PIV measurements carried out. In terms of the thermal diffusion time scale $H^2/\kappa$ ($\simeq2.5 \times 10^{-2}$ thermal diffusion time units). The flow fields are recorded for 10 minutes ($\simeq 4.1 \times 10^{-3}$ thermal diffusion time units) at 20 fps providing 12,000 images for every horizontal plane at a fixed $\Omega$ and $\Delta \! T_{\rm z}$. This frequency was found sufficient to resolve all flow timescales of interest in this study. Temperature readings were acquired at a frequency of \SI{0.1}{Hz} using PicoLog Data logging software, corresponding to 1.18 and 0.173 in terms of free fall units ($t_{\mathrm {ff}} = \sqrt{H/\alpha g \Delta T}$) for $\Delta T$ = 0.5\textsuperscript{o}C and 23\textsuperscript{o}C, respectively. 
\begin{table}
	\centering
	\small	
	\caption{\normalsize Range of non-dimensional control parameters in LEE2 and previous experiments involving TCs: experiment by \cite{aurnouetal2003}, LEE1 \citep{aujogueetal2018}, and estimated Earth’s core parameters \citep{schubert2011pepi}. Here, $\Gamma$ = $2R_{\rm TC}/H$, $\Gamma_{\rm Vol} = Vol_{\rm InTC}/Vol_{\rm OutTC}$, $\chi$ = $R_{TC}/R_D$, $\Prt=\nu / \kappa $, $\Ek=\nu/ 2 \Omega H^2 $ and $\Ra=g \alpha \Delta \! T_{\rm z} H^3 / \kappa \nu$.}
	\renewcommand{\arraystretch}{0.65}
	\begin{tabular}{ccccc}
		\hline
		\parbox{1.1cm}{Control\\ Parameters}&LEE2 & LEE1& \parbox{2cm}{\centering Aurnou et al.}&\parbox{0.8cm}{\centering Earth’s core}\\
		\hline
		\parbox{1.1cm}{$\Gamma$} &1.05 &0.83&0.86&1.06	\\
		\hline
		\parbox{1.1cm}{{\color{black}$\Gamma_{\rm Vol}$}}  &0.776&0.213&0.147&0.146	\\
		\hline
		\parbox{1.1cm}{{\color{black}$\chi$ }} &0.68&0.36&0.35&0.35	\\
		\hline
		\parbox{1.1cm}{$\Prt$} &6.4&7&7&10\textsuperscript{-1}	\\
		\hline
		\parbox{1.1cm}{$\Ek$} &\parbox{1.4cm}{\centering $[0.35, 4.2]$ $\times$ 10\textsuperscript{-5}}&\parbox{1.1cm}{\centering $[0.06, 2.25]$ $\times 10\textsuperscript{-5}$}&\parbox{0.9cm}{\centering $[0.5, 45]$ $\times 10\textsuperscript{-5}$}&  10\textsuperscript{-15}\\
		\hline
		\parbox{1.1cm}{$\Ra$} &\parbox{1.4cm}{\centering $[0.021, 1.1]$ \\ $\times$ 10\textsuperscript{9}} &\parbox{1.1cm}{\centering $[0.014, 2.9]$ \\ $\times$ 10\textsuperscript{9}} &\parbox{0.9cm}{\centering $[0.003, 30]$ \\ $\times$ 10\textsuperscript{9}} & $>$10\textsuperscript{22}\\
		\hline
	\end{tabular}	
	\label{table:system_paremeters}
\end{table}

\subsection{System parameters}
The non-dimensional control parameters are the Rayleigh number $\Ra$ and Ekman number $\Ek$, the Prandtl number $\Prt$ and aspect ratio of the TC $\Gamma$. $\Prt$ and $\Gamma$ are fixed for all the experiments we conducted at $\Prt =\nu/\kappa= 6.4$ and $\Gamma$ = 1.05. In LEE2, $\Ra$ and $\Ek$ are defined as 
\[ \Ra=\frac{g\alpha \Delta \! T_{\rm z} H^3}{\nu\kappa} \quad {\rm and  \quad} \Ek=\frac{\nu}{2\Omega H^2}.
\]
\begin{table*}
	\centering
	\footnotesize
	\caption{\normalsize Non-dimensional control parameters $\Ek, \Ra$, Froude number $\Fr$ and levels of criticality based on $Ra_{c}$, $Ra^{\rm wfs}_{c}$ and $Ra^{\rm wns}_{c}$ for all values of the Ekman numbers achieved in our experimental runs. \textcolor{black}{The critical Rayleigh numbers for the onset of planar layer and wall-mode convections based on LEE1 experimental conditions \citep{aujogueetal2018} are also provided.} Estimates for Earth's conditions ($\Prt \approx 0.1$, $\Ek \approx$ $10^{-15}$, $\Ra \approx10^{22}$) are also provided for comparison with LEE2. $\Ra_{c}$ for the Earth's core is based on the critical Rayleigh number for low $\Prt$, given by $\Ra_c \approx 17.4 (\Ek/\Prt)^{-4/3}$ \citep{kunnen2021}. $\Fr$ for the Earth's core is based on maximum radius $R_{max}$ ($ = 6.4 \times 10^6$ m) near the equator.}
	\label{table:criticality}
	\renewcommand{\arraystretch}{0.6}
		\begin{tabular}{ccccccccc}
			\hline		
			\parbox{1.3cm}{\centering$\Ek$ } &\parbox{1.4cm}{\centering $\Ra_{c}$ } &\parbox{1.5cm}{\centering $\Ra^{\rm wns}_{c}$  } &\parbox{1.6cm}{\centering $\Ra^{\rm wfs}_{c}$ }  & \parbox{1.4cm}{\centering $\Ra$}&$\Rt$&  $\Rt_{ws}$ &  $\Rt_{wf}$&$\Fr$	\\
			\hline
			LEE2: 4.2$\times10^{-5}$   &0.59$\times10^{7}$& 0.80$\times10^{6}$& 0.74$\times10^{6}$  &$[2.4, 114]$ $\times10^{7}$ &$[4.0, 191]$ &$[30, 1432]$ & $[32, 1544]$ &0.002\\			
			\hline
			LEE2: 1.4$\times10^{-5}$   &2.58$\times10^{7}$& 2.35$\times10^{6}$& 2.23$\times10^{6}$ &$[2.4, 114]$ $\times10^{7}$ &$[0.93, 44.2]$ &$[10, 484]$& $[11, 511]$ &0.019\\
			\hline
			LEE2: 0.71$\times10^{-5}$   &6.38$\times10^{7}$& 4.61$\times10^{6}$& 4.42$\times10^{6}$ &$[2.4, 114]$ $\times10^{7}$ &$[0.37, 17.8]$ &$[5.2, 247]$& $[5.4, 258]$ &0.075\\
			\hline
			LEE2: 0.47$\times10^{-5}$   &11.1$\times10^{7}$& 6.93$\times10^{6}$& 6.69$\times10^{6}$ &$[2.4, 114]$ $\times10^{7}$ &$[0.22, 10.3]$ &$[3.5, 164]$& $[3.5, 171]$ &0.17\\
			\hline
			LEE2: 0.35$\times10^{-5}$   &16.4$\times10^{7}$& 9.29$\times10^{6}$& 8.99$\times10^{6}$ &$[2.4, 114]$  $\times10^{7}$&$[0.15, 6.9]$ &$[2.6, 123]$& $[2.7, 127]$ &0.30\\
			\hline
		\textcolor{black} {LEE1:} 1.15$\times10^{-5}$   &3.35$\times10^{7}$& 2.86$\times10^{6}$&2.72$\times10^{6}$ && && &\\
			\hline
				\textcolor{black} {LEE1:} 3.36$\times10^{-6}$   &1.73$\times10^{8}$& 9.67$\times10^{6}$&9.36$\times10^{6}$ && && &\\
			\hline
			Earth: $\Ek$ $\approx$ $10^{-15}$, $  \Prt \approx 0.1$   &$\approx$ 8.07$\times10^{19}$& & &$\approx10^{22}$ &$\approx 124$  &&&0.003\\
			\hline
		\end{tabular}	
		
	\end{table*}

The regimes of rotating convection are probed by varying  $\Ek$ through $\Omega$ and the thermal forcing $\Ra$ through $\Delta \! T_{\rm z}$ between \SI{0.5}{K} and \SI{23}{K} for each value of  $\Omega$.
This yields
$\Ra\in[ 2.4\times10^{7}, 1.14\times10^{9}]$. 
A limiting factor on the rotation rate $\Omega$ is the appearance of centrifugal acceleration at higher rotation speeds, which
competes with the gravitational acceleration. The combined acceleration is not vertical any more, which may result in significant alteration of the convective patterns. The Froude number $\Fr = \Omega^2 R_{TC}/g$ quantifies the ratio of centrifugal acceleration to gravitational acceleration \citep{homsyhudson1971}. In large scale geophysical flows, $Fr\ll1$ but centrifugal acceleration may play a role at smaller scales. For example, Earth’s atmosphere experiences at most $\Fr = 3\times10^{-3}$ near the equator but centrifugal forces can play an important role in generating tornadoes. Here, for the purpose of mimicking the large-scale dynamics of the Earth's TC, we shall keep the effect of centrifugal acceleration as small as possible in LEE2. DNS of rotating convection in CSIIW \citep{hornaurnou2018,hornaurnou2019} showed that the transition to centrifugally dominated flow occurs at $\Fr \approx \Gamma/2$. \textcolor{black}{In LEE2, 
$\SI{5}{rpm}\leq\Omega\leq\SI{60}{rpm}$, yield $10^{-6}\leq\Ek\leq4.2\times10^{-5}$ and $2\times10^{-3}\leq\Fr\leq 0.3<
\Gamma/2 = 0.525$, so centrifugal convection is not relevant in this problem}.\\ 
{At the lateral cylindrical boundary of the TC, a Stewartson layer develops. 
 The structure of the Stewartson layer likely follows that found at the TC generated by two concentric spheres rotating at different velocities \citep{proudman1956_jfm,stewartson1957_jfm,stewartson1966_jfm}. These consist of two outer layers, one inside the TC, one outside the TC, separated by a thinner an inner layer. 
Their thicknesses respectively scale as $H\Ek^{2/7}$, $H\Ek^{1/4}$ and $H\Ek^{1/3}$. For the values of $\Ek$ we investigate, these thicknesses respectively span $[2.8,8]\times10^{-3}$ {m}, $[4.5,11.5]\times10^{-3}$ {m} and $[1.43,5]\times10^{-3}$ {m}, so the corresponding shear region is always less than  $0.154R_{TC}$. Hence, the detailed structure of these layers would be challenging to capture with our current PIV setup.}\\
Table \ref{table:system_paremeters} provides the values of the non-dimensional control parameters of our system, along with those of previous experiments involving TCs \citep{aurnou2007,aujogueetal2018} and the Earth's core \citep{schubert2011pepi}. Since the friction at the TC boundary can be expected to be lower than at a no slip wall but higher than at a free-slip boundary, $\Ra$ is rescaled with the critical Rayleigh numbers for the onset of wall-mode convection with both no-slip and free-slip walls, respectively $\Rt_{\rm ws}= \Ra/\Ra^{\rm wns}_{\rm c}$ and $\Rt_{\rm wf}=\Ra/\Ra^{\rm wfs}_c$ \citep{zhangliao2009,liao2006jfm,chang2006gafd} as well as for plane layer convection $\Rt = \Ra/\Ra_{c}$ \citep{chandrasekhar1961}: these levels of criticality are indeed better measures of the intensity of the thermal forcing in rotating convection than $\Ra$ alone \citep{julienetal2012}. A large range of values of $\Ra$ is explored in this study, covering around four orders of magnitude in $\Rt\in[0.15, 191]$, summarised in table \ref{table:criticality}, along with all achieved values of the control parameters. 
\section{Regimes of rotating convection based on velocity fields}
\label{sec:Regimes of convection}
We first identify the different regimes of convection encountered in our setup. Following \cite{julienetal2012}, we use the Reduced Rayleigh number $\Rt=\Ra/\Ra_c$ as the main control parameter on the basis that flows with comparable values of $\Rt$ obtained for different values of $\Ek$ and $\Ra$ display qualitatively similar topologies even though detailed properties such as the dominating length scale may differ. We identify changes in the flow topology as $\Rt$ increases. In particular, we shall seek how these regimes map to the regimes of convection in a CSIIW 
 and how they differ. To help establish the correspondence between the topological changes and the known regimes of rotating convection in a CSIIW, we also analyse the force balance based on the velocity measurements corresponding to the snapshots of $z-$vorticity across 5 planes spanning the TC's height. Figures \ref{fig:Flow_Snapshots_lowRa}a to \ref{fig:Flow_Snapshots_lowRa}d illustrate the development of convective patterns through snapshots obtained for $ 0.84 \leq \Rt \leq 4.2$, and figures \ref{fig:Flow_Snapshots_highRa}a to \ref{fig:Flow_Snapshots_highRa}d for criticality between $ 7.2 \leq \Rt \leq 160$. Videos are available as supplementary material for each of these cases.
\subsection{Force Balance}
\label{subsec:Force Balance}
We start by analysing the global force balance based on our velocity measurements, so as to obtain a first appraisal of the boundaries between rotation-dominated, rotation-influenced and buoyancy-dominated regimes, before we identify the specific states of convection within them. 
The non-dimensional Navier–Stokes and mass conservation equations in the Boussinesq approximation are 
\begin{eqnarray}
	\Ro\left(\frac{\partial \vel}{\partial t}+\vel \cdot \nabla \vel\right) + \nabla p &=&\vel\times\unitz + \Ek \Delta \vel +   \Ro T \unitz,\\	 
	\nabla\cdot\vel&=&0.
\end{eqnarray}	
Here, velocity $\vel$, temperature $T$, pressure $p$ and time $t$ are the non-dimensional quantities normalized using the height above the heater $H$, the free-fall velocity $U_{\rm ff}$ and $\Delta \! T_{\rm z}$. The pressure gradient is normalised by $2\rho U_{\rm ff}\Omega$ to reflect the quasi-geostrophic balance of rapidly rotating flows. Since our PIV system only provides access to the horizontal velocity components $\up$, we calculate the horizontal component of the inertial, viscous and Coriolis forces. Following \cite{aguirre-guzman2021jfm}, all forces are renormalised by the Coriolis force, so in non-dimensional form, these are expressed as:
\begin{eqnarray}
	\Fi&=&  - \up \cdot \nabla \up,\\	 
	\Fv&=&\Ek\Ro^{-1} \Delta \up,\\
	\Fc&=&\Ro^{-1}\up\times  \unitz.	 	 
\end{eqnarray}
The global magnitude of these forces is estimated from their average in time and over each horizontal plane inside the TC of their local Euclidean norm:
\begin{equation}
\|\boldsymbol F_{\rm N}\|=\frac1{\pi\Gamma^2}\left\langle \int \left[(\boldsymbol F\cdot e_x)^2 + (\boldsymbol F\cdot e_y)^2\right]^{1/2} dS \right\rangle_t,
\end{equation}
where $\rm N$ stands for $\rm I$ (inertial force), $\rm V$ (viscous force), or $\rm C$ (Coriolis force), and $\langle \cdot \rangle_t$ stands for time average over the acquisition time. Since the PIV does not provide access to the vertical velocity component, nor the temperature field, there are few caveats in how these terms are evaluated. First, the inertial force is missing the $u_z\partial_z \up$ term. By virtue of mass conservation, however, it is expected to be of the same order of magnitude as $\up\cdot\nabla \up$. Second, we do not have access to the temperature field either. Hence, we cannot evaluate the buoyancy directly. Since however, the buoyancy force is perpendicular to the Coriolis force, as long as the Oberbeck-Boussinesq approximation remains valid \citep{spiegel1960apj,gray1976ijhmt}, it is the ratio of inertia to the Coriolis force alone that measures whether the flow is buoyancy- or rotation-dominated, or anything in-between.

\begin{figure}
	\centering
	\begin{overpic}[width=\columnwidth]					
		{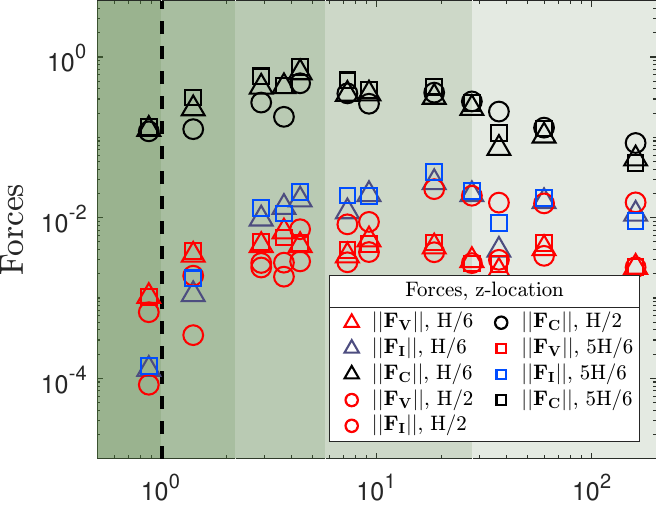} 		
		\put(18,71){\normalsize SR}
		\put(29,71){\normalsize CR}
		\put(39,71){\normalsize CTC}
		\put(58.5,71){\normalsize P}
		\put(80,71){\normalsize GT}	
	\end{overpic}
	\caption{Force balance at $z$ = H/6, H/2 and 5H/6, as a function of supercriticality $\Rt$ for $\Ek = 4.2\times10^{-5}, 1.4\times10^{-5}, 0.71\times10^{-5}, 0.47\times10^{-5}, 0.35\times10^{-5}$. The shaded regions in the background and the acronyms on top of the figures highlight the regimes of rotating convection, SR (subcritical regime), CR (cellular regime), CTC (Convective Taylor Columns), P (plumes and disrupted columns) and GT (Geostrophic turbulence) as observed in LEE2. It should be noted that we only identify regimes where GT is possible but that we cannot be certain that this regime is actually achieved. The vertical black dashed lines indicate the end of the subcritical regime, $\Rt = 1$. While the boundaries delineate distinct regimes of rotating convection observed in LEE2, they are derived from discrete experimental data and should therefore be regarded as indicative.}
\label{fig:forcebalance}
\end{figure}

\begin{figure}
	\centering
		
	\begin{subfigure}[t]{\linewidth}
		\centering
		\begin{overpic}[width=\linewidth]{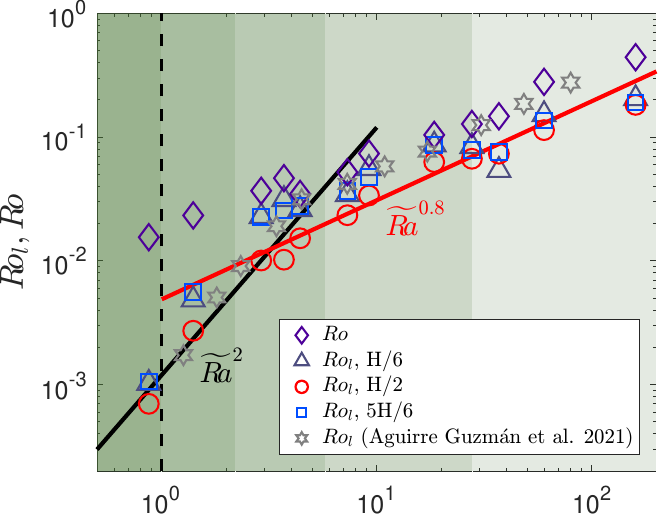}
			\put(5,78){\large (a)}
			\put(18,71){\normalsize SR}
			\put(29,71){\normalsize CR}
			\put(39,71){\normalsize CTC}
			\put(58.5,71){\normalsize P}
			\put(80,71){\normalsize GT}	
		\end{overpic}
	\end{subfigure}
		
	\begin{subfigure}[t]{\linewidth}
		\centering
		\begin{overpic}[width=\linewidth]{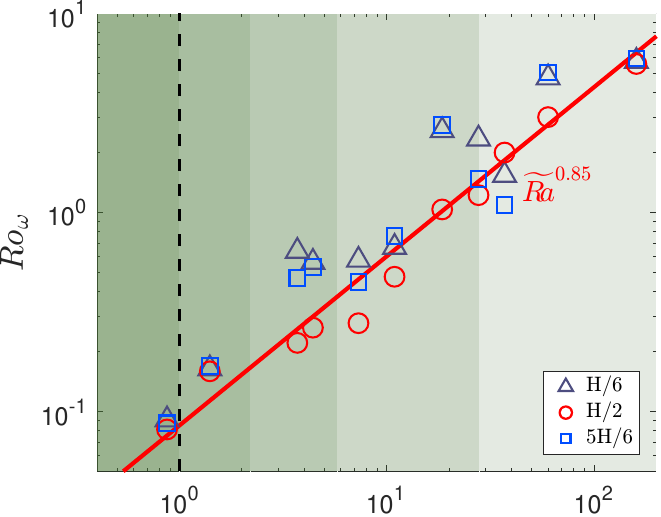}
			\put(5,78){\large (b)}
					\put(18,71){\normalsize SR}
			\put(29,71){\normalsize CR}
			\put(39,71){\normalsize CTC}
			\put(58.5,71){\normalsize P}
			\put(80,71){\normalsize GT}	
			\put(50,-5){\Large ${\widetilde{R\!a}}$}
		\end{overpic}
	\end{subfigure}
	\vspace{0.1cm}
	\caption{(a) Rossby numbers based on global quantities averaged over the TC, and (b) Rossby numbers based on top 10\% of local, instantaneous values of $\omega$ at $z = H/6, H/2$ and $5H/6$, as a function of supercriticality $\Rt$ for $\Ek = 4.2\times10^{-5}, 1.4\times10^{-5}, 0.71\times10^{-5}, 0.47\times10^{-5}, 0.35\times10^{-5}$. Solid black line in (a) highlights the scaling of $\Rt^2$ and $\Rt^{0.8}$, respectively, whereas solid red line in (b) indicates the scaling of $\Rt^{0.85}$. The vertical black dashed lines indicate the end of the subcritical regime, $\Rt = 1$. The background colours identify regimes as on figure \ref{fig:forcebalance}. }
	\label{fig:RossbyNumbers}
\end{figure}

The norms of the inertial, viscous and Coriolis forces at $z = H/6, H/2$ and $5H/6$ are shown in figure \ref{fig:forcebalance}. Similar variations of the force balance are observed in LEE2 and in the rapidly rotating convection in a plane layer with no-slip upper and lower walls and periodic lateral boundary conditions at $\Prt=5$ studied numerically by \cite{aguirre-guzman2021jfm}. The inertial and viscous forces remains several orders of magnitude below the Coriolis force in the subcritical regime(figure \ref{fig:forcebalance}a), \emph{i.e.} for $\Rt < 1$. At around $\Rt \simeq2\pm1$, there is a crossover between these two forces, with inertial force dominating the viscous force at higher $\Rt$. This crossover, however, occurs at lower criticality in LEE2 than in the plane layer where it occurs at $\Rt \approx 5$ \citep{aguirre-guzman2021jfm}. Bearing in mind that our estimate for the inertial forces misses one term, this difference is still significant and may suggest a difference in the nature of the convection around the crossover point between the TC and the plane layer geometries: we shall further probe into this feature using the instantaneous vorticity fields in section \ref{subsec:Rotation-dominated regime}. Since, however, inertia always remains well below the Coriolis force, even up to $\Rt=191$, the convection is always at least rotation-influenced, and never buoyancy-dominated.
\begin{figure*}
\centering
\vspace{2cm}
\hspace*{4em}
\begin{overpic}[height=2.3\columnwidth]{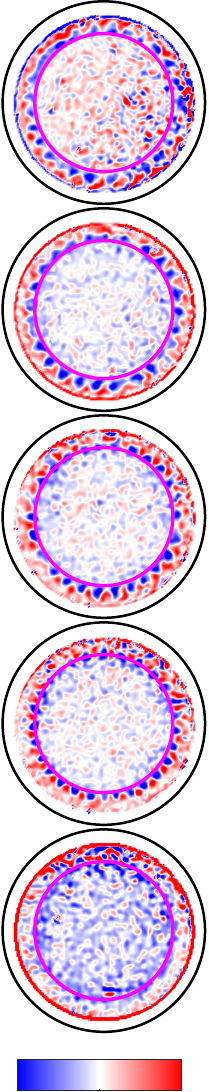}%
\put(-3.0,106){\normalsize (a) $\!\Ek = 4.7 \times 10^{-6}$, $\Rt = 0.86$}
\put(-2,103.5){\normalsize $\Ra^*_q = 2.2 \times 10^{-9}$, $\Ro = 0.018$}
\put(0,101){\parbox{0.43\columnwidth}{\centering\normalsize Subcritical}}
\put(0,3.5){-0.01}
\put(0,3.5){\parbox{0.43\columnwidth}{\centering 0}}
\put(15.5,3.5){0.01}
\put(-5,14){\normalsize $H/6$}
\put(-5,33){\normalsize $H/3$}
\put(-5,52){\normalsize $H/2$}
\put(-5,71){\normalsize $2H/3$}
\put(-5,90){\normalsize $5H/6$}
\put(3.8,-2){\normalsize $\omega_z(x,y,t_i)/\Omega$}
\end{overpic}
\hfill
\begin{overpic}[height=2.3\columnwidth]{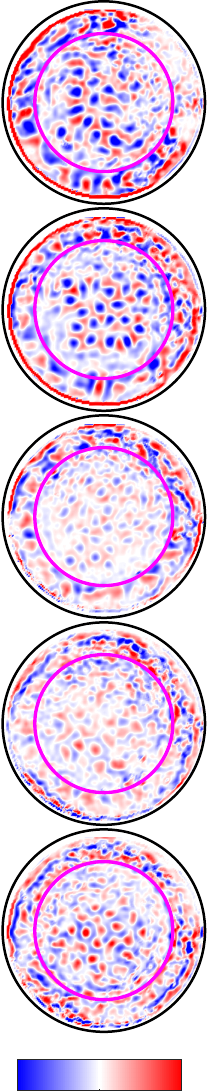}%
\linethickness{1.5pt}
\put(8.5,70){\color{black}\vector(1,-1){8.3}}
\put(-3.0,106){\normalsize (b) $\Ek = 7.1 \times 10^{-6}$, $\Rt = 1.5$}
\put(-2,103.5){\normalsize $\Ra^*_q = 1.2 \times 10^{-8}$, $\Ro = 0.027$}
\put(0,101){\parbox{0.43\columnwidth}{\centering\normalsize Cellular}}
\put(0,3.5){-0.05}
\put(0,3.5){\parbox{0.43\columnwidth}{\centering 0}}
\put(15.5,3.5){0.05}

\put(3.8,-2){\normalsize $\omega_z(x,y,t_i)/\Omega$}
\put(17,60){\includegraphics[width=0.5cm]{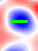}}
\end{overpic}
\hfill
\begin{overpic}[height=2.3\columnwidth]{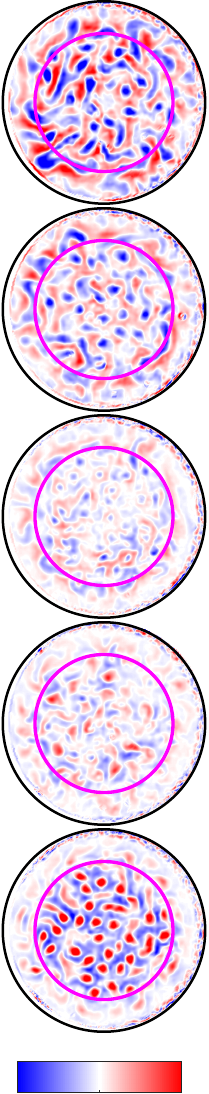}%
\put(-3.0,106){\normalsize (c) $\Ek = 7.1 \times 10^{-6}$, $\Rt = 2.9$}
\put(-2,103.5){\normalsize $\Ra^*_q = 5.7 \times 10^{-8}$, $\Ro = 0.038$}
\put(0,101){\parbox{0.43\columnwidth}{\centering\normalsize Columnar}}
\put(0,3.5){-0.2}
\put(0,3.5){\parbox{0.43\columnwidth}{\centering 0}}
\put(15.5,3.5){0.2}

\linethickness{1.5pt}
	\put(4.5,88.5){\color{black}\vector(-1,-1){6}}
\put(-3,81.5){\normalsize $j_{ctc}$}
\put(3.8,-2){\normalsize $\omega_z(x,y,t_i)/\Omega$}
\end{overpic}
\hfill
\begin{overpic}[height=2.3\columnwidth]{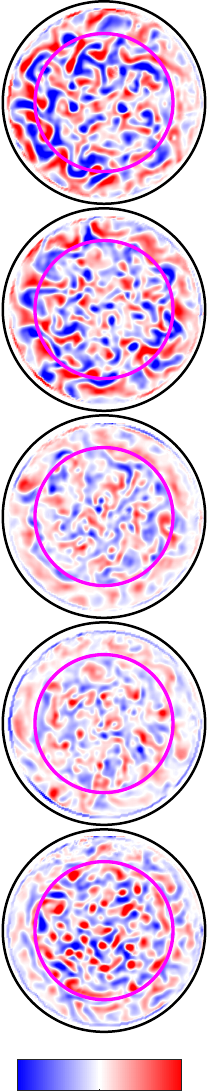}%
\put(-3.0,106){\normalsize (d) $\Ek = 4.7 \times 10^{-6}$, $\Rt = 4.3$}
\put(-2,103.5){\normalsize $\Ra^*_q = 1.3 \times 10^{-7}$, $\Ro = 0.04$}
\put(0,101){\parbox{0.43\columnwidth}{\centering\normalsize Plumes}}
\put(0,3.5){-0.2}
\put(0,3.5){\parbox{0.43\columnwidth}{\centering 0}}
\put(15.5,3.5){0.2}

\linethickness{1.5pt}
\put(5.7,95){\color{black}\vector(-1,-1){6.5}}
\put(-2,87.5){\normalsize $j_{p}$}  
\put(3.8,-2){\normalsize $\omega_z(x,y,t_i)/\Omega$}
\end{overpic}
\vspace{1em}

\caption{Snapshot of the vorticity field in the horizontal plane $z$ = $H/6$, $H/3$, $H/2$, $2H/3$ and $5H/6$ (bottom to top) for different level of criticality, (a) $Ek$ = $4.7\times$ 10\textsuperscript{-6}, $\Rt$ = 0.86, (b) $Ek$ = $7.1\times$ 10\textsuperscript{-6}, $\Rt$ = 1.5, (c) $Ek$ = $7.1\times$ 10\textsuperscript{-6}, $\Rt$ = 2.9, and (d) $Ek$ = $4.7\times$ 10\textsuperscript{-6}, $\Rt$ = 4.3. The black line represents the boundary of the glass cylinder, and the purple line represents the position of the heater that defines the TC. The inset in figure (b) for $z = 2H/3$ highlights the length scale at the onset of the rotating convection in LEE2 is similar to as predicted by \cite{chandrasekhar1961} for rotating convection in a cylinder (shown using solid green line). \textcolor{black}{$j_{ctc}$ and $j_p$ in figures \ref{fig:Flow_Snapshots_lowRa}c and \ref{fig:Flow_Snapshots_lowRa}d at $z$ = $5H/6$ highlight example of jets in the Convective Taylor columns and plumes regimes, respectively.}}
\label{fig:Flow_Snapshots_lowRa}
\end{figure*}

To validate these estimates and quantify how the convective Rossby number $\Ro$ (\textcolor{black}{input parameter}) reflects them, we now measure the role of rotation from these forces using the local Rossby number $\Ro_l$ (\textcolor{black}{output parameter}) as the ratio of the inertial to Coriolis forces $\Ro_l = \|\Fi\|/\|\Fc\|$, following \cite{aguirre-guzman2021jfm}. 
The variations of $\Ro_l$ with $\Rt$ are shown on figure \ref{fig:RossbyNumbers}a.
First, $\Ro_l$ reaches at most $\simeq0.2$ across all $\Rt$, suggesting that the convection is strongly rotationally constrained across all regimes (figure \ref{fig:RossbyNumbers}a) and that a buoyancy-dominated regime is probably not reached, even at the highest levels of supercriticality we have investigated ($\Rt=191$). 
The rotational constraint is especially strong as the values of $\Ro_l$ are lower than those in the plane layer for a similar range of criticality (see figure 3b in \cite{aguirre-guzman2021jfm}), perhaps as a result of the slightly higher Prandtl number we are considering (6.4 \emph{v.s.} 5). Despite this difference, the variations of $\Ro_l$ with $\Rt$ follow a very similar pattern to that observed in plane layer rapidly rotating convection by \cite{aguirre-guzman2021jfm}: $\Ro_l$ follows a scaling of $\Ro_l \sim \Rt^2$ for $\Rt \lesssim 4$ (figure \ref{fig:RossbyNumbers}a). This scaling is indicative of a Viscous-Archimedean-Coriolis force balance (VAC) with cellular and columnar structures \citep{kingetal2013}, which, unlike \cite{aguirre-guzman2021jfm}, we could not capture from the analysis of individual forces, for lack of a precise measure of the buoyancy force, especially at low $\Rt$. 
From $\Rt\gtrsim2\pm1$, the variations of $\Ro_l$ are closer to $\Ro_l\sim \Rt^{0.8}$, similar to the findings of \cite{aguirre-guzman2021jfm}, who found a crossover between the two scalings around $\Rt\gtrsim6$. Hence, 
inertia becomes larger than viscous forces in LEE2 at half the criticality where it does so in plane layer.

\textcolor{black}{A caveat in this analysis is that the Rossby numbers and forces on figure \ref{fig:RossbyNumbers}(a) are global quantities averaged over the TC when the TPC constraint can be violated locally. What matters in this case is the local inertia associated to the velocity patterns that cross the TC. 
{We define the Rossby number $Ro_{\omega}=\langle (F_{10}(z)\omega_z)^2\rangle^{1/2}/\Omega$ associated to these patterns using spatial variations of vorticity, that capture better the smaller scales. Additionally, to focus on the most intense events, we use a filter $F_{10}(z)$ to retain only the 10\% most intense values in the plane of height $z$.}
This quantity is plotted for each case on a new figure \ref{fig:RossbyNumbers}(b) against $\widetilde{Ra}$. It shows that local inertia due to the 10\% most intense isolated structures is at least one order of magnitude greater than suggested by the value of $Ro_l$ and that it even becomes comparable to the Coriolis force very low criticality. For $\widetilde{Ra}\simeq10$ the average inertia associated to the top 10\% most intense values $\omega$ even exceeds the Coriolis force. This shows that it is local inertia that breaks the TPC at the TC boundary.}

To summarize, the global force balance analysis informs us that the convection we investigate up to $\Rt \approx 191$ is always rotation-dominated or at least rotation-influenced, never buoyancy-dominated. At low criticality we reach the regime where viscous force dominates inertial force and $\Ro_l \sim \Rt^2$ and at higher criticality inertial force dominates over viscous force and $\Ro_l \sim \Rt^{0.8}$. In the plane layer rotating convection these scalings reflected the presence of cellular and columnar structures  ($\Ro_l \sim \Rt^2$) and plumes ($\Ro_l \sim \Rt^{0.8}$). \textcolor{black}{Nevertheless, inertia can still exceed the Coriolis force locally and temporarily, even at low criticality.}  We shall now identify and characterize the regimes that underpin these scalings in LEE2, using time-resolved velocity fields. 
\subsection{Subcritical regime, $\Rt<1$}
\label{subsec:subcritical_regime}
In the subcritical regime, illustrated by a snapshot at $\Rt = 0.86$  on figure \ref{fig:Flow_Snapshots_lowRa}a, the vorticity signal inside the TC is close to noise level, \emph{i.e.} typically an order of magnitude lower than the vorticity outside the TC, and nearly two orders of magnitude lower than inside the TC for $\Rt\geq1$. This means that within the precision of our measurements, we cannot detect convection inside the TC. This is consistent with the subcritical regime inside a CSIIW with one very important difference: we see no evidence of wall modes along the inside boundary of the TC, even in regimes far beyond their theoretical onset for a cylinder with either solid or free-slip side walls \citep{zhangliao2009}: $\Rtwns = 13.8$ on figure \ref{fig:Flow_Snapshots_lowRa}a). While wall modes are normally fainter and set in at higher thermal forcing with free-slip than with no-slip boundary conditions \citep{zhangliao2009}, they are very resilient to changes at the cylinder boundaries \citep{eckeshishkina2023,favierknobloch2020}. They may be partially suppressed local thermally conducting inserts at a cylindrical wall \citep{eckeshishkina2023} or totally suppressed by thin fins of thickness $\mathcal{O}(\Ek^{1/3})$ protruding from flat walls \citep{terrienetal2023}. By contrast with these studies, there are no wallmodes in the presence of the TC in the subcritical regime. {This is consistent with the results of LEE1 where wallmodes were not observed in the subcritical regime either. 
Both experiments feature TCs with similar aspect ratios, and in both cases, measurements in subcritical regimes were made at values of $Ra$ well above the predicted onset of wallmodes for both free-slip and no-slip boundary conditions \citep{vasil2025_jfm} (The friction at the TC boundary is somewhere between these two ideal cases). However, unlike in LEE1, wallmodes are not detected for $\Rt>1$ either.
The most obvious difference to explain this difference is the presence of inertia outside the TC in LEE2, which is absent at low criticality in LEE1.} 

{Indeed, the proximity of the vertical cold outer wall next to the TC in LEE2's cylindrical vessel incurs a much higher level of baroclinicity than in LEE1's hemispherical vessel.} Because of it, a base toroidal flow always exists outside the TC region in LEE2, regardless how low the thermal forcing. This is visible on the time- and azimuthally averaged radial profiles of radial velocities and axial vorticities on figure \ref{fig:urRMS_subcritical}. {The baroclinic nature of this flow is established by comparing its $z-$variation to the predicted scalings for thermal winds \citep{maxworthy1994_jpo,aurnouetal2003,aujogueetal2018}, see appendix \ref{sec:zonal_flow}.
\textcolor{black}{Co-rotating vortices exist in this region outside the TC in the subcritical regime (figure \ref{fig:Flow_Snapshots_lowRa}a) whose proximity with the TC boundary may interfere with the region where wall modes would form and so potentially suppress their onset}. No such flow was detected at low criticality  LEE1.}

\begin{figure}
	\centering
	\begin{subfigure}[t]{0.49\linewidth}
		\centering
		\begin{overpic}[width=\linewidth]{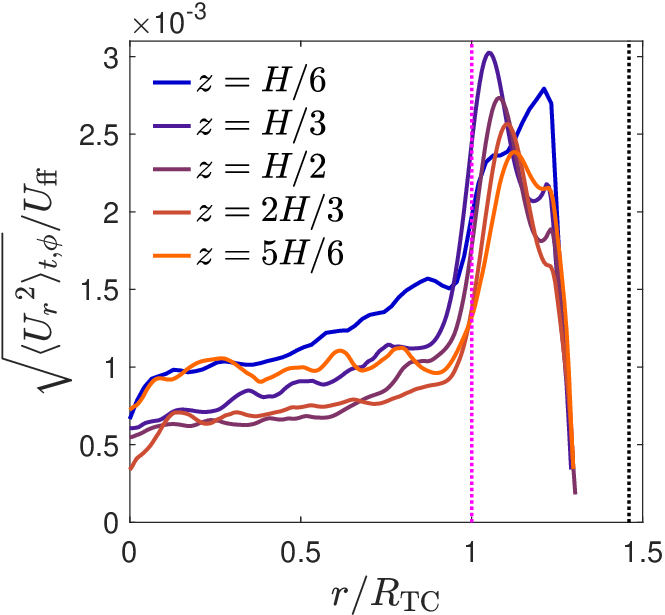}
				\put(2,84){\normalsize (a)}

		\end{overpic}
	\end{subfigure}
	\hfill
	\begin{subfigure}[t]{0.49\linewidth}
		\centering	
				\begin{overpic}[width=\linewidth]{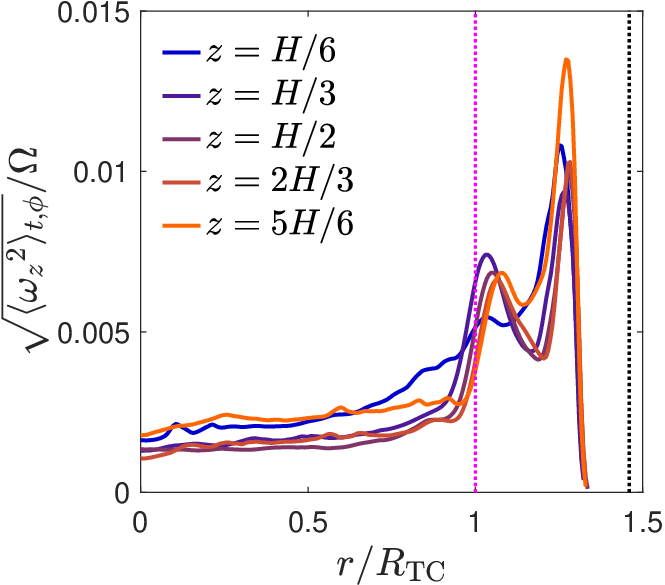}
			\put(0,84){\normalsize (b)}	
		\end{overpic}
	\end{subfigure}
	\caption{Time- and azimuthal-averaged radial profiles of r.m.s of (a) radial velocities ($\sqrt{\langle U_r^2 \rangle_{t,\phi}}$) normalized by the free-fall velocity ($U_{\rm ff}$) and (b) axial vorticity ($\sqrt{\langle {\omega_{z}}^2 \rangle _{t,\phi}}$) normalized by rotation rate ($\Omega$) for $\Ek = 4.7 \times 10^{-6}$, $\Rt = 0.86$, (snapshots on figure \ref{fig:Flow_Snapshots_lowRa}a, corresponding to $\Ra^*_q = 2.2 \times 10^{-9}$ and $\Ro = 0.018$). Dotted vertical purple and black lines indicate the edge of the TC and the solid wall, respectively. We can see a high value of the $\sqrt{\langle U_r^2 \rangle_{t,\phi}}$ and $\sqrt{\langle {\omega_{z}}^2 \rangle _{t,\phi}}$ outside the TC suggesting a presence of baroclinically driven flow in the subcritical regime.}
	\label{fig:urRMS_subcritical}
\end{figure}

These vortices are similar to the rim instability observed earlier in a TC embedded in a hemispherical vessel \citep{aurnou2007}. 
{These vortices are more intense} near the top plane at $z=5H/6$ and almost no trace of them subsists in the lowest PIV plane (figure \ref{fig:Flow_Snapshots_lowRa}a, $z=H/6$ and figure \ref{fig:urRMS_subcritical}b). {This suggests that baroclinicity is stronger near the top.}
{Their occurrence can be undestood from the similarity between the region outside the TC with an outer cold wall and hotter TC boundary, and a differentially heated rotating anulus. Despite the differences in mechanical and thermal boundary conditions at the TC wall, both confiurations are driven by the  baroclinic convection due to the radial temperature gradient \citep{fowlis1965_jas,hide1975_ap,scolan2017_ef,hignett1985_qjrms}. Indeed the base axisymmetric flow shares the same topology in both cases \citep{lewis2004_gafd}. In the annulus, the flow becomes unstable to a practically steady wave pattern above a critical value of the thermal Rossby number built on the free-fall velocity based on the gap $D$ width $Ro_t=(g\alpha \Delta T_z H/{\Omega ^2 D}= (\Ro H/D)$. From the ($Ro_t,\Omega$) stability maps established by \cite{fowlis1965_jas} (fig. 3) for an experiment in water of similar dimensions to the annulus outside LEE2's TC, 
the vortices we observe outside the TC at $\Ro=0.18$, $\Ek=4.7\times10^{-6}$ (\emph{i.e.} $\Ro_t=5.2$ and $\Omega=\SI{0.54}{rad/s}$) are right in regime predicted for the occurence of steady waves. The experiments only report the onset wave number of 8, which is smaller than the $\sim$ 24 vortices we observe, bearing in mind the differences between the setups and the fact that our observations are well beyond the onset.}  
\subsection{Rotation-dominated regime}
\label{subsec:Rotation-dominated regime}
\subsubsection{Cellular regime}
\label{subsubsec:Cellular regime}
At $\Rt=1.5$, vorticity contours in the five PIV planes reveal columnar vortices, that are mostly steady (figure \ref{fig:Flow_Snapshots_lowRa}b). 
Their typical size is consistent with the onset length scale of plane rotating Rayleigh-B\'enard convection \citep{chandrasekhar1961}, shown as inset for $z = 2H/3$ for $\Rt=1.5$ (figure \ref{fig:Flow_Snapshots_lowRa}b). 
The vorticity in each structure changes sign between the top and bottom of the cylinder indicating a progressive reversal of rotation along the cylinder axis, approximately symmetric about the middle plane (figures \ref{fig:Flow_Snapshots_lowRa}b). 
\begin{figure*}
\vspace{2cm}
\hspace*{4em}
	\begin{overpic}[height=2.3\columnwidth]{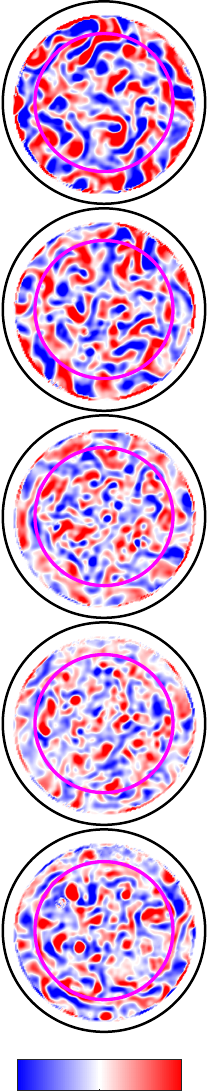}
		\put(-3.0,106){\normalsize (a) $\Ek = 7.1 \times 10^{-6}$, $\Rt = 7.4$,}
		\put(-0.8,103.5){\normalsize $\Ra^*_q = 5.0 \times 10^{-7}$, $\Ro = 0.06$}
		\put(0,101){\parbox{\columnwidth}{\centering\normalsize Disrupted columns}}
		
		\linethickness{1.5pt}
		\put(16.5,90){\color{black}\vector(2,-1){4.5}}
		
		\put(-5,14){\normalsize $H/6$}
		\put(-5,33){\normalsize $H/3$}
		\put(-5,52){\normalsize $H/2$}
		\put(-5,71){\normalsize $2H/3$}
		\put(-5,90){\normalsize $5H/6$}
		
		\put(3.2,-1.7){\normalsize $\omega_z(x,y,t_i)/\Omega$}
\put(0,3.5){-0.2}
\put(0,3.5){\parbox{0.43\columnwidth}{\centering 0}}
\put(15.5,3.5){0.2}
		
	\end{overpic}
	\hfill
	\begin{overpic}[height=2.3\columnwidth]{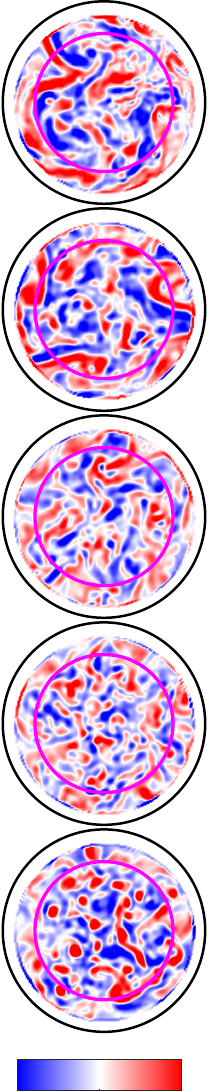}
		\put(-3.0,106){\normalsize (b) $\Ek = 1.4 \times 10^{-5}$, $\Rt = 19$,}
		\put(-0.8,103.5){\normalsize $\Ra^*_q = 4.3 \times 10^{-6}$, $\Ro = 0.12$}
		\linethickness{1.5pt}
		\put(6,93){\color{black}\vector(-1,-1){6}}
		\put(-1.5,86){\normalsize $j_{dc}$}
		\put(3.2,-1.7){\normalsize $\omega_z(x,y,t_i)/\Omega$}
\put(0,3.5){-0.5}
\put(0,3.5){\parbox{0.43\columnwidth}{\centering 0}}
\put(15.5,3.5){0.5}

	\end{overpic}
	\hfill
	\begin{overpic}[height=2.3\columnwidth]{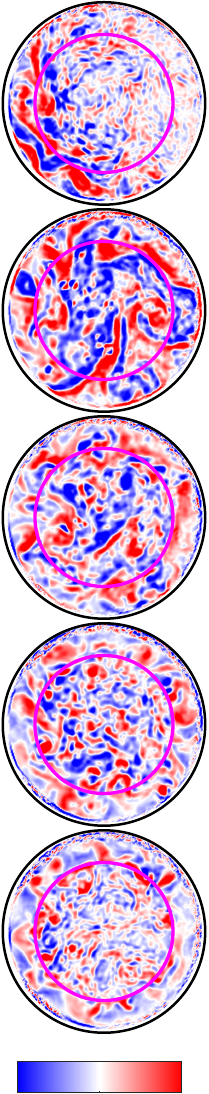}
		\put(-3.0,106){\normalsize (c) $\Ek = 1.4 \times 10^{-5}$, $\Rt = 37$,}
		\put(-0.8,103.5){\normalsize $\Ra^*_q = 1.1 \times 10^{-5}$, $\Ro = 0.17$}
		\put(0,101.3){\parbox{\columnwidth}{\centering\normalsize Rotation-influenced regime}}
		
		\put(3.2,-1.7){\normalsize $\omega_z(x,y,t_i)/\Omega$}
\put(0,3.5){-0.5}
\put(0,3.5){\parbox{0.43\columnwidth}{\centering 0}}
\put(15.5,3.5){0.5}
		
	\end{overpic}
	\hfill
	\begin{overpic}[height=2.3\columnwidth]{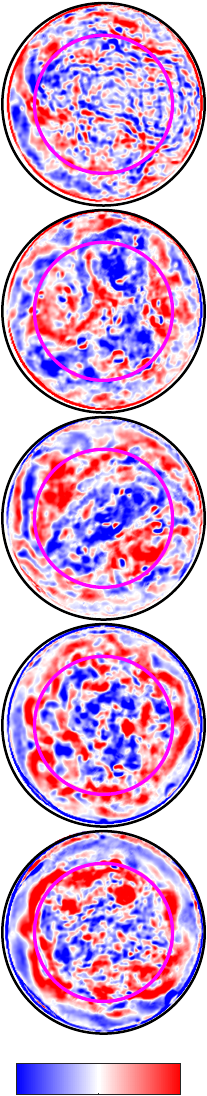}
		\put(-3.0,106){\normalsize (d) $\Ek = 4.2 \times 10^{-5}$, $\Rt = 160$,}
		\put(-0.8,103.5){\normalsize $\Ra^*_q = 2.8 \times 10^{-4}$,  $\Ro = 0.52$}
		
\put(0,3.5){-1}
\put(0,3.5){\parbox{0.43\columnwidth}{\centering 0}}
\put(15.5,3.5){1}
		\put(3.2,-1.7){\normalsize $\omega_z(x,y,t_i)/\Omega$}
		
	\end{overpic}
	\vspace{1em}
	\caption{Snapshots of the vorticity field in the horizontal plane $z$ = $H/6$, $H/3$, $H/2$, $2H/3$ and $5H/6$ (bottom to top) for different level of criticality, (a) $Ek$ = $7.1\times$ 10\textsuperscript{-6}, $\Rt$ = 7.4, (b) $Ek$ = $1.4\times$ 10\textsuperscript{-5}, $\Rt$ = 19, (c) $Ek$ = $1.4\times$ 10\textsuperscript{-5}, $\Rt$ = 37, and (d) $Ek$ = $4.2\times$ 10\textsuperscript{-5}, $\Rt$ = 160. The black line represents the boundary of the glass cylinder, and the purple line represents the position of the heater that defines the TC. $j_{dc}$ in figures \ref{fig:Flow_Snapshots_highRa}a, b at $z$ = $5H/6$ highlights example of jets in the `disrupted column' regime.}
	\label{fig:Flow_Snapshots_highRa}
\end{figure*}
In the region near the inner TC boundary, however, the pattern tends to be suppressed. This phenomenon is common in CSIIW \citep{Zhong1993} and is usually attributed to two intertwined phenomena: first, curved solid walls suppress the columns where they cross the lattice. Second, strong wall modes may develop that supersede columns within lattice 
\citep{hornschmid2017}. In LEE2, however, wall modes are not present. Also, 
the columnar cells are strongly suppressed in the upper two planes where baroclinicity in the region outside the TC is strongest. By contrast, it is less effective near the inner TC boundary in the bottom two planes where baroclinicity is weaker. The middle plane exhibits the faintest pattern everywhere in the TC since it captures the point where the vorticity changes sign. These differences suggest that here again, the baroclinically-driven flow in the region outside the cylinder, which is also stronger in the upper part of the vessel, may contribute to the suppression of the flow near the inner side of the TC boundary. \textcolor{black}{Appendix \ref{sec:zonal_flow} shows an example of how the measured radial temperature difference compares to a radial temperature difference estimated from the intensity of the zonal flow, assuming the later is driven by a thermal wind mechanism. Measured and estimated values match well over a range of criticalities from $\Rt \approx 1.5$ to 9.2. Below this range, large baroclinic vortices dominate the region outside the TC and drift more slowly than the zonal flow and beyond this range, inertia becomes sufficiently strong to invalidate the thermal wind balance.} 

Inspection of the region outside the TC brings further support to this hypothesis. At the top plane ($z=5H/6$), the anticyclonic vortices observed in the subcritical regimes have become larger and more intense. They are separated by layers of cyclonic vorticity and both of them extend radially into a region closer to the outer vessel solid wall, 
These structures are less pronounced as $z$ decreases and give way to faint, smaller structures in the bottom two planes ($z=H/3$ and $z=H/6$). 
{In summary, the suppression of the convective columns near the inner TC boundary increases with the intensity of the vortices outside the TC. 
This suggests that the suppression may result from the inertia induced by these vortices rather than from the presence of the TC boundary itself.}

\subsubsection{Convective Taylor columns}
\label{subsubsec:Taylor Columns regime}
At $\Rt=2.9$, columns inside the TC are still identifiable, with the same individual features as at $\Rt=1.5$ but they are no longer static \citep{sakai1997}. They evolve in a chaotic way, evidenced by their different positions in all five planes (figure \ref{fig:Flow_Snapshots_lowRa}c). This corresponds to the regime of \emph{convective Taylor columns} \citep{julienetal1996,julienetal2012,kunnen2021,eckeshishkina2023}.
Unlike in the cellular regime, the flow near the inner boundary of the TC is no longer suppressed in the upper two planes ($z=2H/3$ and $z=5H/6$). This is caused by thin vorticity filaments crossing the TC boundary there, and linking up to larger structures inside the TC that touch its boundary. This indicates a local breakdown of the TPC allowing a radial flow through the TC boundary. 
Indeed, the vortices in the region outside the TC now evolve more chaotically than at $\Rt=1.5$. The higher value of the global Rossby number $\Ro=0.038$, and the fact that these vortices carry the highest level of vorticity in the top plane suggest that their higher inertia allows them to break the TPC there to induce these local jets \textcolor{black}{($j_{ctc}$, shown in figure \ref{fig:Flow_Snapshots_lowRa}c at $z = 5H/6$)}. The phenomenon is less pronounced at $z\leq2H/3$, and does not generate enough inertia to break the TPC in these lower planes ($z\leq H/2$).
\subsubsection{Plumes and disrupted columns}
\label{subsec:Plumes and disrupted columns}
At $\Rt=4.3$ (figure \ref{fig:Flow_Snapshots_lowRa}d), and even more so at $\Rt=7.4$ (figure \ref{fig:Flow_Snapshots_highRa}a), vorticity snapshots exhibit the signature of thermal plumes found in several numerical studies at slightly higher levels of criticality (at $\Rt= 7.5$ for $Pr=5.5$ by \cite{aguirre-guzman2021jfm}): Almost circular spots of intense anticyclonic (\emph{resp.} cyclonic) vorticity are visible in the bottom plane at $z=H/6$, (\emph{resp.} top plane at $z=5/6$), but these are practically absent in the mid-plane, suggesting that they do not extend over the entire cylinder height any more. Hence the columns are now disrupted and give way to plumes.

Unlike earlier studies with either adiabatic or periodic boundary conditions at their side boundary, inertia generated outside LEE2's TC boundary is mostly responsible for the disruption of  the columns: at $\Rt=4.3$, the jets induced by the vortices there penetrate deep into the TC in the top plane, leaving only a few plumes undisturbed near its centre \textcolor{black}{($j_{p}$, shown in figure \ref{fig:Flow_Snapshots_lowRa}d at $z = 5H/6$)}. This phenomenon is even more pronounced at $\Rt=7.4$ \textcolor{black}{($j_{dc}$, shown in figure \ref{fig:Flow_Snapshots_highRa}a at $z$ = $5H/6$)}. At $\Rt=19$, there is no trace of the plumes in any of the four top planes where the baroclinic source of inertia is strongest, but a few remain in the bottom plane where it is weakest. In the top two planes, jets crossing the TC boundary have grown to the point where they connect the inside of the TC to the outer vessel cold wall, and so effectively spread the upward heat flux to the upper part of the side wall boundary \textcolor{black}{($j_{dc}$, shown in figure \ref{fig:Flow_Snapshots_highRa}b at $z$ = $5H/6$)}. Whilst most obvious in the top two planes at $\Rt=7.4$, this effect takes place in all but the bottom one at $\Rt=19$, thereby considerably increasing heat transfer. 

Accordingly, the TPC is clearly violated in 4 of the 5 PIV planes at $\Rt=4.3$, with structures straddling the TC boundary in all four planes and similar levels of vorticity on either side of it. Only at the lowest plane $z=H/6$ does the TC boundary separate the intense convective plumes within the TC from levels of vorticity an order of magnitude smaller outside it. At $\Rt=7.4$, vorticity levels are balanced either side of the TC boundary in all planes but only a few weak jets straddle it in the lower plane, where the TPC remains locally influential.

As for plane layer rotating convection at $\Prt=5$ \citep{aguirre-guzman2021jfm}, the transition from convective Taylor columns to plumes occurs when inertia becomes dominant over viscous forces, at the transition between scalings $\Ro_l\sim \Rt^2$ and $\Ro_l\sim \Rt^{0.8}$, albeit at a lower criticality. Hence, the convective states underlying these two scaling regimes are qualitatively the same in both cases but differ in two ways: First, as the TC boundary breaks, inertia in the upper region outside the TC locally enhances inertia within the TC.
Second, the simulations by \cite{aguirre-guzman2021jfm} are conducted at a slightly lower Prandtl number than our experiments. It is unclear, however whether this difference is significant, and if it is, lower Prandtl numbers rather favour higher flow velocities, which translates into higher inertia in horizontal planes \citep{abbate2023gafd}.
Hence baroclinically-induced inertia at the TC boundary is the more likely explanation as to why the transition to plumes occurs at lower criticality in the TC than in the plane layer. \textcolor{black}{Indeed, while baroclinicity or baroclinic instabilities are not inertial in nature, baroclinicity drives a motion whose finite velocity has inertia associated to it.}   
\subsection{Turbulence and large scales in the rotation-influenced regime}
\label{subsec:buoyancy dominated regime}
For $\Rt\geq37$, no clear trace of plumes is left and all four planes show small-scale vorticity fluctuations down to the level of PIV resolution (figures \ref{fig:Flow_Snapshots_highRa}c, d). The finer scales are more visible in the upper and lower planes at $z = 5H/6$ and $z = H/6$, indicating that they are clearly three-dimensional. At the other end of the spectrum of length scales, larger patches of vorticity appear. While the successive recordings in the different planes do not permit to conclude whether these patches extend across the entire height, their number between 2 and 4 is consistent between the planes. Their horizontal structure resembles the large scale vortices observed at the transition between the rotation-dominated and rotation-influenced regime in classical RRBC \citep{aguirre-guzman2021jfm}. This is also consistent with the regime of still strong rotational constraint in which they exist. Indeed, the force balance in section \ref{subsec:Force Balance} shows that even at $\Rt\geq37$ the Coriolis force largely dominates the inertial force ($\Ro_l \lesssim 0.2$), which is expected for the large-scale vortices (LSV) or geostrophic turbulence \citep{aguirre-guzman2021jfm}. However, in the absence of an estimate of the \textcolor{black}{pressure force} $\Fp$, we cannot check whether $\Fp$ and $\Fc$ are in still in balance as would be expected within a geostrophic regime. 
Additionally, the snapshot at $z=2H/3$ suggests that the formation of these large structures may be driven by the reduced number of strong vorticity sheets extending from the vessel cold walls through the TC boundary up to the centre of the TC. 

\textcolor{black}{It is difficult to identify the geostrophic turbulence regime, and especially LSVs with horizontal PIV planes only because these do not show the $z-$dependence of the turbulent structure. Additionally, the heat and momentum fluxes at the TC boundary affect the diffusivity-free scaling in LEE2, so those may not be a reliable signature of the regime of geostrophic turbulence either.
Hence, to identify regimes where GT is likely,  we simply assessed whether the flow was turbulent from its chaotic nature, and whether at least the large scales are quasi-geostrophic from the global values of $\Ek$ and $\Ro_l$. 
The basis for this idea is that structures for which $\Ek\ll1$ and $Ro\ll1$ are necessarily $z-$independent in the bulk (to $\mathcal O(\Ek,\Ro)$) due to the Taylor-Proudman constraint.} Further work would be necessary to verify whether these states indeed correspond to geostrophic turbulence and whether these structures are indeed the LSV predicted by \cite{aguirre-guzman2021jfm}. Nevertheless, it is undeniable that such large-scale structures and the fine-structure turbulence observed at $\Rt=37$ are usually observed at higher criticality ($\Rt\simeq80$ in \cite{aguirre-guzman2021jfm}) than in LEE2. Hence the breakup of the TPC at the TC boundary induced by baroclinicity seems to favour transition to turbulent regimes and the appearance of large structures at lower criticality than in classical rotating Rayleigh-B\'enard Convection. We shall seek further indications of this effect in the heat transfer signature of these regimes. 

\section{Heat Transfer Scalings}
\label{sec:Heat Transfer Scalings}
\subsection{$\Nu$ \emph{vs.} $\Ra$ laws}
\begin{figure}
	\centering
	\begin{minipage}[t]{\columnwidth}
		\centering
		\begin{overpic}[width=\columnwidth]{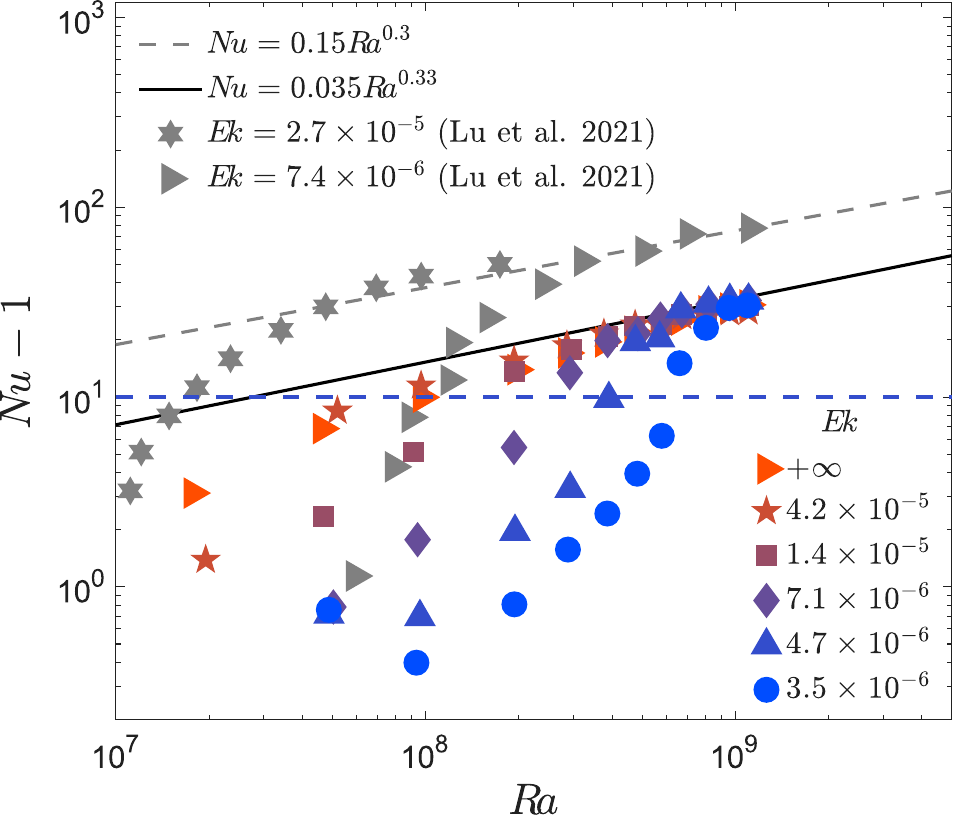}
			
							\put(1,82){\normalsize (a)}
		\end{overpic}

		\label{fig:overpic1}
	\end{minipage}
	\hfill
	\vspace{0.2cm}
	\begin{minipage}[t]{\columnwidth}

		\begin{overpic}[width=\columnwidth]{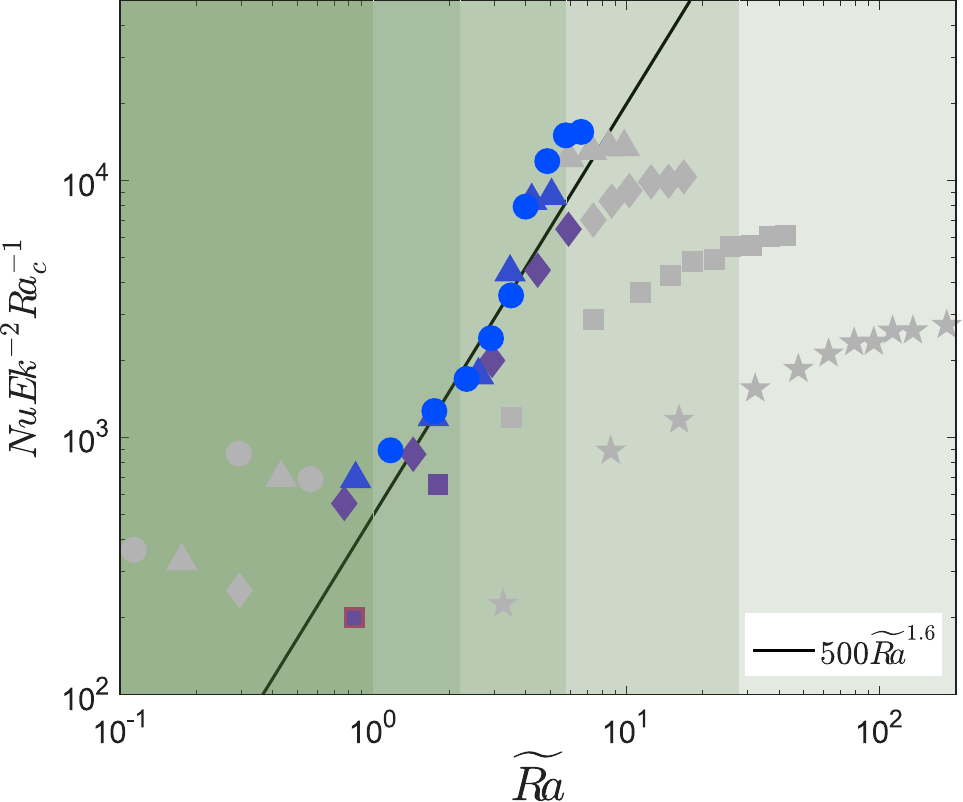} 
				\put(1,82){\normalsize (b)}
				\put(28,78.0){\normalsize SR}
				\put(42,78.0){\normalsize CR}
				\put(49,78.0){\normalsize CTC}
				\put(66,78.0){\normalsize P}
				\put(85,78.0){\normalsize GT}	
		\end{overpic}

		\label{fig:overpic2}
	\end{minipage}
	
	\caption{a) Variations of the Nusselt number $\Nu - 1$ with $\Ra$ for six value of $\Ek$ in LEE2. Black solid line and grey dashed line in (a) indicate $\Nu - 1 = 0.15\Ra^{0.3}$ and $\Nu - 1 = 0.15\Ra^{0.3}$, respectively. Grey star and triangle symbols show \cite{luetal2021}'s experimental data for rotating convection in a CSIIW with $\Gamma = 3.8$, $\Ek = 2.7 \times 10^{-5}$ and  $\Gamma = 2.0$, $\Ek = 7.4 \times 10^{-6}$, respectively. (b) Variation of $\Nu \Ek^{-2} \Ra_c^{-1}$ vs $\Rt$. In figure (b) different coloured symbols show data for same $\Ek$ as in (a). The background colours identify regimes as on figure \ref{fig:forcebalance}.
}
	\label{fig:Nu_Ra_Classical}
\end{figure}
Temperatures recorded simultaneously with the velocity enable us to quantify heat transfer with the classical Nusselt number $\Nu$, which represents the ratio of the heat flux to a conductive heat flux,
{
\begin{equation}	
	\Nu = \frac{P}{\pi R_{\rm TC}^2Q_c}.
\end{equation}
Here, the conductive heat flux $Q_c$ is calculated by solving the Laplace equation for the temperature on a meridional section of the fluid vessel assuming zero velocity in the entire domain, axisymmetry, $T=\Delta T$ at the heater surface, $T=0$ at the top and side walls and adiabaticity on all other boundaries. The solution is obtained numerically using the second-order finite elements method implemented in the MATLAB PDE toolbox, to a precision better than $1\%$ on $Q_c$. However, unlike in cylinders with adiabatic side walls, the conductive state cannot be achieved in the experiment because of the baroclinic convection driven at arbitrary low $\Delta T$. By contrast to the definition of the Nusselt number for CSIIW where the heat flux is normalised by the axial conductive flux per unit surface $\kappa\Delta \!T_{{\rm z}}/H$ only, 
$Q_c$ incorporates axial and lateral heat fluxes. As a result, the values of $\Nu$ are lower than for CSIIW by a factor $Q_c/(H\kappa\Delta \!T_{{\rm z}})\simeq 5.05$.
The dimensional experimental heat flux is calculated from the electrical power supplied to the resistance inside the heating element assumed entirely passed through the top surface of the heater}. Therefore it represents the `total heat' transferred across the entire convection cell. 
The variations of $\Nu$ with $\Ra$ are reported in figure \ref{fig:Nu_Ra_Classical}. 

In the high $\Ra$ limit, the heat transfer becomes independent of the rotation (\emph{i.e.} of $\Ek$) and follows the buoyancy-dominated scaling of the form $\Nu \sim \Ra^{1/3}$. This regime is only reached for the highest thermal forcing, even though the values of the Rossby number for these cases ($\Ro_l\lesssim0.2$) indicate a strong rotational constraint on the large scales: this suggest that a significant part the heat flux is carried by flow structures that are less rotationally constrained, such as those \textcolor{black}{induced} by baroclinicity near the TC boundaries that break the TPC there. 

{For lower $\Ra$ values, the Nusselt number follows the scaling $\Nu \sim \Ek^2\Ra^{3/2}$ 
In a CSIIW, the rapid rise in $\Nu$ from the conduction value is attributed to the nonlinear growth from onset within the quasi-geostrophic regime where rotation dominates \citep{eckeshishkina2023}. This tendency is recovered in LEE2, 
but in the rapidly rotating regime, the enhanced heat transfer at low $\Ra$ due to baroclinic convection outside the TC lifts the tail of the shingles and this hides the lower part of the usual $\Nu \sim \Ek^2\Ra^{3/2}$(figure \ref{fig:Nu_Ra_Classical}b)}}.

The most noticeable difference with plane layers and CSIIW, takes place in the subcritical regime: In CSIIW, wall modes are present \citep{Zhong1991,eckeetal2022}. Their contribution to heat transport incurs a moderate increase of the Nusselt number in this regime, typically $\Nu\lesssim3$ \citep{rossby1969,Zhong1993}. Therefore, experiments in CSIIW commonly shows a sharp increase in $\Nu$ values from the conduction case ($\Nu = 1$) once the onset for bulk-mode convection is reached, with relatively limited effect of wall-mode-driven heat transfer in this regime \citep{chengetal2015,chengetal2018}. In LEE2 by contrast, we observe {some enhanced heat transfer in the subcritical regime, with $Nu>1$}. 

This is visible left of the inflection point in figure \ref{fig:Nu_Ra_Classical}. While no wall modes are present in the TC's subcritical regimes, the \textcolor{black}{toroidal} baroclinically-driven flow just outside the cylinder destabilises into a ring of vortices (figure \ref{fig:Flow_Snapshots_lowRa}a). Such vortices were also observed by \cite{aurnou2007}, who noted their helical nature, and associated them to a `rim instability'. They were found to distribute the heat in the interior and exterior of the TC. The presence of baroclinic instability has also been observed in other tangent cylinder-like geometries \citep{jacobsivey1998,cuistreet2001}. Both the base baroclinic flow and the vortices occurring as the result of the rim instability feature a strong convective motion capable of efficiently transporting heat. Hence, we argue that the increased value of $\Nu$ in the subcritical regime in LEE2, compared to the values in CSIIW, results from the baroclinically-driven flow near the TC boundary. 
{Note that except for the subcritical regime, the variations of $\Nu$ when corrected for the heat flux contributing to the overall heating of the fluid volume show the same overall behaviour, only with a shift in value (see Appendix \ref{appendix:Nu_Correction}). This indicates that the small fraction of heat lost in this way does not affect the overall heat transfer mechanisms.}
\subsection{Relative Heat flux $\Nu/\Nu_0$ \emph{vs.} $\Ro$}
\begin{figure}
	\centering					
	\begin{overpic}[width=\columnwidth]{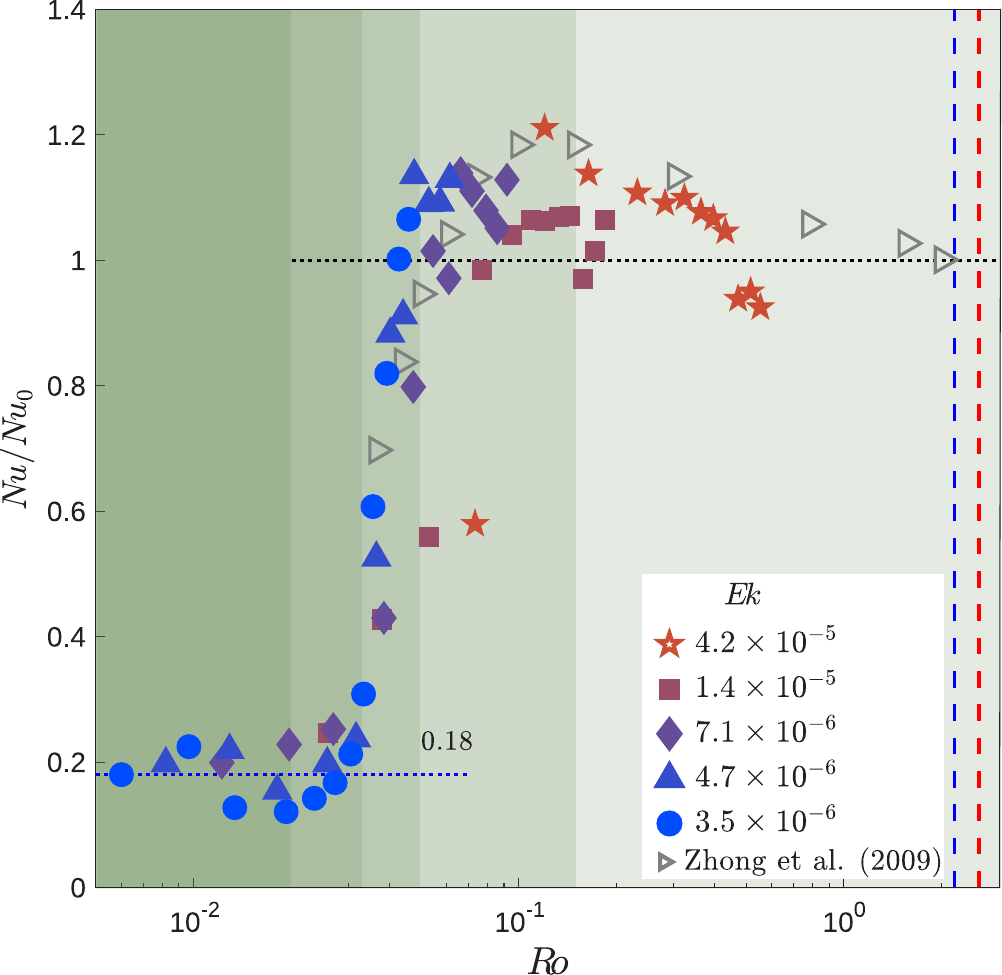}
		
		\put(20,90){\normalsize SR}
		\put(31,90){\normalsize CR}
		\put(37,90){\normalsize CTC}
		\put(48,90){\normalsize P}
		\put(67,90){\normalsize GT}	
		 \end{overpic}						
	\caption{Ratio of $\Nu$ to $\Nu_0$ versus $\Ro$ for different $\Ek$. The blue and black horizontal dotted lines indicate $\Nu/\Nu_0$ = 0.18 and 1 respectively. Open grey triangles correspond to the DNS data for rotating convection in a CSIIW at $\Gamma = 1$, $\Prt = 6.4$ and $\Ra = 1 \times 10^8$ from \cite{zhongetal2009}. The vertical dashed red and blue dashed lines indicate the critical Rossby number $\Ro_c = 1/(a \Gamma (1+b \Gamma))$ for the bifurcation point marking the end of the heat-enhancement range for the aspect ratio of the TC and the outer vessel \citep{weiss2010prl}. Given our similar value of $\Prt=6.4$, we use \cite{weiss2010prl}' numerical values for $\Prt = 4.38$: $a = 0.381$ and $b = 0.061$ to obtain this rough estimate. The background colours identify regimes as on figure \ref{fig:forcebalance}. }
\label{fig:NubyNu0}
\end{figure}
The convective Rossby number $\Ro$ is classically used to characterize the regimes of rotating convection \citep{zhongetal2009,aurnou2020prf,eckeshishkina2023}. For CSIIW, the rotation-dominated regime appears for $\Ro\ll1$, the buoyancy-dominated regime for $\Ro\gg1$, and the rotation-affected regime for $\Ro \sim 1$ \citep{eckeshishkina2023}. To probe the regime change, $\Nu$ is normalized by the non-rotating value $\Nu_0$. For $\Prt \gtrsim 1$, this ratio exhibits an enhancement of heat transfer by up to 30$\%$ in the rotation-affected region in the form of an overshoot of $\Nu/\Nu_0$ reaching up to around 1.3 \citep{zhongetal2009}. This enhancement is generally associated with the Ekman pumping \citep{hartetal2002,kunnenetal2006,zhongetal2009}. In LEE2, the variations of $\Nu/\Nu_0$ \emph{vs.} $\Ro$ reveal the same three regimes, but with important differences with the CSIIW (figure \ref{fig:NubyNu0}). First, there is a distinct plateau in the bottom-left side of the graph 
that corresponds to the subcritical regime ($\Rt<1$), where $\Nu/\Nu_0$ remains around 0.18. In CSIIW, the moderate heat transfer due to wall-modes in the subcritical regime yields very low values of $\Nu/\Nu_0$, that are usually not even shown on such plots \citep{zhongetal2009,eckeshishkina2023}. For LEE2, by contrast, this constant value of the $\Nu/\Nu_0$ in the subcritical regime is due to the baroclinicity outside the TC, which drives a significant flow even when convection in the TC is absent (see section \ref{subsec:subcritical_regime}). 

The transition from the subcritical to rotation-dominated regime occurs around $\Ro \approx 0.02$, with a sharp increase in the Nusselt number for all rotation rates (figure \ref{fig:NubyNu0}). There is a good agreement in this regime between the CSIIW \citep{zhongetal2009} and LEE2, with an overshoot of $\Nu/\Nu_0$ up to around 1.2. For CSIIW, the transition from rotation-dominated to rotation affected occurs once $\Nu/\Nu_0$ starts to decrease from its maximum value, and the rotation affected region continues until it reaches unity \citep{kunnen2021}. 
For CSIIW, however, the rotation-affected regime expands up to $\Ro \approx 1$ whereas in LEE2, $\Nu/\Nu_0 = 1$ is reached at a much lower $\Ro$, especially for low $\Ek$, i.e. faster rotation: at $\Ro \approx 0.5$ ($\Rt \approx 160$,fig. \ref{fig:Flow_Snapshots_highRa}d) for  $\Ek = 4.2 \times 10^{-5}$, and $\Ro \approx 0.1$ ($\Rt \approx 7.2$, figure \ref{fig:Flow_Snapshots_highRa}a) for $\Ek = 7.1 \times 10^{-6}$, both marked in fig.  \ref{fig:NubyNu0}.
These values are significantly lower than \cite{weiss2010prl}'s prediction for the bifurcation point at the end of the heat transfer enhancement regime, for either the TC's aspect ratio or that of the outer vessel. 
On the one hand, the decreasing values of the overshoot at lower $\Ek$ are consistent with expectations in RRBC
\citep{zhongetal2009}; These stem from the suppression of Ekman pumping in the $\Ek\rightarrow0$ limit. On the other hand, the seemingly early transition to the $\Nu/\Nu_0 = 1$ value is also consistent with the velocity fields observation showing 
disrupted columns around $\Rt \sim 7-18$, due to inertia generated outside the TC (section \ref{subsec:Plumes and disrupted columns}), which eventually leads to geostrophic turbulence with large structures at lower $\Rt$ than for CSIIW or plane layers (section \ref{subsec:buoyancy dominated regime}). 
\subsection{Diffusivity-free scalings}
\begin{figure}
	\centering					
	\begin{overpic}[width=\columnwidth]		
		{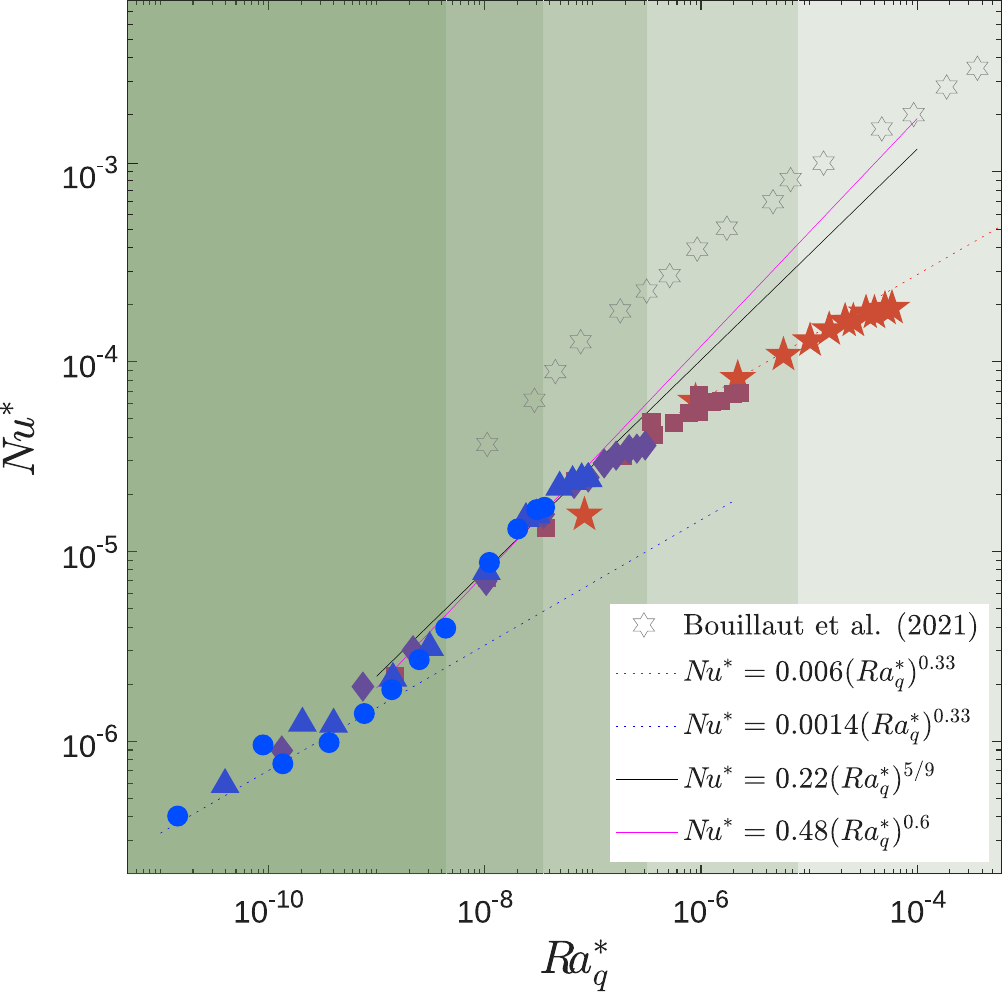} 	
		
		\put(22,92){\normalsize SR}
		\put(38,92){\normalsize CR}
		\put(48,92){\normalsize CTC}
		\put(67,92){\normalsize P}
		\put(83,92){\normalsize GT}	
		  \end{overpic}			
	\caption{Heat transfer expressed in diffusivity-free parameters, $\Nu^*$ and $\Ra^*_{q}$ for all $\Ek$ in LEE2. Open grey star symbols represent the experimental data from the master curve shown by \cite{bouillautetal2021} for rotating convection driven by volumetric heating in a cylindrical geometry with $\Ra$ between $\approx$ $2.3\times 10^8$ and $6.6 \times 10^8$, and $\Ek$ between $\approx$ $4 \times 10^{-6}$ and $3.5 \times 10^{-5}$. Dotted and solid lines represent scaling laws (see legend) whose prefactor is fitted based on the good collapse of the corresponding segment of the datasets. The background colours identify regimes, {and symbols identify Ekman numbers} as on figure \ref{fig:NubyNu0}. 
} 
\label{fig:NuStarbyRaqStar}
\end{figure}
At high thermal forcing, the heat transfer 
is controlled by the bulk flow where turbulence dominates \citep{britoetal2004} and becomes independent of viscous and thermal diffusivities. The question raises whether turbulence reaches this state.
Scaling laws based on diffusivity-free parameters \citep{christensen2002,christensenaubert2006} are obtained by normalising the temperature difference $\Delta T$ and heat flux by diffusivity-free quantities. In rotating convection, normalising these quantities without using diffusivities leads to the definition of 
the modified flux Rayleigh number, $\Ra^*_{q}$, and 
the modified Nusselt number, $\Nu^*$
\begin{eqnarray}
	\Ra^*_{q} &=& \Ra\Nu^*\Ek^2\Prt^{-1},\\
	\Nu^* &=&\Nu\Ek\Prt^{-1}.
\end{eqnarray}
\cite{aurnou2007} and \cite{chengaurnou2016} showed that scalings in these variables of the form $\Nu \sim \Ra^{\alpha}$ are related to scalings of the form $\Nu^* \sim (\Ra^*_{q})^{\beta}$ through $\beta = \alpha/(1+\alpha)$. In this sense the diffusivity-free parameters contain the same physics as the classical ones, but their value lies in how they tend to collapse data into single curves, for different rotation parameters. 
The graph of $\Nu^*$ \emph{vs.} $\Ra^*_{q}$ (figure \ref{fig:NuStarbyRaqStar}) indeed collapses into a single curve, and does so all the more precisely as $\Ra^*_q$ is high, which indicates independence of diffusivities. 
There are three distinct regions along that curve. For $\Ra^*_{q} \gtrsim 10^{-6}$, $\Nu^* \sim (\Ra^*_{q})^{1/3}$ (figure \ref{fig:NuStarbyRaqStar}). In this scaling, $\Ek$, hence $\Omega$, effectively cancels out which reflects a regime of diffusivity-free, buoyancy-dominated convection. 
This scaling is similar to the one obtained by \cite{bouillautetal2021} for $\Ra^*_{q} > 10^{-5}$ in a rotating cylinder where the fluid was heated radiatively directly in the bulk so as to ``bypass" the diffusion in the boundary layers. \cite{aurnouolson2001} obtained this scaling too, albeit at a much higher $\Ra^*_{q} \approx 1$ for rotating-convection in CSIIW using liquid metal as the working fluid. These differences with LEE2 hint that this scaling {may reflect 
a different diffuisivity-free state in LEE2, driven by the additional source of buoyancy due to baroclinicity.}
We shall come back to this point in section \ref{sec:radial_heat_transfer} with the insight of how much of the heat flux escapes radially \emph{vs.} axially.

In the range $10^{-8} \lesssim \Ra^*_{q} \lesssim 10^{-6}$, the data is closest to the scaling of $\Nu^* \sim (\Ra^*_{q})^{3/5}$. There are two known scalings close to this value: \cite{aurnou2007} argues that the common $\beta = 5/9\simeq0.56$ scaling is indicative of a rapidly rotating regime and \cite{christensen2002} arrived at the same scaling for $ 10^{-7} < \Ra^*_{q} < 10^{-3}$ with the same interpretation, using numerical simulations in a spherical shell geometry at $\Prt=1$. This suggests that this scaling is a property of the rotating convection that is robust to changes in geometry. \cite{bouillautetal2021} obtained  the $\beta=3/5$ scaling too and argue, as others, that it is the signature of geostrophic turbulence \citep{hanasogeetal2012,aurnou2020prf}. In LEE2, this interpretation is consistent with the other measures of heat transfer ($\Nu$ and $\Ro$) presented in this section as well as with the intense turbulence observed in the velocity fields in a high rotational constraint $\Ro_l\lesssim1$ in section \ref{sec:Regimes of convection}.

For $\Ra^*_{q} \lesssim 10^{-8}$, we observe $\Nu \sim (\Ra)^{1/3}$, similar to the buoyancy-dominated regime, however with a different pre-factor (figure \ref{fig:NuStarbyRaqStar}). These low values of $\Ra^*_{q}$ correspond to the subcritical flow regime, {which is specific to LEE2, because} heat transfer is enhanced by the baroclinically-driven flow. In this sense, they are buoyancy-dominated and so explain the rotation-independent scaling $\Ra^*_{q}\sim(\Nu^*)^{1/3}$, {despite different underlying mechanism physics to CSIIW. However, a word of caution must given regarding the value of the exponent in this regime: first, it is observed over a small range only. Second, the values of $\Nu$ in this low range are also overestimated because of the heat captured by the fluid (see figure \ref{fig:DiffusivityFreeScalings_Appendix}).}
\section{A measure of the Taylor-Proudman constraint at the TC boundary} 
\label{sec:tpc}
Until now, we have characterised the convection inside the TC. We found how the succession of regimes therein was progressively altered by inertia that progressively loosens the TPC at the TC boundary. What now remains to understand is the nature of the thermal and kinematic boundary conditions at the TC boundary in these different regimes.

\begin{figure*}
	\centering
		\begin{subfigure}[t]{0.49\linewidth}
		\centering
		\begin{overpic}[width=\linewidth]{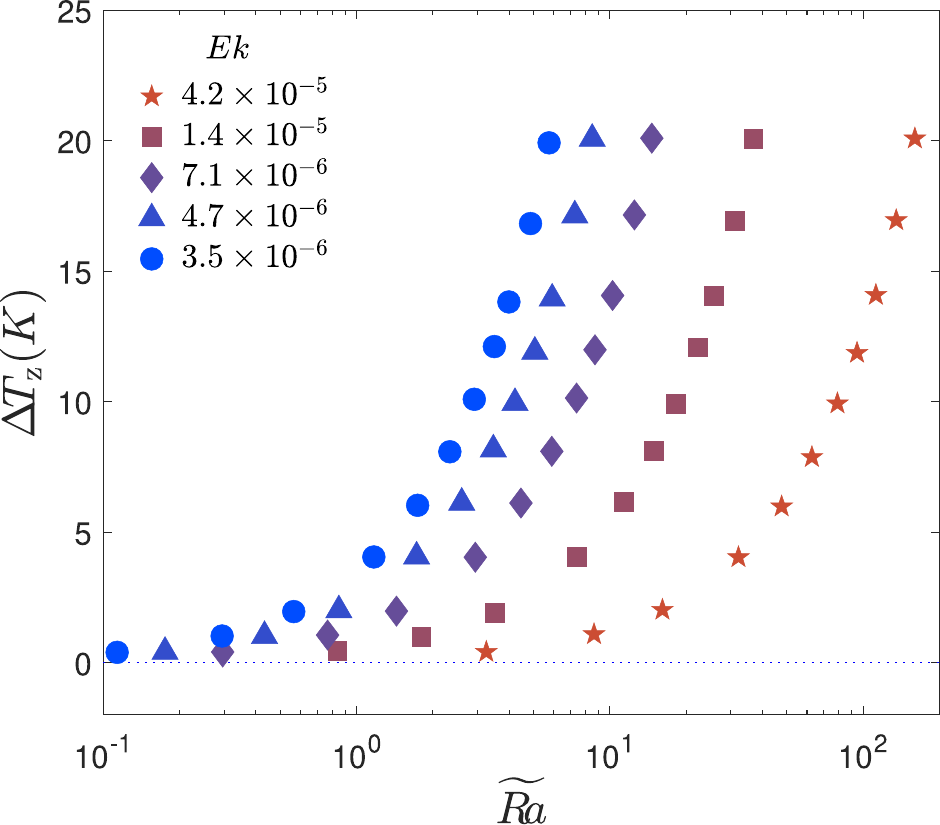}
			\put(2,88){\normalsize (a)} 
		\end{overpic}
		\end{subfigure}\hfill		
	\begin{subfigure}[t]{0.49\linewidth}
		\centering
		\begin{overpic}[width=\linewidth]{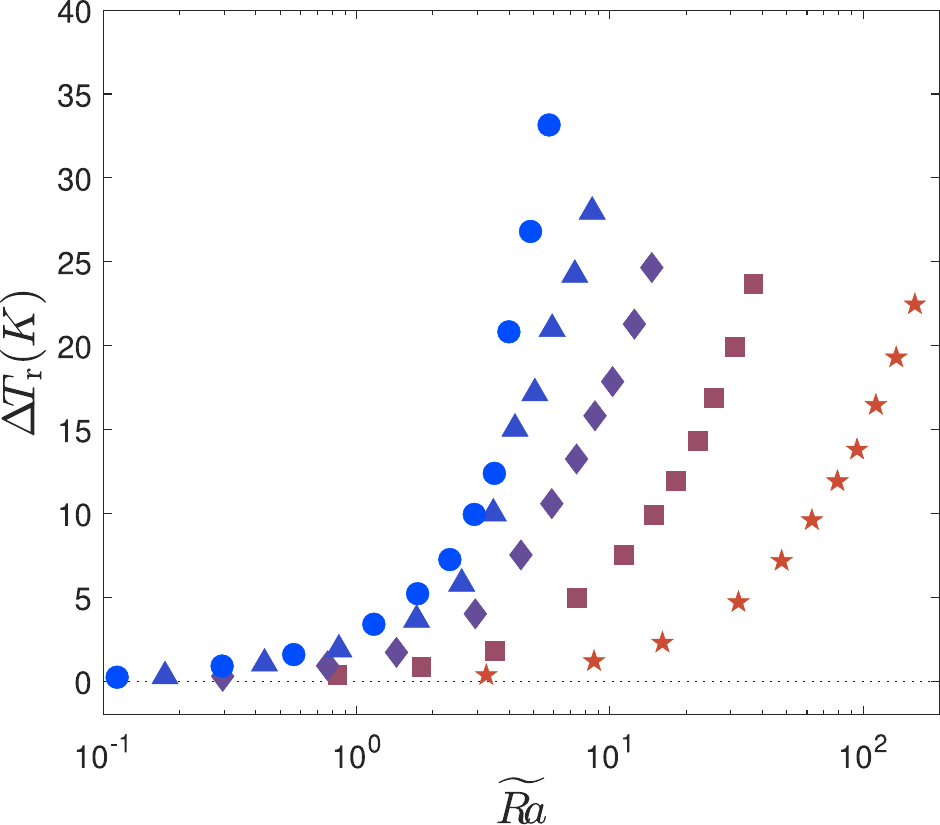}
			\put(2,88){\normalsize (b)}
		\end{overpic}
	\end{subfigure}
	
\vspace{0.3cm}

	\begin{subfigure}[t]{0.49\linewidth}
		\centering
		\begin{overpic}[width=\linewidth]{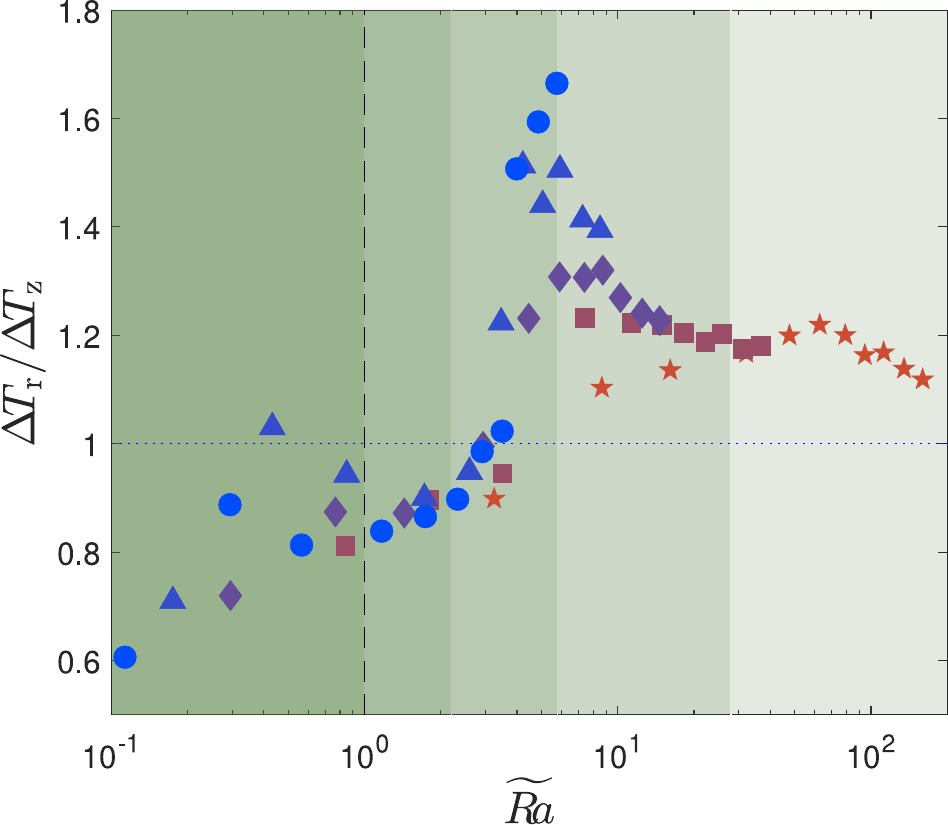}
			\put(2,88){\normalsize (c)}
		\put(22,81){\normalsize SR}
		\put(38,81){\normalsize CR}
		\put(48,81){\normalsize CTC}
		\put(67,81){\normalsize P}
		\put(83,81){\normalsize GT}	

		\end{overpic}
	\end{subfigure}\hfill	
	\begin{subfigure}[t]{0.49\linewidth}
		\centering
		\begin{overpic}[width=\linewidth]{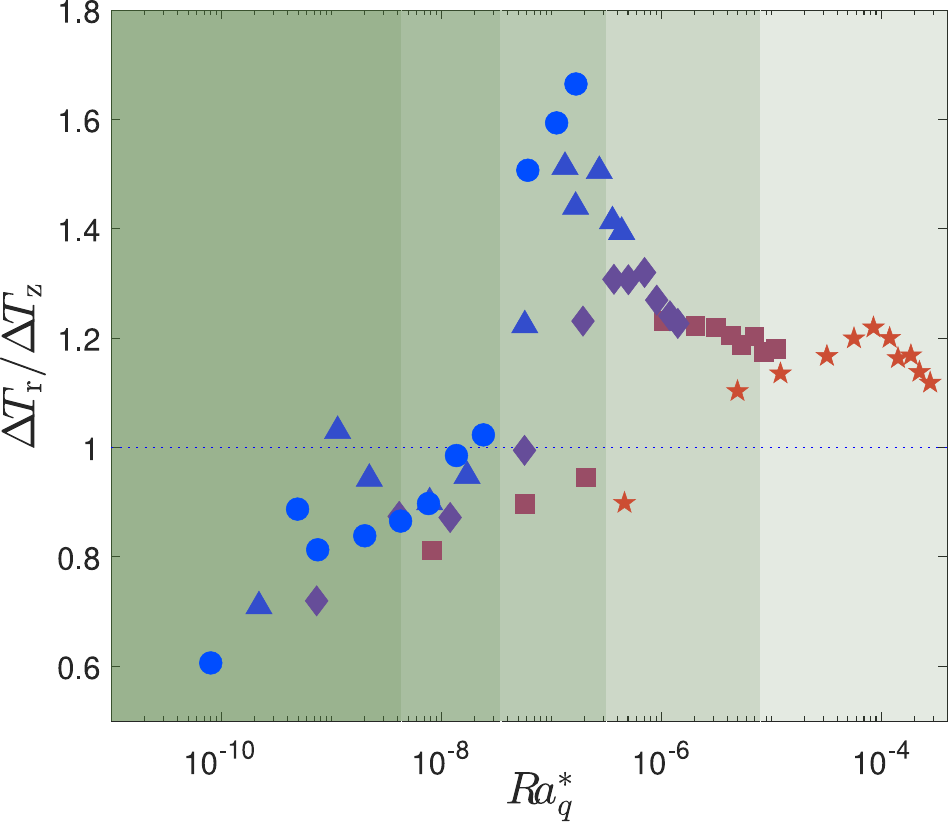}
			\put(2,88){\normalsize (d)}
		\put(22,81){\normalsize SR}
		\put(43,81){\normalsize CR}
		\put(53,81){\normalsize CTC}
		\put(67,81){\normalsize P}
		\put(83,81){\normalsize GT}	
			
		\end{overpic}
	\end{subfigure}
	
	\caption{(a) Axial temperature difference ($\Delta \! T_{\rm z}$) and (b) radial temperature difference ($\Delta \! T_{\rm r}$) inside the cylinder  as a function of $\Rt$. The ratio of the radial and axial temperature differences, $\Delta \! T_{\rm r}/\Delta \! T_{\rm z}$ as a function of (c) $\Rt$ and (d) $Ra^*_{q}$. In (b), (c) and (d) different symbols represent the same Ekman number as in (a). The background colours identify regimes as on figure \ref{fig:forcebalance}. The vertical black dashed line in (c) indicate the end of the subcritical regime, $\Rt = 1$.}
	\label{fig:TempDiff}
\end{figure*}

\subsection{Axial and Radial heat transfer}
\label{sec:radial_heat_transfer}
We start with the thermal aspects. Since it is not currently possible to directly measure the heat flux at the TC boundary in LEE2, we use the temperature difference between the heater and the inner surface of the side wall as proxy: Since the temperature at the side wall is not part of the heater's control loop (unlike the temperature difference between the inner surface of the top wall and the heater, which is kept constant), it measures directly the amount of heat transported radially for an imposed axial temperature difference. Since the side of the heater is thermally insulating, any increase in temperature at the side wall can only result from a heat flux through the TC boundary. Given the presence of baroclinic flow outside the TC, however, the heat may potentially cross the TC boundary anywhere along the TC, and be subsequently transported to the location of the sensor by this flow. Hence, $\Delta T_{\rm r}$ gives a measure of the global heat flux across the TC boundary.

The axial and radial temperature differences inside the cylinder are shown in fig. \ref{fig:TempDiff}. Plotting the dimensional temperature differences without normalizing (figs. \ref{fig:TempDiff}a, b) shows that for $\Rt\gtrsim3$, the radial temperature difference $\Delta \! T_{\rm r}$ is higher than the axial temperature difference $\Delta \! T_{\rm z}$, indicating that the heat flux is preferentially axial. This effect is all enhanced by rotation: Hence, despite the radial transport due to the baroclinic flow outside the TC, the confinement imposed at the TC boundary channels the heat flux axially. The fact that $\Delta \! T_{\rm r}$ grows all the fastest with $\Rt$ as rotation is fast confirms this phenomenology: at higher rotation, axial heat transfer is impeded and the regulation must raise the heater temperature higher at higher $\Rt$ to keep $\Delta \! T_{\rm z}$ to the set value. The rapid rise in $\Delta \! T_{\rm r}$ with $\Rt$ indicates that the inner temperature of the outer side wall remains low as the heater temperature increases. Hence heat does not accumulate outside the TC, because the TC boundary suppresses the radial heat flux much more efficiently than rotation suppresses the axial heat flux inside the TC.

The variations of the ratios of the radial to axial temperature differences, $\Delta T_{\rm r}/\Delta T_{\rm z}$ with $\Rt$ and $\Ra^*_{q}$ (fig. \ref{fig:TempDiff}c, d) highlight how this phenomenology varies across the flow regimes.  
There is a sharp transition 
at $\Rt \approx 3$ (fig. \ref{fig:TempDiff}c). For $\Rt \lesssim 3$, $\Delta \! T_{\rm r}$ remains below $\Delta \! T_{\rm z}$ suggesting a greater radial than axial heat transfer. 
In the subcritical regime ($\Rt<1$), this is attributed to the baroclinicity outside the TC that enhances transfer across the TC, while axial heat transfer inside the TC is purely conductive, hence very low. This confirms our previous conclusion that the higher values of $\Nu$ 
in LEE2, compared to CSIIW with adiabatic side walls, are due to radial heat transfer by the baroclinic flow outside the TC. For $1\leq\Rt \lesssim 3$, the radial temperature difference remains smaller than the axial temperature difference, so the axial heat flux remains comparatively low in the cellular regime. 
For $\Rt \gtrsim 3$, however, there is a distinct $\Ek$ dependence in the variations of $\Delta \! T_{\rm r}/\Delta \! T_{\rm z}$: For low $\Ek$, there is an overshoot in the values of this ratio  before it reaches a plateau at a value around 1.2. This overshoot  corresponds to the regime of convective Taylor columns. It reaches a maximum of around 1.6 for the fastest rotation, $\Ek = 3.5 \times 10^{-6}$ at $\Rt \approx 4$. This high value of the ratio suggests that the Taylor-Proudman constraint, which is still well enforced along most of the TC boundary, efficiently suppresses the radial heat transfer while at the same time, the columns 
favour axial heat transfer. Again, the increasing value of the overshoot with rotation speaks in favour of this phenomenology. 
For the slowest rotation, $\Ek = 4.2 \times 10^{-5}$, no such overshoot is observed.
This suggest that TC constraint is not strong enough to block the radial heat flux in the columnar regime, despite the relatively low value of $\Ek$.

\textcolor{black}{After the overshoot the ratio decreases towards 
1.2 at $\Rt\simeq 40$ (figure \ref{fig:TempDiff}c) and $\Ra^*_{q} \approx \times 10^{-6}$ (figure \ref{fig:TempDiff}d).} All curves for different $\Ek$ follow ``shingles" rejoining a curve where 
they all collapse reasonably well: This collapsed curve starts with a steep decrease, followed by a transition to a plateau at around $\Rt \approx 20$ and $\Ra^*_{q} \approx 2 \times 10^{-5}$. 
The steeper part of the curve correspond to the geostrophic regime where with diffusivity-free scaling $\Nu^* \sim (Ra^*_{q})^{3/5}$, 
whereas the plateau part of the curve (up to $\Rt\simeq 40$) spans the transition to the non-rotating diffusivity-free scaling $\Nu^* \sim (Ra^*_{q})^{1/3}$. The case at the lower rotation at $\Rt\gtrsim80$ reaches further into this latter regime,  where the ratio $\Delta \! T_{\rm r}/\Delta \! T_{\rm z}$ further decreases. In this regime the TPC is broken so the radial heat flux is less and less suppressed. The radial transport by the baroclinic flow becomes sufficiently efficient to homogenize the temperature along the upper and the side wall, leading $\Delta \! T_{\rm r}/\Delta \! T_{\rm z}$ to decrease towards 1. Hence, it is likely that the non-rotating diffusivity-free scaling $\Nu^* \sim (Ra^*_{q})^{1/3}$ at high $Ra^*_{q}$ reflects the heat transfer due to the baroclinic flow {, rather than a change of regime towards bouyancy-dominated convection within the TC}.
This is consistent with the still high rotational constraint in this regime, evidenced by the low values of $\Ro_l$ in the region of 0.2. 

\subsection{Velocities across the TC boundary}
\begin{figure}
	\centering
	\begin{overpic}[width=\columnwidth]					
		{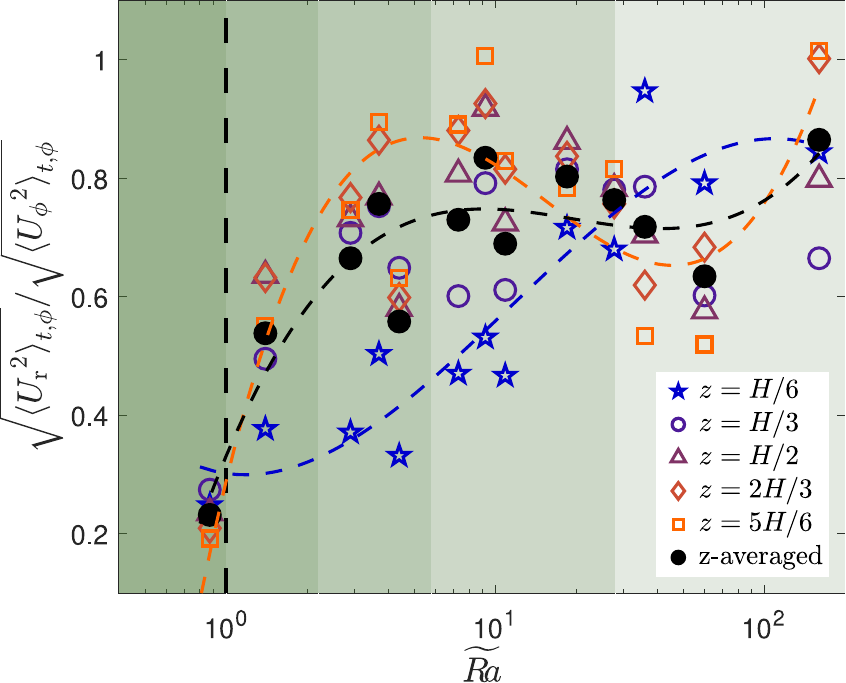} 		
		
		\put(18,74){\normalsize SR}
		\put(29,74){\normalsize CR}
		\put(39,74){\normalsize CTC}
		\put(60,74){\normalsize P}
		\put(80,74){\normalsize GT}
	\end{overpic}
	\caption{Quantifying the Taylor-Proudman constraint on the radial velocity at the TC boundary using the ratio of the time and azimuthal averaged radial and azimuthal velocity intensities, $\sqrt{\langle U^2_{r}\rangle_{t,\phi}}$/$\sqrt{\langle U^2_{\phi}\rangle_{t,\phi}}$. Dashed interpolated lines (calculated using least square fit to third-order polynomials) with same colours as the corresponding symbols are shown for $z = H/6, 5H/6$ and $z$- averaged data to guide the eye. The background colours identify regimes as on figure \ref{fig:forcebalance}. The vertical black dashed lines indicate the end of the subcritical regime, $\Rt = 1$.
	}		
	\label{fig:UrRMSbyUthetaRMS}
\end{figure}

We now turn to the radial and azimuthal velocities at the TC boundary. We use the time- and azimuthally averaged radial and azimuthal velocity intensities at the TC, i.e $r = R_{\rm TC}$. In practice, these quantities are averaged over all PIV points within the range $0.95 R_{\rm TC} \leq r\leq 1.05 R_{\rm TC}$, and plotted against $\Rt$ on figure \ref{fig:UrRMSbyUthetaRMS}. 
There is a trade-off between the higher statistical convergence of data obtained over a thicker radial range and capturing points located further away from the TC boundary, that are less constrained by the TPC. Because of it, the ratio of \emph{rms} radial to azimuthal velocity does not converge to zero but to a low, finite value in the limit where the TPC is fully enforced (\emph{i.e.} when $u_r(r = R_{\rm TC})=0$).

In the subcritical regime ($\Rt\leq1$), the \emph{rms} radial velocities remain very low, compared to their azimuthal counterparts as $\sqrt{\langle U^2_{r}\rangle_{t,\phi}}$/$\sqrt{\langle U^2_{\phi}\rangle_{t,\phi}}\simeq 0.2$ in all five PIV planes along the cylinder height, as shown on figure \ref{fig:UrRMSbyUthetaRMS}. Hence the TPC imposes that the baroclinic flow outside the TC is confined outside it.
The TPC is progressively broken by inertia in the cellular and Taylor Column regime ($1 \lesssim \Rt \lesssim 4$) as there is an increase in the ratio $\sqrt{\langle U^2_{r}\rangle_{t,\phi}}$/$\sqrt{\langle U^2_{\phi}\rangle_{t,\phi}}$, reaching around 0.5. The process starts at the higher planes, 
which is consistent with the the vorticity snapshots in figures \ref{fig:Flow_Snapshots_lowRa}c and \ref{fig:Flow_Snapshots_lowRa}d. For $\Rt = 2.9$, at $z = H/6$, 
little interaction between the inside and outside of TC takes place. However, at $z = 5H/6$, there is a significant flow outside the TC and vorticity sheets clearly straddle the TC boundary, linking the regions inside and outside the TC. At higher $\Rt$, in the plumes and disrupted columns regime ($4 \lesssim \Rt \lesssim 36$), the  ratio $\sqrt{\langle U^2_{r}\rangle_{t,\phi}}$/$\sqrt{\langle U^2_{\phi}\rangle_{t,\phi}}$ reaches a saturation around 0.8. There still exists a similar $z$-dependence in the breakdown of the TP constraint values with $\Rt$, also consistent with the vorticity snapshots in figures \ref{fig:Flow_Snapshots_highRa}a and \ref{fig:Flow_Snapshots_highRa}b. For $\Rt \simeq 36$, in the geostrophic turbulence regime, $\sqrt{\langle U^2_{r}\rangle_{t,\phi}}$/$\sqrt{\langle U^2_{\phi}\rangle_{t,\phi}}$ remains around 0.8 and no significant dependency on the $\Rt$ is observed. There is no significant $z$-dependence on $\sqrt{\langle U^2_{r}\rangle_{t,\phi}}$/$\sqrt{\langle U^2_{\phi}\rangle_{t,\phi}}$, either suggesting that the TP constraint is broken along the entire cylinder height. 
The variations of $\sqrt{\langle U^2_{r}\rangle_{t,\phi}}$/$\sqrt{\langle U^2_{\phi}\rangle_{t,\phi}}$ confirm and quantify the progressive breakup of the TPC at the TC boundary observed in the vorticity snapshot. The main result here, is that even though the flow remains strongly rotationally constrained throughout the regimes investigated up to $\Rt=191$ ($\Ro_l\lesssim0.2$), the TPC starts breaking up locally as early as $\Rt\simeq3$ until the TC boundary becomes pervasive to momentum over its entire height at $\Rt\simeq 40$. 
\section{Discussion}
\label{sec:discussion}
The regimes of rotating convection in a Tangent Cylinder (TC) were  identified using time-resolved velocity measurements for a criticality varying in the range $\Rt\in[0.15,191]$. 
Cross-analysing the averaged force balance in 5 horizontal planes with the topology of the axial vorticity field showed that convection in the TC follows broadly the same sequence of regimes as those of observed in plane layers or cylindrical vessels with impermeable, thermally insulating walls, with increasing criticality:
first a subcritical regime without bulk convection below the predicted onset of plane Rotating RBC, then a cellular regime, then convective Taylor columns, plumes, followed by geostrophic turbulence, potentially with Large-Scale Vortices. 
The values of the effective Rossby number based on the measured ratio of inertial to Coriolis forces $\Ro_l\lesssim0.2$ show that even near $\Rt=191$, the turbulent convection was still rotation-dominated. Hence the buoyancy-dominated regime observed in other studies at higher criticality was never reached. The similarity with plane layers and CSIIW was reflected in the heat transfer associated to these regimes showing a sharp increase in the ratio of the Nusselt number to the non-rotating Nusselt number $\Nu/\Nu_0$ around $Ro\simeq4\times10^{-2}$ to overshoot values above 1 followed by a progressive decrease to 1. 

There were, however, significant differences, some {of which are imporant} for our understanding of the convection in the Earth's outer core. {These differences stem from inertia due to the flow outside the TC. In LEE2, this flow is due to baroclinicity near the vertical cold outer wall next to the TC, but in the Earth'core, the convection in the equatorial region would act as the source of inertia there. However, the resulting structures are very similar in topology and in the way they interact with the TC. Hence, for our purpose of experimentally modelling the Earth's TC, the baroclinic flow outside LEE2's TC acts as a good proxy for the convection in the equatorial region of the Earth's core.} 

First, in the subcritical regime below the onset of bulk convection, wall-modes, generally observed in CSIIW were absent. The absence of walls modes continued into the supercritical regime. {This differs from the TC within a hemispherical geometry with less baroclinicity in LEE1, where wall modes were observed in the supercritical regimes (but not in the subcritical regime \citep{aujogueetal2018}).}
In the subcritical regime, the flow outside the TC increases the heat transfer compared to a CSIIW for which wallmodes transfer heat less efficiently. 

Second, inertia due to the baroclinic flow incurs a local breakdown of the TP constraint in the columnar regime near the top of the TC. At higher criticality, in the plumes and disrupted columns regime, the TP constraint is further violated over almost the entire height of the cylinder (except near the heater). At, $\Rt \geq 36$, no trace of plumes is observed. {Based on the high rotational constraint and the chaotic nature of the flow, the flow enters the geostrophic turbulence regime. Large structures emerge too.
The corresponding heat transfer follows the usual diffusivity-free scaling of $\Nu^* \sim (\Ra^*_{q})^{3/5}$ for geosgrophic turbulence. Unlike classical RBC configurations where this scaling is found, the heat transfer here also accounts for heat transfer through the side walls. 
}
Overall, baroclinically-driven inertia triggers transition to the geostrophic turbulence regime at much lower criticality than in CSIIW or plane layers ($\Rt \geq 80$). This suggests that a TC geometry may offer an easier pathway to the geostrophic turbulence regime in rotating-convection, that tends to be elusive in experiments. 
The question of the nature of the large scale flow structures observed in this regime remains however open: in particular, more work is needed to find out whether these correspond to the Large Scale Vortices identified in numerical simulations by \cite{aguirre-guzman2021jfm}.

Finally, at the highest criticalities, diffusivity-free heat transfer follows the same scaling as in the ultimate regime of non-rotating convective turbulence. Here, however, the flow is still rotationally constrained within the TC and the scaling is indicative of heat transfer by the baroclinically-driven convection, that becomes at least comparable to the axial heat transfer by Rotating Rayleigh-B\'enard convection within the TC.
\begin{figure*}
	\centering
	\begin{overpic}[width=1\linewidth]{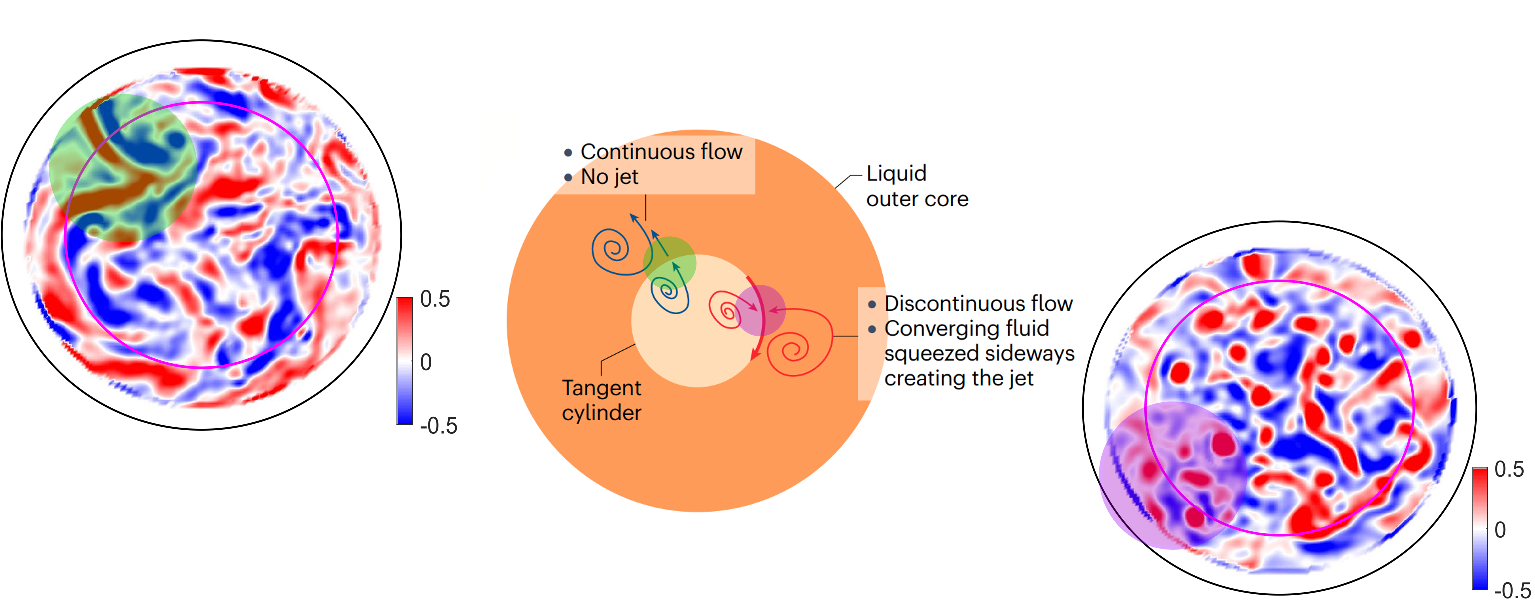}
		\put(33,29){\normalsize (a)}
		\put(1,33){\normalsize (c)}
		\put(70,22){\normalsize (b)}

 	  \put(27.2,10){\makebox(25,20)[bl]{\color{white}\rule{5mm}{18mm}}}
		\put(27.5,11){\normalsize $-0.5$}		
		\put(27.5,15){\normalsize $0$}		
		\put(27.5,19){\normalsize $0.5$}	  		
		\put(21.5,9.2){\normalsize $\omega_z(x,y,t_i)/\Omega$}

		\put(97,-0.5){\makebox(25,20)[bl]{\color{white}\rule{5mm}{18mm}}}
		\put(97.5,0.5){\normalsize $-0.5$}		
		\put(97.5,4.5){\normalsize $0$}		
		\put(97.5,8.5){\normalsize $0.5$}	  		
		\put(91,-1.5){\normalsize $\omega_z(x,y,t_i)/\Omega$}

	\end{overpic}	
	\vspace{0.1cm}	
	\caption{(a) Sketches of flow patterns inferred from the SWARM data  by \cite{finlay2023nat}, figure 3(d), showing local regions of the TC where the TPC is broken leading to jets crossing the TC boundary, and where the TPC locally holds, where jets are deflected along it. Magnification of the snapshots on figure \ref{fig:Flow_Snapshots_highRa}(b) at (b) $z=H/6$ and (c) $z=5H/6$, showing exactly this phenomenology in LEE2 for $\Rt=19$, $\Ek=1.4\times10^{-5}$. The shaded green and magenta regions in the figures highlight the similarities between the instantaneous flow structures obtained in LEE2 and the sketches of \cite{finlay2023nat}. } 
	\label{fig:finlay_snapshots}
\end{figure*}

These results highlight several mechanisms to help us understand the role of the TC boundary in the Earth's outer core. The main process is the progressive breakup of the TPC due to the inertia of the flow outside the TC. In LEE2, inertia is driven by the baroclinicity due to the isothermal boundary condition at the side wall of the vessel located close to the TC boundary itself. Although no such boundary exists in the Earth core, the presence of geostrophic columns driven by radial convection in the outer core plays a similar role: by virtue of the quasi-geostrophy there, the convection driven by the radial temperature gradient in the equatorial region drives inertia all along the entire TC boundary. Since, in this region, these structures are aligned with the Earth's rotation and not with the thermal gradient between the CMB and the ICB, they produce baroclinicity too, which also drives fluid motion and adds up to the inertia in this region. The main point is that important inertia is present just outside the TC boundary both in LEE2 and in Earth's the outer, even though it is generated by different mechanisms. \textcolor{black}{Inertia is also expected to be generated by the nonlinear form of the Rossby waves that are thought to exist near the TC boundary (produced experimentally in a spherical shell \cite{olson2011pepi,cardin1994pepi}.} Regardless how inertia is generated, $\Ro_l$ provides a measure of its effect on the flow dynamics inside the TC. In this respect, the low values of $\Ro_l$, in the range $10^{-2}\lesssim\Ro_l\lesssim10^{-1}$ in the regime of geostrophic turbulence are higher but consistent with the values expected for the Earth's core, for criticalities in the range $40\lesssim\Rt\lesssim200$ also consistent with estimates for the Earth's core. In this sense the level of inertia and thermal forcing in LEE2 are representative of the Earth, albeit for much slower rotation ($3.5\times10^{-6}\leq\Ek\leq4.2\times10^{-5}$). This phenomenology therefore shows that even under strong rotational constraint, inertia driven outside the TC may still break the TPC at the scale of the TC to the point where the practically no part of the TC boundary acts as a solid boundary any more. The thermally insulating and the impermeable character of the TC boundary both break down at much lower criticality. Hence, these results show that despite the very strong rotational constraint near the TC boundary, inertia outside the Earth's TC alone is sufficient to prevent it from behaving like a solid, thermally insulating boundary. This is indeed consistent with the behaviour of the main gyre wrapped around the TC. As an illustration of this process, \cite{finlay2023nat} infer two possible flow patterns near the TC boundary depending on whether the TPC is broken or not: one where impinging flows are deflected along the TC boundary to form jets, and one where these flows cross the TC boundary. Figure \ref{fig:finlay_snapshots} indeed shows the striking resemblance between these authors' simplified representation of these patterns and the flow patterns observed in LEE2 in cases where the TPC breaks up near the top boundary but still holds near the heater, at $\Rt=19$. 

Another interesting aspect of our results, from the perspective of the Earth's core concerns the emergence of large flow structures within the TC. At this stage, much remains to be understood regarding their nature and their origin, and this shall be the subject of future work with LEE2. However, it is rather clear from previous studies that such structures would only appear in well-developed geostrophic turbulence \citep{aguirre-guzman2021jfm}. Such regimes have proved elusive in classical convection experiments using cylindrical vessels. The additional inertia outside the TC however, violates the TPC and favours the occurrence of this regime at lower forcing. Since the same ingredients are present in the Earth's outer core, it is possible that large scales are produced in the vicinity of the TC at moderate criticality through a similar process. We saw that such vortices start straddling the TC boundary at criticalities as low as $\Rt\simeq40$, \emph{i.e.} within the range expected for the Earth's TC. While in LEE2, they are confined there by the solid vessel wall close to the TC, the looser confinement imposed by the hemispherical shape of the CMB would allow them a wider range of action, consistent with the location of the large patches of magnetic field observed in geomagnetic data \citep{finlay2023nat}.

One of the main difficulties in assessing what rotating convection experiments tell us about it the Earth arises out assessing the effect of the Earth's magnetic field, especially on length scales \citep{cardin1994pepi,shew2005pepi,horn2022prsa,horn2023prsa,potherat2024crphys}. We wish to finish this study by offering a few thoughts on how the processes outlined in this discussion may be affected by it. Studies of MHD turbulence and MHD convection in plane layers, cylindrical or rectangular vessels bounded by solid walls give us some ideas of how the dipolar magnetic field in the TC affects the convection and the turbulence therein \citep{sommeria1982jfm,potherat2014jfm,horn2023prsa,baker2018prl}: In these configurations, MHD turbulence tends to favour the emergence of large scales 
through nonlinear energy transfer and through scale-selective dissipation 
\citep{kolesnikov1974fd,eckert2001ijhff,potherat2015jfm,baker2018prl,potherat2017prf}. This may offer an alternative mechanism for the generation of large structures in the TC, but it is not clear how the two mechanisms would coexist or compete. For example, even though background rotation and magnetic fields individually promote two-dimensional velocity fields, the combination of both may promote a three-dimensional one \citep{cao2018pnas,potherat2024prl}! 

In rotating magneto-convection, the onset instability that feeds the turbulence may favour large scales too, when the Coriolis and Lorentz force are of comparable order of magnitude, relatively close to onset and at $\Prt\gtrsim1$ \citep{chandrasekhar1961,eltayeb1975jfm,aujogue2015pf,horn2022prsa}. At the lower $\Prt$ of the Earth's outer core, oscillatory modes alter the picture significantly, so it is not currently clear how the dipolar magnetic field influences the size of the larger scales at Earth-like criticalities \citep{horn2023prsa}.

The Earth's magnetic field also alters the constraint at the TC boundary in several ways: First, Magneto-Coriolis Waves provide a macroscopic mechanism for breaking the classical TPC that is consistent with the large-scale shape of the gyre wrapped around the TC \citep{gillet2010nat, finlay2023nat}, but does not necessarily explain its smaller scale meandering in and out of it \citep{gillet2022pnas}. Second, the time-averaged component of the magnetic field {may considerably alter the shear region near the TC boundary \citep{hollerbach1994_pf, hollerbach1996_gafd, dormy1998_epsl, soward2000_jfm}. It also} imposes a \emph{Magnetic} Taylor-Proudman Constraint (MTPC), instead of the classical TC. This constraint may drive a large scale meridional flow through the TC boundary controlled by the morphology of the azimuthal flow there \citep{potherat2024prl}.  This too provides an alternative mechanism for breaking the classical TPC at the TC boundary. The MTPC however relies on 
the same line of reasoning as the TPC, except that it applies to the combined Lorentz and Coriolis forces instead of the Coriolis force only. Hence, it is reasonable to expect that sufficiently strong local buoyancy-driven inertia would break the MTPC near the TC boundary in the same way as it breaks the TPC in non-magnetic LEE2.

\section*{ACKNOWLEDGMENTS}
The authors are grateful to Professor Susanne Horn for sharing her expertise of Rotating Convection during numerous discussions on the interpretation of the results. This work was sponsored by Research Project Grant RPG-017-421 from the Leverhulme Trust.

\section*{Data availability}
The data that support the ﬁndings of this study are available from the corresponding author upon reasonable request.

\bibliographystyle{gji}

\appendix

\section{Nusselt number correction}
\label{appendix:Nu_Correction}

\begin{figure}
	\centering
	\begin{minipage}[t]{\columnwidth}
		\centering
		\begin{overpic}[width=\columnwidth]{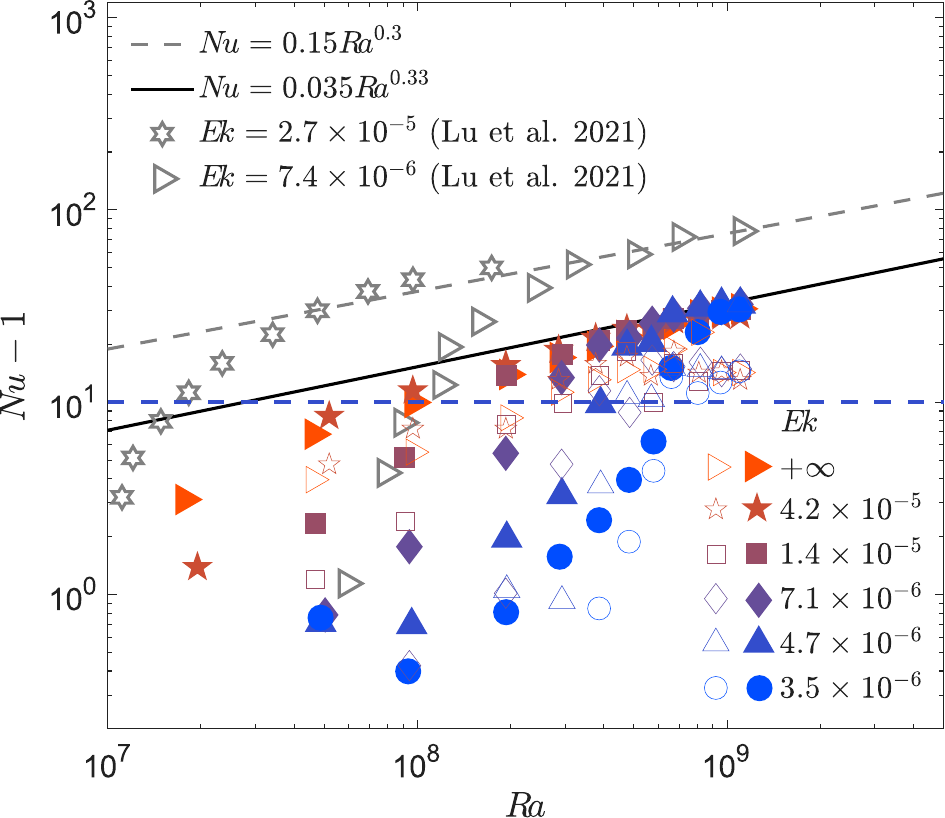}
			
			\put(1,82){\normalsize (a)}
			
		\end{overpic}
			\end{minipage}
	\hfill
	\vspace{0.2cm}
	\begin{minipage}[t]{\columnwidth}
		\begin{overpic}[width=\columnwidth]{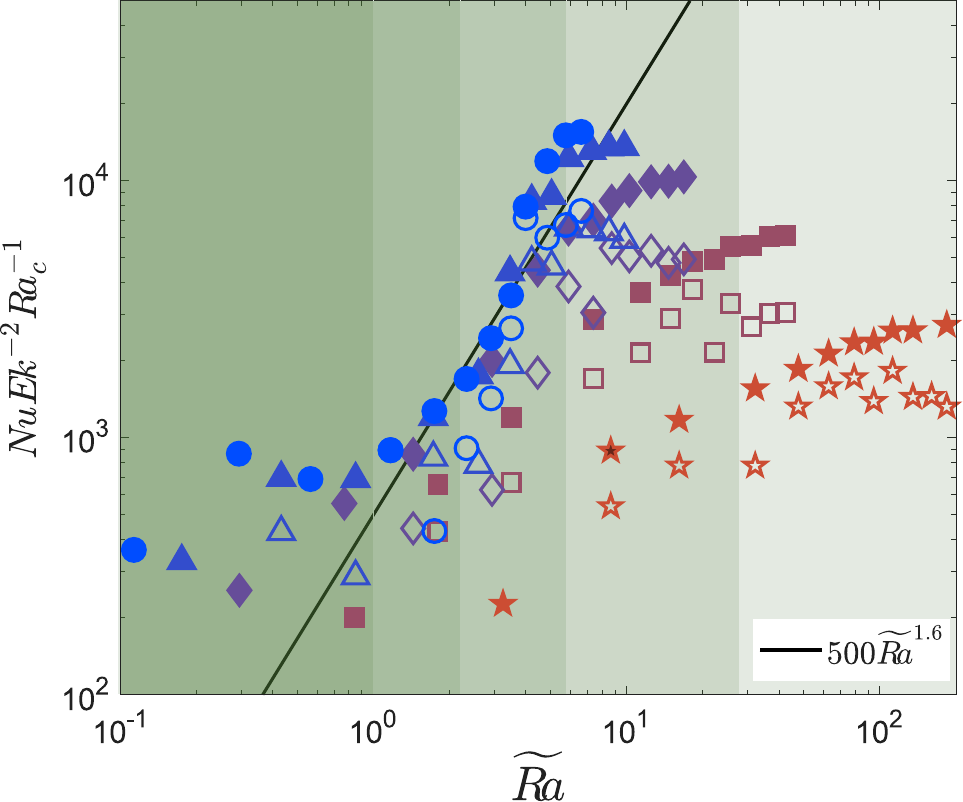} 
			\put(1,82){\normalsize (b)}
			\put(28,78.0){\normalsize SR}
			\put(42,78.0){\normalsize CR}
			\put(49,78.0){\normalsize CTC}
			\put(66,78.0){\normalsize P}
			\put(85,78.0){\normalsize GT}	
		\end{overpic}
		
	\end{minipage}
	
	\caption{a) Variations of the Nusselt number $\Nu - 1$ with $\Ra$ for six value of $\Ek$ in LEE2. Black solid line and grey dashed line in (a) indicate $\Nu - 1 = 0.15\Ra^{0.3}$ and $\Nu - 1 = 0.15\Ra^{0.3}$, respectively. Grey star and triangle symbols show \cite{luetal2021}'s experimental data for rotating convection in a CSIIW with $\Gamma = 3.8$, $\Ek = 2.7 \times 10^{-5}$ and  $\Gamma = 2.0$, $\Ek = 7.4 \times 10^{-6}$, respectively. (b) Variation of $\Nu \Ek^{-2} \Ra_c^{-1}$ vs $\Rt$. In figure (b) different coloured symbols show data for same $\Ek$ as in (a). The background colours identify regimes as on figure \ref{fig:forcebalance}.  
{Open symbols indicate values corrected to remove the heat stored in the fluid as the mean temperature drifts.}
		\label{fig:Nu_Ra_Classical_Appendix}}
\end{figure}
\begin{figure}
	\centering					
	\begin{overpic}[width=1\linewidth]{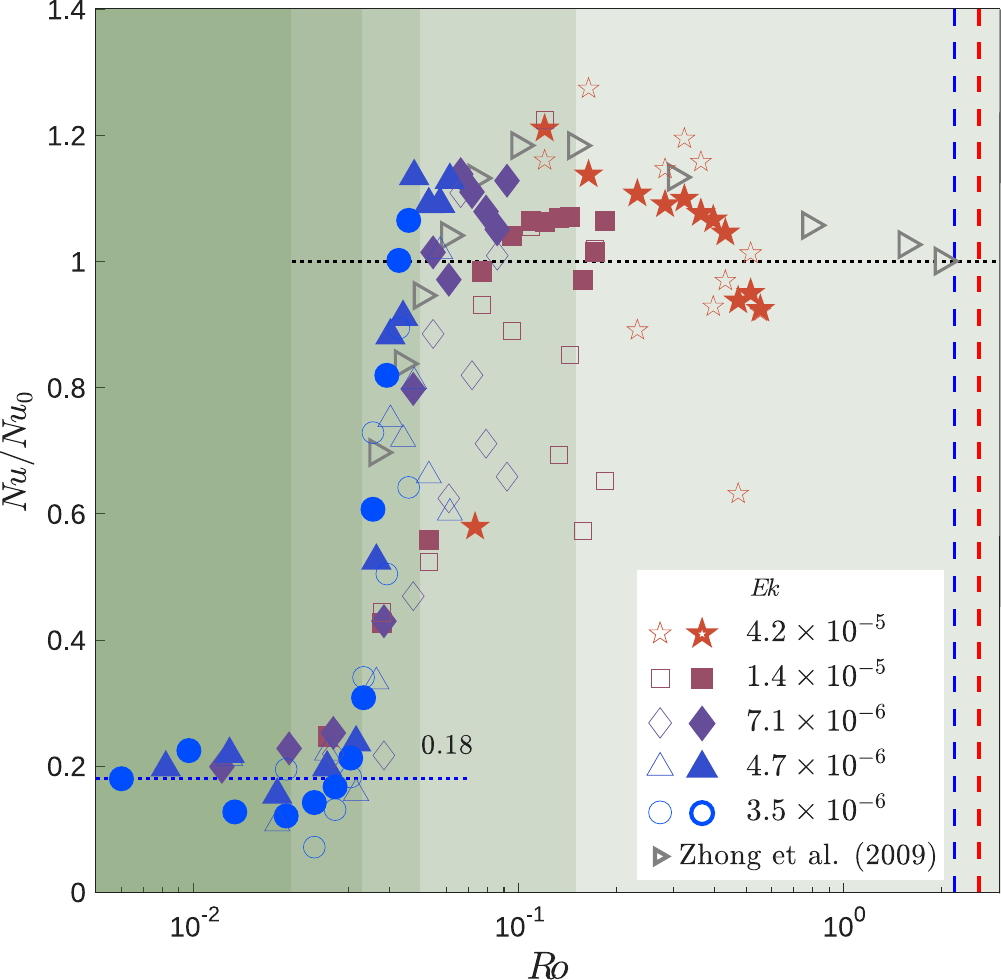}
		
	\put(20,90){\normalsize SR}
	\put(31,90){\normalsize CR}
	\put(37,90){\normalsize CTC}
	\put(48,90){\normalsize P}
	\put(67,90){\normalsize GT}	
		
	\end{overpic}						
	\caption{Ratio of $\Nu$ to $\Nu_0$ versus $\Ro$ for different $\Ek$. The blue and black horizontal dotted lines indicate $\Nu/\Nu_0$ = 0.18 and 1 respectively. Open grey triangles correspond to the DNS data for rotating convection in a CSIIW at $\Gamma = 1$, $\Prt = 6.4$ and $\Ra = 1 \times 10^8$ from \cite{zhongetal2009}. The vertical dashed red and blue dashed lines indicate the critical Rossby number $\Ro_c = 1/(a \Gamma (1+b \Gamma))$ for the bifurcation point marking the end of the heat-enhancement range for the aspect ratio of the TC and the outer vessel \citep{weiss2010prl}. Given our similar value of $\Prt=6.4$, we use \cite{weiss2010prl}' numerical values for $\Prt = 4.38$: $a = 0.381$ and $b = 0.061$ to obtain this rough estimate. The background colours identify regimes as on figure  \ref{fig:forcebalance}. For LEE2 data, solid filled symbols indicate heat transfer data obtained from total power, $P$ and empty symbols indicate data obtained from effective power $P_{\rm conv}$.
{Open symbols indicate values corrected to remove the heat stored in the fluid as the mean temperature drifts.}
	}
	\label{fig:NubyNu0_Appendix}
\end{figure}

\begin{figure}
	\centering					
	\begin{overpic}[width=\columnwidth]		
		{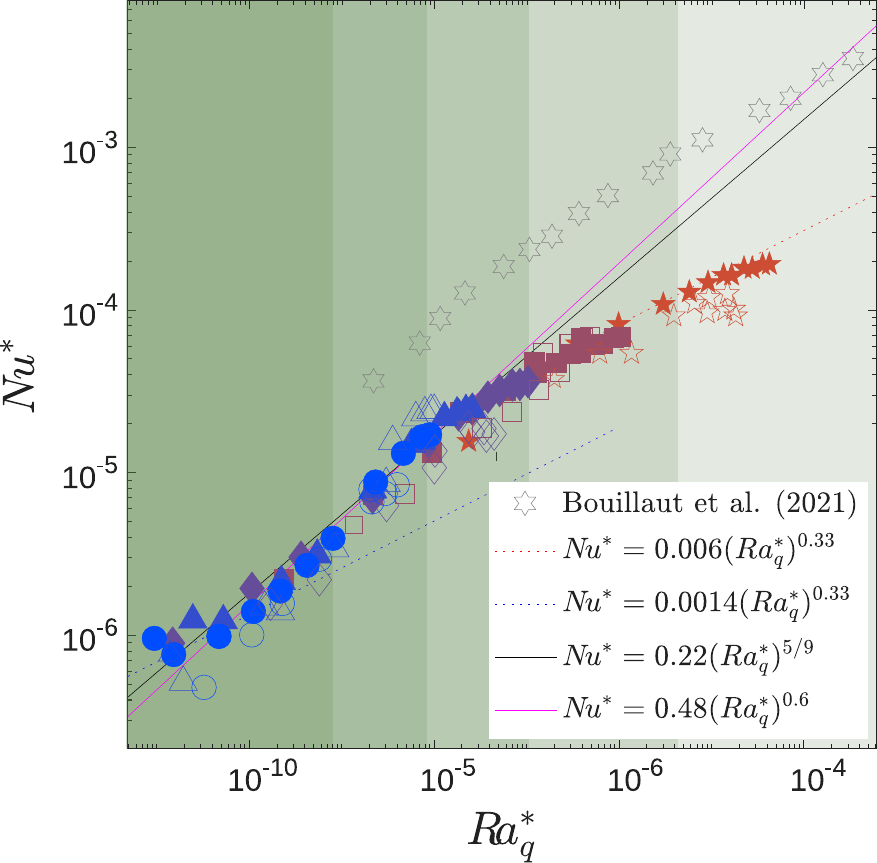} 	
		
		\put(22,92){\normalsize SR}
		\put(38,92){\normalsize CR}
		\put(48,92){\normalsize CTC}
		\put(67,92){\normalsize P}
		\put(83,92){\normalsize GT}

	\end{overpic}			
	\caption{Heat transfer expressed in diffusivity-free parameters, $\Nu^*$ and $\Ra^*_{q}$ for all $\Ek$ in LEE2. Open grey star symbols represent the experimental data from the master curve shown by \cite{bouillautetal2021} for rotating convection driven by volumetric heating in a cylindrical geometry with $\Ra$ between $\approx$ $2.3\times 10^8$ and $6.6 \times 10^8$, and $\Ek$ between $\approx$ $4 \times 10^{-6}$ and $3.5 \times 10^{-5}$. Dotted and solid lines represent scaling laws (see legend) whose prefactor is fitted based on the good collapse of the corresponding segment of the datasets. The background colours identify regimes as on figure \ref{fig:forcebalance}.  For LEE2 data, solid filled symbols indicate heat transfer data obtained from total power, $P$ and empty symbols indicate data obtained from effective power $P_{\rm conv}$. Symbols represent Ekman numbers as on figure \ref{fig:NubyNu0_Appendix}.
{Open symbols indicate values corrected to remove the heat stored in the fluid as the mean temperature drifts.}
} 
	
	\label{fig:DiffusivityFreeScalings_Appendix}
\end{figure}

\section{Estimation of the radial temperature difference between the heater and the inner side wall needed to drive experimentally observed zonal flows}
\label{sec:zonal_flow}
The radial temperature difference that would be needed to drive the observed zonal flow through a thermal wind mechanism was estimated as follows: In axisymmetric geometry, the balance between buoyancy and Coriolis forces gives rise to azimuthal motion as, 

\begin{equation}
	\frac{\partial U_{\phi}}{\partial z} \sim - \frac{g \beta}{2 \Omega} \frac{\partial T'}{\partial r}.
	\end{equation}
The maximum value of time- and azimuthal-averaged azimuthal velocity ($U_{\phi}$) at a height $z$, given by $U_{\phi, z_i, max}$ is utilized. Central differencing scheme is used to calculate the derivative,
\begin{equation}
	\frac{U_{\phi, z_{i+1}, max} - U_{\phi, z_{i-1}, max} }{\Delta z} = - \frac{g \beta}{2 \Omega} \frac{\Delta T}{\Delta R}|_{z_i},
\end{equation}

\begin{equation}
	\frac{\Delta T}{\Delta R}|_{z_i} = - \frac{2 \Omega}{g \beta} 	\frac{U_{\phi, z_{i+1}, max} - U_{\phi, z_{i-1}, max} }{\Delta z}. 
\end{equation}
The temperature difference is calculated outside and inside the TC:
\begin{equation}
	\Delta T_{In}|_{z_i} = - {\Delta R_{In}} \bigg(\frac{2 \Omega}{g \beta} \bigg)	\frac{U_{\phi, In, z_{i+1}, max} - U_{\phi, In, z_{i-1}, max} }{\Delta z}, 
\end{equation}
\begin{equation}
	\Delta T_{Out}|_{z_i} = - {\Delta R_{Out}} \bigg(\frac{2 \Omega}{g \beta} \bigg)	\frac{U_{\phi, Out, z_{i+1}, max} - U_{\phi, Out, z_{i-1}, max} }{\Delta z}. 
\end{equation}
Here, $\Delta R_{In} = 0.075$m and $\Delta R_{Out} = 0.035$m. We also calculate a cumulative temperature difference given by $\Delta T_{Out}|_{z_i} + \Delta T_{Out}|_{z_i}$. In the subcritical regime, the mechanism may be different: indeed, PIV shows that the main zonal flow is unstable to a baroclinic instability generating large vortices that dominate the zonal flow. These appear to drift more slowly than the thermal wind scaling. This was not observed in LEE1, suggesting that they may contribute to the suppression of the wall modes inside the TC.

\begin{figure*}
	\centering
	\begin{subfigure}[t]{0.32\linewidth}
		\centering
		\begin{overpic}[width=\linewidth]{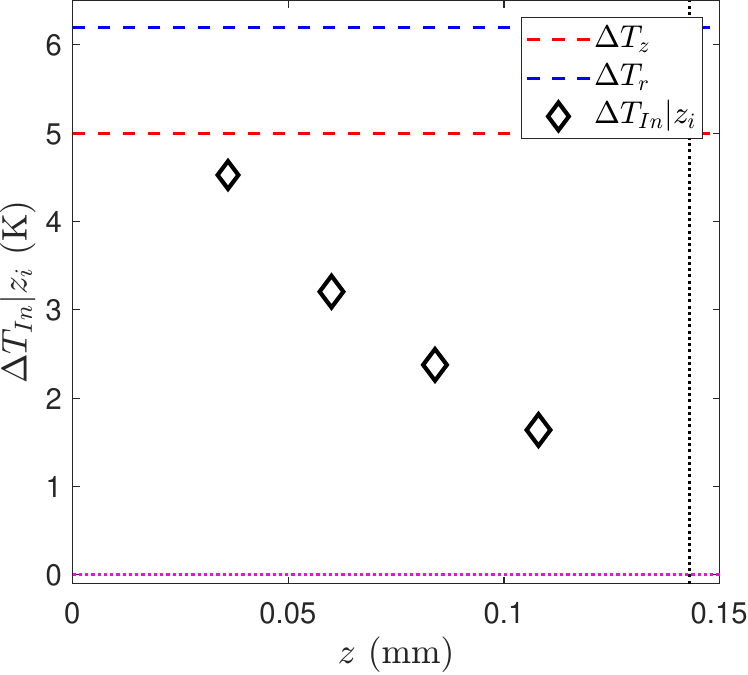}
			\put(2,88){\normalsize (a)} 
		\end{overpic}
	\end{subfigure}
	\hspace{0.1cm}
	\hfill
	\begin{subfigure}[t]{0.32\linewidth}
		\centering
		\begin{overpic}[width=\linewidth]{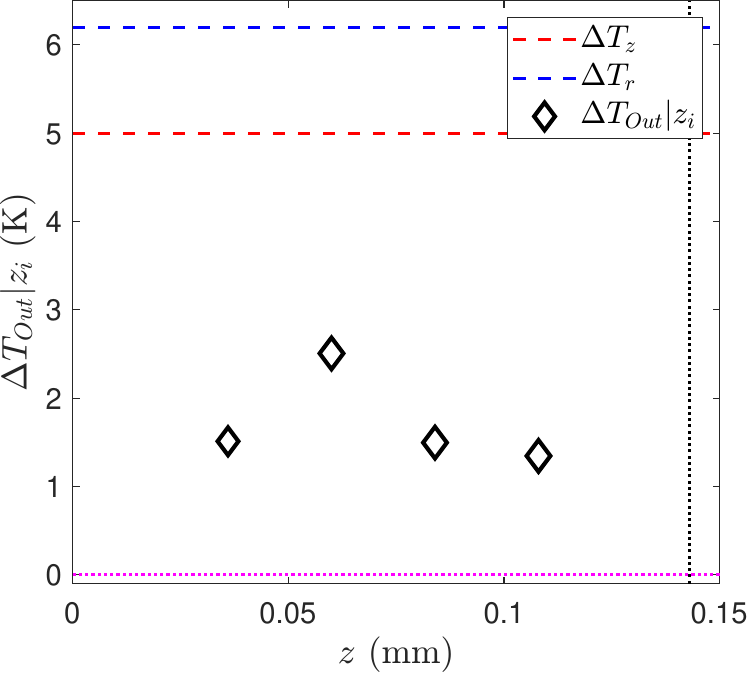}
			\put(2,88){\normalsize (b)}
		\end{overpic}
	\end{subfigure}
\hspace{0.1cm}
\hfill
	\begin{subfigure}[t]{0.32\linewidth}
		\centering
		\begin{overpic}[width=\linewidth]{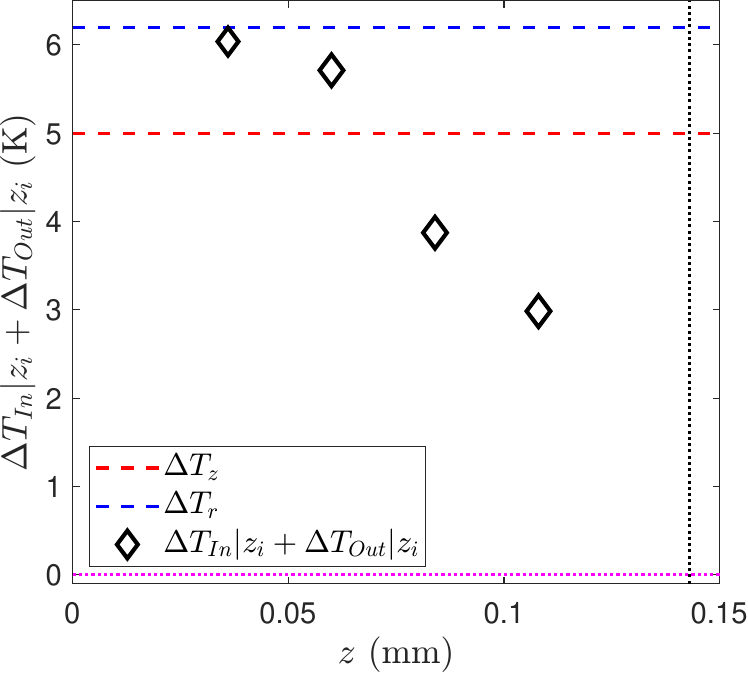}
			\put(2,88){\normalsize (c)}
		\end{overpic}
	\end{subfigure}\hfill

	\caption{Estimation of the radial temperature difference (a) inside the TC, (b) outside the TC and (c) cumulative of inside and outside the TC, using a thermal wind balance for $\Ek = 7.1 \times 10^{-6}$, $\Rt = 2.9$. }
	\label{fig:Appendix_1_ThermalWindBalance_Omega30rpm_DeltaT5K}
\end{figure*}

\bibliography{References}

@article{vasil2025_jfm,
  title={Rapidly rotating wall-mode convection},
  author={Vasil, Geoffrey M and Burns, Keaton J and Lecoanet, Daniel and Oishi, Jeffrey S and Brown, Benjamin and Julien, Keith},
  journal={Journal of Fluid Mechanics},
  volume={1017},
  pages={A37},
  year={2025},
  publisher={Cambridge University Press}
}

@article{hide1975_ap,
  title={Sloping convection in a rotating fluid},
  author={Hide, R and Mason, PJ},
  journal={Advances in Physics},
  volume={24},
  number={1},
  pages={47--100},
  year={1975},
  publisher={Taylor \& Francis}
}

@article{proudman1956_jfm,
	author={Proudman, I.},
       year=1956,
       title={The almost-rigid rotation of viscous fluid between concentric spheres},
       journal={J. Fluid Mech.},
       volume=1,
       number=5, 
       page={505-516}
}

@article{stewartson1957_jfm,
	author={Stewartson, K.},
       year=1957,
       title={On almost rigid rotations},
       journal={J. Fluid Mech.},
      volume=3,
      number=1, 
      pages={17-26}
}

@article{stewartson1966_jfm,
	author={Stewartson, K.},
       year=1966,
       title={On almost rigid rotations. Part2},
       journal={J. Fluid Mech.},
      volume=26,
      number=1,
      pages={131-144}
}

@article{gillet2006_jfm,
  title={The quasi-geostrophic model for rapidly rotating spherical convection outside the tangent cylinder},
  author={Gillet, N and Jones, CA},
  journal={Journal of Fluid Mechanics},
  volume={554},
  pages={343--369},
  year={2006},
  publisher={Cambridge University Press}
}

@article{lin2021_jfm,
  title={Large-scale vortices and zonal flows in spherical rotating convection},
  author={Lin, Yufeng and Jackson, Andrew},
  journal={Journal of Fluid Mechanics},
  volume={912},
  pages={A46},
  year={2021},
  publisher={Cambridge University Press}
}

@article{hulot2002_nat,
  title={Small-scale structure of the geodynamo inferred from Oersted and Magsat satellite data},
  author={Hulot, Gauthier and Eymin, C{\'e}line and Langlais, Beno{\^\i}t and Mandea, Mioara and Olsen, Nils},
  journal={Nature},
  volume={416},
  number={6881},
  pages={620--623},
  year={2002},
  publisher={Nature Publishing Group UK London}
}

@article{hollerbach1996_gafd,
  title={Magnetohydrodynamic shear layers in a rapidly rotating plane layer},
  author={Hollerbach, Rainer},
  journal={Geophysical \& Astrophysical Fluid Dynamics},
  volume={82},
  number={3-4},
  pages={237--253},
  year={1996},
  publisher={Taylor \& Francis}
}

@article{hollerbach1994_pf,
  title={Imposing a magnetic field across a nonaxisymmetric shear layer in a rotating spherical shell},
  author={Hollerbach, Rainer},
  journal={Physics of Fluids},
  volume={6},
  number={7},
  pages={2540--2544},
  year={1994},
  publisher={American Institute of Physics}
}

@article{soward2000_jfm,
  title={Non-axisymmetric magnetohydrodynamic shear layers in a rotating spherical shell},
  author={Soward, Andrew M and Hollerbach, Rainer},
  journal={Journal of Fluid Mechanics},
  volume={408},
  pages={239--274},
  year={2000},
  publisher={Cambridge University Press}
}

@article{dormy1998_epsl,
  title={MHD flow in a slightly differentially rotating spherical shell, with conducting inner core, in a dipolar magnetic field},
  author={Dormy, Emmanuel and Cardin, Philippe and Jault, Dominique},
  journal={Earth and Planetary Science Letters},
  volume={160},
  number={1-2},
  pages={15--30},
  year={1998},
  publisher={Elsevier}
}

@article{scolan2017_ef,
  title={A rotating annulus driven by localized convective forcing: a new atmosphere-like experiment},
  author={Scolan, H{\'e}l{\`e}ne and Read, Peter L},
  journal={Experiments in Fluids},
  volume={58},
  number={6},
  pages={75},
  year={2017},
  publisher={Springer}
}

@article{lewis2004_gafd,
  title={Linear stability analysis for the differentially heated rotating annulus},
  author={Lewis, Gregory M and Nagata, Wayne},
  journal={Geophysical \& Astrophysical Fluid Dynamics},
  volume={98},
  number={2},
  pages={129--152},
  year={2004},
  publisher={Taylor \& Francis}
}

@article{fowlis1965_jas,
  title={Thermal convection in a rotating annulus of liquid: effect of viscosity on the transition between axisymmetric and non-axisymmetric flow regimes},
  author={Fowlis, WW and Hide, R},
  journal={Journal of Atmospheric Sciences},
  volume={22},
  number={5},
  pages={541--558},
  year={1965}
}

@article{hignett1985_qjrms,
  title={A comparison of laboratory measurements and numerical simulations of baroclinic wave flows in a rotating cylindrical annulus},
  author={Hignett, By P and White, AA and Carter, RD and Jackson, WDN and Small, RM},
  journal={Quarterly Journal of the Royal Meteorological Society},
  volume={111},
  number={467},
  pages={131--154},
  year={1985},
  publisher={Wiley Online Library}
}

@article{maxworthy1994_jpo,
author="Maxworthy, T. and Narimousa, S.",
year=1994, 
title="Unsteady turbulent convection into a Homogeneous rotating fluid with oceanic applications",
journal="J. Phys. Ocean.",
volume=24,
pages="865-887"
 }

@article{weiss2010prl,
title={{Finite-Size Effects Lead to Supercritical Bifurcations in Turbulent Rotating Rayleigh-Bénard Convection}},
journal={Phys. Rev. Lett.},
author={Weiss, S. and Stevens, R.J.A.M and Zhong, J.-Q. and Clerx, H.J.H. and Lohse, D. and Ahlers, G.},
year=2010,
pages={224501},
volume=105
}

@article{maffei_et_al_2021,
	title={On the inverse cascade and flow speed scaling behaviour in rapidly rotating Rayleigh--B{\'e}nard convection},
	author={Maffei, Stefano and Krouss, Mitchell J and Julien, Keith and Calkins, Michael A},
	journal={Journal of Fluid Mechanics},
	volume={913},
	pages={A18},
	year={2021},
	publisher={Cambridge University Press}
}

@article{spiegel1960apj,
  title={On the Boussinesq approximation for a compressible fluid.},
  author={Spiegel, Edward A and Veronis, G},
  journal={Astrophys. J.},
  volume={131},
  pages={442},
  year={1960}
}

@article{gray1976ijhmt,
  title={The validity of the Boussinesq approximation for liquids and gases},
  author={Gray, Donald D and Giorgini, Aldo},
  journal={Int. J. Heat Mass Transf.},
  volume={19},
  number={5},
  pages={545--551},
  year={1976},
  publisher={Elsevier}
}

@article{chang2006gafd,
  title={{Convection in rotating annular channels heated from below. Part 2. Transitions from steady flow to turbulence}},
  author={Chang, Yingli and Liao, Xinhao and Zhang, Keke},
  journal={Geophys. Astrophys. Fluid Dyn.},
  volume={100},
  number={3},
  pages={215--241},
  year={2006},
  publisher={Taylor \& Francis}
}

@article{liao2006jfm,
  title={On boundary-layer convection in a rotating fluid layer},
  author={Liao, X and Zhang, K and Chang, Y},
  journal={J. Fluid Mech.},
  volume={549},
  pages={375--384},
  year={2006},
  publisher={Cambridge University Press}
}

@article{eckert2001ijhff,
  title = {{MHD Turbulence Measurements in a Sodium Channel Exposed to a Transverse Magnetic Field}},
  author = {Eckert, S and Gerbeth, G and Witke, W and Langenbrunner, H},
  pages = {358--364},
  year = 2001,
  journal = {Int. J. Heat Fluid Flow},
  number = 22
}

@Article{eltayeb1975jfm,
  author    = {Eltayeb, IA},
  title     = {Overstable hydromagnetic convection in a rotating fluid layer},
  journal   = {J. Fluid Mech.},
  year      = {1975},
  volume    = {71},
  number    = {01},
  pages     = {161--179},
  publisher = {Cambridge Univ Press},
}

@article{kolesnikov1974fd,
  title = {Experimental Investigation of Two-Dimensional Turbulence Behind a Grid},
  author = {Kolesnikov, Y. B. and Tsinober, A. B.},
  pages = {621--624},
  year = 1974,
  journal = {Fluid Dyn.},
  number = 4,
  volume = 9
}

@article{potherat2015jfm,
  title = {{The decay of wall-bounded MHD turbulence at low $Rm$}},
  author = {Poth\'erat, A and Kornet, K.},
  pages = {605--636},
  year = 2015,
  journal = {J. Fluid Mech.},
  doi = {doi:10.1017/jfm.2015.572},
  volume = 783,
}

@article{potherat2017prf,
            title="Do magnetic fields enhance turbulence at low magnetic {R}eynolds number?",
                author="Poth\'erat, A. and Klein, R.",
                year=2017,
                volume=2,
                number=6,
                pages=063702,
                journal="Phys. Rev. Fluids",
                doi="doi:10.1103/PhysRevFluids.2.063702",
}

@Article{potherat2014jfm,
  author  = {A. Poth\'erat and Klein, R.},
  title   = {{Why, how and when MHD turbulence at low $Rm$ becomes three-dimensional}},
  journal = {J. Fluid Mech.},
  year    = {2014},
  volume  = {761},
  pages   = {168-205},
  doi     = {10.1017/jfm.2014.620},
}

@article{sommeria1982jfm,
  title = {{Why, How and When MHD Turbulence Becomes Two-Dimensional}},
  author = {Sommeria, J. and Moreau, R.},
  pages = {507--518},
  year = 1982,
  journal = {J. Fluid Mech.},
  volume = 118
}

@article{shew2005pepi,
  author    = {Shew, Woodrow L and Lathrop, Daniel P},
  title     = {Liquid sodium model of geophysical core convection},
  journal   = {Phys. Earth Planet. Inter.},
  year      = {2005},
  volume    = {153},
  number    = {1-3},
  pages     = {136--149},
  publisher = {Elsevier}
}

@Article{cardin1994pepi,
 author    = {Cardin, Philippe and Olson, Peter},
 title     = {Chaotic thermal convection in a rapidly rotating spherical shell: consequences for flow in the outer core},
 journal   = {Phys. Earth Planet. Inter.},
 year      = {1994},
 volume    = {82},
 number    = {3-4},
 pages     = {235--259},
 publisher = {Elsevier},
}

@article{abbate2023gafd,
  title={{Rotating convective turbulence in moderate to high Prandtl number fluids}},
  author={Abbate, Jewel A and Aurnou, Jonathan M},
  journal={Geophys. Astrophys. Fluid Dyn.},
  volume={117},
  number={6},
  pages={397--436},
  year={2023},
}

@article{horn2022prsa,
author={Horn, Susanne and Aurnou, Jonathan M.},
year=2022,
title={The {E}lbert range of magnetostrophic convection. {I}. Linear theory},
journal={Proc. R. Soc. A.},
pages={47820220313.20220313},
doi={10.1098/rspa.2022.0313},
}

@article{horn2023prsa,
        author={Horn, S. and Aurnou, J.},
        title={The {E}lbert range of rotating magnetoconvection {II}: comparison between linear theory and turbulent {DNS}},
	volume=481,
       	number=2310,
       pages={20240016},
        journal={Proc. Roy. Soc. A},
        year=2025,
}

@article{sakuraba2002gafd,
        author = {Ataru Sakuraba},
        title = {Linear Magnetoconvection in Rotating Fluid Spheres Permeated by a Uniform Axial Magnetic Field},
        journal = {Geophys. Astrophys. Fluid Dyn.},
        volume = {96},
        number = {4},
        pages = {291-318},
        year = {2002},
        publisher = {Taylor \& Francis},
        doi = {10.1080/03091920290024234}
}

@article{hotta2018apjl,
journal={Astrophys. J. Lett.},
volume=860,
pages={L24},
number={2},
year=2018,
doi={doi:10.3847/2041-8213/aacafb},
title={Breaking {T}aylor-{P}roudman Balance by Magnetic Fields in Stellar Convection Zones},
author={Hotta, H.}
}

@article{baker2018prl,
        author={Baker, N.T. and Poth\'erat, A. and Davoust, L. and Debray, F.},
        title={Inverse and direct energy cascades in Three-Dimensional Magnetohydrodynamic turbulence at low Magnetic {R}eynolds Number},
        journal={Phys. Rev. Lett.},
        volume=120,
        pages=224502,
        year=2018,
        doi={doi:10.1103/PhysRevLett.120.224502}
}

@article{aujogue2015pf,
   author = "Aujogue, K. and Poth\'erat, A. and Sreenivasan, B.",
   title = "Onset of plane layer magnetoconvection at low {E}kman number",
   journal = "Phys. Fluids",
   year = "2015",
   volume = "27",
   number = "10",
   eid = 106602,
   pages = "106602",
   url = "http://scitation.aip.org/content/aip/journal/pof2/27/10/10.1063/1.4934532",
   doi = "http://dx.doi.org/10.1063/1.4934532"
}

@book{chandrasekhar1961,
  title = {Hydrodynamic and Hydromagnetic Stability},
  author = {Chandrasekhar},
  publisher = {Clarendon},
  year = 1961
}

@article{gillet2022pnas,
author = {Nicolas Gillet  and Felix Gerick  and Dominique Jault  and Tobias Schwaiger  and Julien Aubert  and Mathieu Istas },
title = {{Satellite magnetic data reveal interannual waves in Earth’s core}},
journal = {Proc. Natl. Acad. Sci.},
volume = {119},
number = {13},
pages = {e2115258119},
year = {2022},
}

@Article{gillet2010nat,
  author={Gillet, Nicolas and Jault, Dominique and Canet, Elisabeth and Fournier, Alexandre},
  title   = {Fast torsional waves and strong magnetic field within the Earth’s core},
  journal = {Nature},
  year    = {2010},
  volume  = {465},
  pages   = {74--77},
  doi     = {10.1038/nature09010},
}

@article{pais2008gji,
  title={{Quasi-geostrophic flows responsible for the secular variation of the Earth's magnetic field}},
  author={Pais, MA and Jault, Dominique},
  journal={Geophys. J. Int.},
  volume={173},
  number={2},
  pages={421--443},
  year={2008},
  publisher={Blackwell Publishing Ltd Oxford, UK}
}

@article{pais2014gji,
  title={Variability modes in core flows inverted from geomagnetic field models},
  author={Pais, Maria A and Morozova, Anna L and Schaeffer, Nathana{\"e}l},
  journal={Geophys. J. Int.},
  volume={200},
  number={1},
  pages={402--420},
  year={2014},
  publisher={Oxford University Press}
}

@article{gillet2019gji,
  title={A reduced stochastic model of core surface dynamics based on geodynamo simulations},
  author={Gillet, Nicolas and Huder, Loic and Aubert, Julien},
  journal={Geophys. J. Int.},
  volume={219},
  number={1},
  pages={522--539},
  year={2019},
  publisher={Oxford University Press}
}

@article{gillet2015jgr,
  title={{Planetary gyre, time-dependent eddies, torsional waves, and equatorial jets at the Earth's core surface}},
  author={Gillet, N and Jault, D and Finlay, CC},
  journal={J. Geophys. Res. Solid Earth},
  volume={120},
  number={6},
  pages={3991--4013},
  year={2015},
  publisher={Wiley Online Library}
}

@article{cardin1992grl,
  title={{An experimental approach to thermochemical convection in the Earth's core}},
  author={Cardin, Philippe and Olson, Peter},
  journal={Geophys. Res. Lett.},
  volume={19},
  number={20},
  pages={1995--1998},
  year={1992},
  publisher={Wiley Online Library}
}

@article{Ning1993,
  title={{Rotating Rayleigh-B{\'e}nard convection: Aspect-ratio dependence of the initial bifurcations}},
  author={Ning, Li and Ecke, Robert E},
  journal={Phys. Rev. E},
  volume={47},
  number={5},
  pages={3326},
  year={1993},
  publisher={APS}
}

@article{Zhong1991,
  title={{Asymmetric modes and the transition to vortex structures in rotating Rayleigh-B{\'e}nard convection}},
  author={Zhong, Fang and Ecke, Robert and Steinberg, Victor},
  journal={Phys. Rev. Lett.},
  volume={67},
  number={18},
  pages={2473},
  year={1991},
  publisher={APS}
}

@article{Zhong1993,
  title={{Rotating Rayleigh--B{\'e}nard convection: asymmetric modes and vortex states}},
  author={Zhong, Fang and Ecke, Robert E and Steinberg, Victor},
  journal={J. Fluid Mech.},
  volume={249},
  pages={135--159},
  year={1993},
  publisher={Cambridge University Press}
}

@Article{aujogue2016rsi,
  author    = {Aujogue, K{\'e}lig and Poth{\'e}rat, Alban and Bates, Ian and Debray, Fran{\c{c}}ois and Sreenivasan, Binod},
  title     = {{Little Earth Experiment: An instrument to model planetary cores}},
  journal   = {Rev. Sci. Instrum.},
  year      = {2016},
  volume    = {87},
  number    = {8},
  publisher = {AIP Publishing},
}

@article{livermore2017natgeo,
  title={{An accelerating high-latitude jet in Earth’s core}},
  author={Livermore, Philip W and Hollerbach, Rainer and Finlay, Christopher C},
  journal={Nat. Geosci.},
  volume={10},
  number={1},
  pages={62--68},
  year={2017},
  publisher={Nature Publishing Group UK London}
}

@article{terrienetal2023,
author={Terrien, Louise and Favier, Benjamin and Knobloch, Edgar},
year=2023, 
title={{Suppression of Wall Modes in Rapidly Rotating Rayleigh-B\'enard Convection by Narrow Horizontal Fins}}, 
journal={Phys. Rev. Lett.}, 
volume={130}, 
pages=174002
}

@article{hornschmid2017,
author={Horn, S. and Schmid, P.J.}, 
year=2017, 
title={{Prograde, retrograde, and oscillatory modes in rotating Rayleigh-B\'enard convection}}, 
journal={J. Fluid Mech.}, 
volume={831}, 
pages={182-211}
}

@Article{busse1970jfm,
  author    = {Busse, Friedrich H},
  title     = {Thermal instabilities in rapidly rotating systems},
  journal   = {J. Fluid Mech.},
  year      = {1970},
  volume    = {44},
  number    = {3},
  pages     = {441--460},
  publisher = {Cambridge University Press},
}

@article{potherat2024prl,
  title = {Magnetic Taylor-Proudman Constraint Explains Flows into the Tangent Cylinder},
  author = {Poth\'erat, Alban and Aujogue, K\'elig and Debray, Fran\ifmmode \mbox{\c{c}}\else \c{c}\fi{}ois},
  journal = {Phys. Rev. Lett.},
  volume = {133},
  issue = {18},
numpages = {6},
  year = {2024},
  month = {Oct},
  publisher = {American Physical Society},
  doi = {10.1103/PhysRevLett.133.184101},
  url = {https://link.aps.org/doi/10.1103/PhysRevLett.133.184101}
}

@article{potherat2024crphys,
	author={Poth\'erat, A. and Horn, S.},
	title={Seven decades of exploring planetary interiors with rotating convection experiments},
	journal={invited review, Special issue "Geophysical and Astrophysical Fluid Dynamics in the Laboratory", C.R. Phys.},
       volume=25,
      number=S3, 
      pages={21--75},
     year=2024
}

@Article{olson2011pepi,
  author    = {Olson, Peter},
  title     = {Laboratory experiments on the dynamics of the core},
  journal   = {Phys. Earth Planet. Inter.},
  year      = {2011},
  volume    = {187},
  number    = {3-4},
  pages     = {139--156},
  publisher = {Elsevier},
}

@Article{olson2013areps,
  author    = {Olson, Peter},
  title     = {Experimental dynamos and the dynamics of planetary cores},
  journal   = {Annu. Rev. Earth Planet. Sci.},
  year      = {2013},
  volume    = {41},
  pages     = {153--181},
  publisher = {Annual Reviews},
}

@InCollection{cardin2015tg,
  author    = {P. Cardin and P. Olson},
  title     = {8.13 - Experiments on Core Dynamics},
  booktitle = {Treatise on Geophysics (Second Edition)},
  publisher = {Elsevier},
  year      = {2015},
  editor    = {Gerald Schubert},
  pages     = {317-339},
  address   = {Oxford},
  edition   = {Second Edition}, 
}

@Article{Sakai1997,
  author    = {Sakai, Satoshi},
  title     = {The horizontal scale of rotating convection in the geostrophic regime},
  journal   = {J. Fluid Mech.},
  year      = {1997},
  volume    = {333},
  pages     = {85--95},
  publisher = {Cambridge University Press},
}

@article{cao2018pnas,
        journal={Proc. Natl. Acad. Sci.},
        title={Geomagnetic polar minima do not arise from steady meridional circulation},
        author={Cao, H. and Yadav, Rakesh K. and Aurnou, Jonathan M.},
        year=2018,
        volume=115,
        number=44,
       pages={11186-11191},
        doi={doi:10.1073/pnas.1717454115}
}

@article{finlay2023nat,
author={Finlay, C.C. and Gillet, N. and Aubert, J. and Livermore, P.W. and Jault, D. },
title={{Gyres, jets and waves in the Earth’s core}},
journal={Nat. Rev. Earth Environ.},
volume=4,
pages={377-392},
year=2023,
doi={https://doi.org/10.1038/s43017-023-00425-w}
}

@book{greenspan1969,
  title = {The theory of rotating fluids},
  author = {Greenspan, H. P.},
  publisher = {Cambridge University Press},
  year = 1969
}

@article{proudman1916prsa,
author = {Proudman, J.},
title = {On the motion of solids in a liquid possessing vorticity},
year = {1916},
journal = {Proc. R. Soc. Lond. A},
volume = {92},
number = {642},
pages = {408 – 424},
}

@article{taylor1917prsa,
title={Motion of solids in fluids when the flow is not irrotational},
author={Taylor, Geoffrey Ingram},
journal={Proc. Roy. Soc. Lond. A},
volume={93},
number={648},
pages={99--113},
year={1917},
publisher={The Royal Society London}
}

@article{aguirre-guzman2021jfm, 
title={{Force balance in rapidly rotating Rayleigh–Bénard convection}}, 
volume={928}, 
DOI={10.1017/jfm.2021.802}, 
journal={J. Fluid Mech.}, 
author={Aguirre Guzmán, Andr\'es J. and Madonia, Matteo and Cheng, Jonathan S. and Ostilla-M\'onico, Rodolfo and Clercx, Herman J.H. and Kunnen, Rudie P.J.},
year={2021}, 
pages={A16}
}

@article{hanasogeetal2012,
 author={Hanasoge, Shravan M and Duvall Jr, Thomas L and Sreenivasan, Katepalli R},
year=2012, 
title={Anomalously weak solar convection}, 
journal={Proc. Natl. Acad. Sci.}, 
volume=109,
number=30, 
pages={11  928-11932}
}

@article{zhangliao2009,
author = {Zhang, K. and Liao, X.},
year=2009, 
title={The onset of convection in rotating circular cylinders with experimental boundary conditions},
journal = {J. Fluid Mech.},
volume = {622},
pages = {63-73},
}

@article{hornaurnou2018,
author={Horn, S. and Aurnou, J.M.},
year=2018, 
title={{Regimes of Coriolis-centrifugal convection}},
address = {\textbf{120}, 204502},
journal = {Phys. Rev. Lett.},
volume={120},
number={20},
pages={204502},
year={2018},
}

@article{hornaurnou2019,
author={Horn, S and Aurnou, J.M.}, 
year=2019, 
title={Rotating convection with centrifugal buoyancy: numerical predictions for laboratory experiments},
address = {\textbf{4}, 073501},
journal = {Phys. Rev. Fluids},
volume={4},
number={7},
pages={073501},
year={2019},
}

@article{eckeshishkina2023,
author={Ecke, R.E. and Shishkina, O.}, 
year=2023, 
title={{Turbulent rotating Rayleigh–Bénard convection}},
journal = {Annu. Rev. Fluid Mech.},
volume = {55},
pages = {603-638},
}

@article{zhongetal2009,
author={Zhong, Jin-Qiang and Stevens, Richard JAM and Clercx, Herman JH and Verzicco, Roberto and Lohse, <? format?> Detlef and Ahlers, Guenter},
year=2009, 
title={{Prandtl-, Rayleigh-, and Rossby-number dependence of heat transport in turbulent rotating Rayleigh-Bénard convection}},
volume={102},
number={4},
pages={044502},
year={2009},
journal = {Phys. Rev. Lett.},
}

@article{christensen2002,
author={Christensen, U.R.},
title = {Zonal flow driven by strongly supercritical convection in rotating spherical shells},
journal = {J. Fluid Mech.},
year = {2002},
volume = {470},
pages = {115-133},
}

@article{christensenaubert2006,
author={Christensen, U.R. and Aubert, J.}, 
year=2006, 
title={Scaling properties of convection-driven dynamos in rotating spherical shells and application to planetary magnetic fields},
journal = {Geophys. J. Int.},
volume = {166.1},
pages = {97-114},
}

@article{britoetal2004,
author={Brito, D. and Aurnou, J.M. and Cardin, P.}, 
year=2004, 
title={Turbulent viscosity measurements relevant to planetary core-mantle dynamics},
volume = {141}, 
pages={3–8},
journal = {Phys. Earth Planet. Int.}
}

@article{aurnou2007,
author={Aurnou, J.M.},
title = {Planetary core dynamics and convective heat transfer scaling},
journal = {Geophys. Astrophys. Fluid Dyn.},
year = {2007},
volume = {101.5-6},
pages = {327-345},
}

@article{chengaurnou2016,
author={Cheng, J.S. and Aurnou, J.M.},
year=2016, 
title={Tests of diffusion-free scaling behaviors in numerical dynamo datasets},
journal = {Phys. Earth Planet. Int},
volume = {436},
pages = {121-129},
}

@article{bouillautetal2021,
	author={Bouillaut, Vincent and Miquel, Benjamin and Julien, Keith and Auma{\^\i}tre, S{\'e}bastien and Gallet, Basile},
year=2021, 
title={Experimental observation of the geostrophic turbulence regime of rapidly rotating convection},
volume = {118},
number={44}, 
pages={e2105015118},
journal = {Proc. Natl. Acad. Sci.},
}

@article{homsyhudson1971,
author={Homsy, G.M. and Hudson, J.L.},
year=1971, 
title={Centrifugal convection and its effect on the asymptotic stability of a bounded rotating fluid heated from below},
journal = {J. Fluid Mech.},
volume = {48.3},
pages = {605-624},
}

@article{kunnenetal2006,
author={Kunnen, RPJ and Clercx, HJH and Geurts, BJ},
year=2006, 
title={Heat flux intensification by vortical flow localization in rotating convection},
volume =74,
number=5, 
pages=056306,
journal = {Phys. Rev. E}
}

@article{hartetal2002,
author={Hart, JE and Kittelman, S and Ohlsen, DR},
year=2002, 
title={Mean flow precession and temperature probability density functions in turbulent rotating convection},
journal = {Phys. Fluids},
volume={14},
number={3},
pages={955--962},
year={2002},
}

@article{eckeetal2022,
author={Ecke, Robert E and Zhang, Xuan and Shishkina, Olga},
year=2022, 
title={{Connecting wall modes and boundary zonal flows in rotating Rayleigh-Bénard convection}},
volume = {7},
number=1, 
pages={L011501},
journal = {Phys. Rev. Fluids}
}

@article{chengetal2018,
author={Cheng, Jonathan S and Aurnou, Jonathan M and Julien, Keith and Kunnen, Rudie PJ},
year=2018, 
title={A heuristic framework for next-generation models of geostrophic convective turbulence},
journal = {Geophys. Astrophys. Fluid Dyn.},
volume = {112.4},
pages = {277-300},
}

@article{cuistreet2001,
author={Cui, A. and Street, R.L.}, 
year=2001, 
title={Large-eddy simulation of turbulent rotating convective flow development},
journal = {J. Fluid Mech.},
volume = {447},
pages = {53-84},
}

@article{jacobsivey1998,
author={Jacobs, P. and Ivey, G.N.}, 
year=1998, 
title={Large-eddy simulation of turbulent rotating convective flow development},
journal = {J. Fluid Mech.},
volume = {369},
pages = {23-48},
}

@article{favierknobloch2020,
author={Favier, B. and Knobloch, E.}, 
year=2020, 
title={{Robust wall states in rapidly rotating Rayleigh–Bénard convection}},
volume={895}, 
pages={R1},
journal = {J. Fluid Mech.}
}

@article{julienetal2012,
author={Julien, Keith and Rubio, Antonio M and Grooms, Ian and Knobloch, Edgar},
year=2012, 
title={{Statistical and physical balances in low Rossby number Rayleigh–Bénard convection}},
journal = {Geophys. Astrophys. Fluid Dyn.},
 volume={106},
number={4-5},
pages={392--428},
}

@article{aurnou2020prf,
  title={Connections between nonrotating, slowly rotating, and rapidly rotating turbulent convection transport scalings},
  author={Aurnou, Jonathan M. and Horn, Susanne and Julien, Keith},
  journal={Phys. Rev. Res.},
  volume = {2},
  issue = {4},
  pages = {043115},
  numpages = {13},
  year = {2020},
  month = {Oct},
  publisher = {American Physical Society},
  doi = {10.1103/PhysRevResearch.2.043115}
}

@article{kingetal2013,
author={King, E.M. and Stellmach, S. and Buffett, B.}, 
year=2013, 
title={{Scaling behaviour in Rayleigh–Bénard convection with and without rotation}}, 
journal={J. Fluid Mech.}, 
volume={717}, 
pages={449-471}
}

@article{konoroberts2002,
author={Kono, M and Roberts, PH},
year= 2002, 
title={Recent geodynamo simulations and observations of the geomagnetic field}, 
journal={Rev. G  eophys.}, 
volume=40,
number=4,
pages={4-1}
}

@book{gubbinsemilio2007,
author={Gubbins, D. and Emilio, H.B.}, 
year=2007, 
title={Encyclopedia of geomagnetism and paleomagnetism}, 
publisher={Springer Science \& Business Media}
}

@article{schubert2011pepi,
  title={{Planetary magnetic fields: Observations and models}},
  author={Schubert, G. and Soderlund, K.M.},
  journal={Phys. Earth Planet. Inter.},
  volume={187},
  number={3},
  pages={92--108},
  year={2011},
  publisher={Elsevier}
}

@article{aurnouetal2003,
	title={{Experiments on convection in Earth’s core tangent cylinder}},
	author={Aurnou, Jonathan and Andreadis, Steven and Zhu, Lixin and Olson, Peter},
	journal={Earth Planet. Sci. Lett.},
	volume={212},
	number={1-2},
	pages={119--134},
	year={2003},
	publisher={Elsevier}
}

@article{aurnouolson2001,
	title={{Experiments on Rayleigh--B{\'e}nard convection, magnetoconvection and rotating magnetoconvection in liquid gallium}},
	author={Aurnou, JM and Olson, PL},
	journal={J. Fluid Mech.},
	volume={430},
	pages={283--307},
	year={2001},
	publisher={Cambridge University Press}
}

@article{aujogueetal2018,
	title={Experimental study of the convection in a rotating tangent cylinder},
	author={Aujogue, K{\'e}lig and Poth{\'e}rat, Alban and Sreenivasan, Binod and Debray, Fran{\c{c}}ois},
	journal={J. Fluid Mech.},
	volume={843},
	pages={355--381},
	year={2018},
	publisher={Cambridge University Press}
}

@article{schaefferetal2017,
	title={{Turbulent geodynamo simulations: a leap towards Earth’s core}},
	author={Schaeffer, N. and Jault, D. and Nataf, H.-C. and Fournier, A.},
	journal={Geophys. J. Int.},
	volume={211},
	number={1},
	pages={1--29},
	year={2017},
	publisher={Oxford University Press}
}

@article{kunnen2021,
	title={The geostrophic regime of rapidly rotating turbulent convection},
	author={Kunnen, Rudie PJ},
	journal={J. Turbul.},
	volume={22},
	number={4-5},
	pages={267--296},
	year={2021},
	publisher={Taylor \& Francis}
}

@article{luetal2021,
	title={Heat-transport scaling and transition in geostrophic rotating convection with varying aspect ratio},
	author={Lu, Hao-Yuan and Ding, Guang-Yu and Shi, Jun-Qiang and Xia, Ke-Qing and Zhong, Jin-Qiang and others},
	journal={Phys. Rev. Fluids},
	volume={6},
	number={7},
	pages={L071501},
	year={2021},
	publisher={APS}
}

@article{chengetal2015,
	title={Laboratory-numerical models of rapidly rotating convection in planetary cores},
	author={Cheng, Jonathan S and Stellmach, Stephan and Ribeiro, Adolfo and Grannan, Alexander and King, Eric M and Aurnou, Jonathan M},
	journal={Geophys. J. Int.},
	volume={201},
	number={1},
	pages={1--17},
	year={2015},
	publisher={Oxford University Press}
}

@article{julienetal1996,
	title={{Rapidly rotating turbulent Rayleigh-B{\'e}nard convection}},
	author={Julien, K and Legg, S and McWilliams, J and Werne, J},
	journal={J. Fluid Mech.},
	volume={322},
	pages={243--273},
	year={1996},
	publisher={Cambridge University Press}
}

@article{rossby1969,
	title={{A study of B{\'e}nard convection with and without rotation}},
	author={Rossby, HT},
	journal={J. Fluid Mech.},
	volume={36},
	number={2},
	pages={309--335},
	year={1969},
	publisher={Cambridge University Press}
}
\end{document}